\documentclass[preprint,12pt,sort&compress]{elsarticle}
\usepackage{amssymb}
\usepackage{amsthm}
\usepackage{amsfonts}
\usepackage{amsmath,bm} 
\usepackage{gensymb}	
\usepackage{graphicx}
\usepackage{xcolor}
\usepackage{dcolumn}
\usepackage{booktabs}

\usepackage{lmodern}           
\usepackage[T1]{fontenc}       
\usepackage[cp1250]{inputenc}  
\usepackage{setspace}
\usepackage{mdframed}
\usepackage{enumitem}		   
\usepackage{subcaption}		   
\usepackage{hyperref}
\usepackage{changes}		   
\journal{Carbon}

\usepackage[demo]{adjustbox}
\usepackage{xcolor}
\usepackage{atbegshi}
\AtBeginDocument{\AtBeginShipoutNext{\AtBeginShipoutDiscard}}	
\PassOptionsToPackage{hyphens}{url}\usepackage{hyperref} 

\begin{document}

	\pagestyle{empty} 
\begin{titlepage}
	\color[rgb]{.4,.4,1}
	\hspace{5mm}

	\bigskip
	
	\hspace{15mm}
	\begin{minipage}{10mm}
		\color[rgb]{.7,.7,1}
		\rule{1pt}{226mm}
	\end{minipage}
	\begin{minipage}{133mm}
		\vspace{10mm}        
		\color{black}
		\sffamily
		\LARGE\bfseries Deep learning framework \\[-0.3\baselineskip] for carbon nanotubes: \\[-0.3\baselineskip] mechanical properties and \\[-0.3\baselineskip] modeling strategies     \\[-0.3\baselineskip] 
		
		\vspace{5mm}
		{\large {Preprint of the article published in \\[-0.4\baselineskip] Carbon (2021) }} 
		
		\vspace{10mm}        
		{\large Marko \v{C}ana\dj{}ija } 
		
		\large
		
		\vspace{40mm}
		\vspace{5mm}
		
		\small
		\url{https://doi.org/10.1016/j.carbon.2021.08.091}
		
		\textcircled{c} 2021. This manuscript version is made available under the CC-BY-NC-ND 4.0 license \url{http://creativecommons.org/licenses/by-nc-nd/4.0/}
		\hspace{30mm} 
		\color[rgb]{.4,.4,1} 
		\includegraphics[width=3cm]{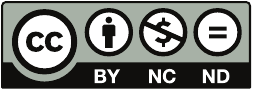}        
	\end{minipage}
\end{titlepage}

\begin{frontmatter}

\title{Deep learning framework for carbon nanotubes: mechanical properties and modeling strategies}

\author[URI]{Marko \v{C}ana\dj{}ija \corref{cor}}
\ead{marko.canadija@riteh.hr}
\cortext[cor]{Corresponding author. Tel.: +385-51-651-496; Fax.: +385-51-651-490}

\address[URI]{University of Rijeka, Faculty of Engineering, Department of Engineering Mechanics, Vukovarska 58, 51000 Rijeka, Croatia}

\begin{abstract}
Tensile tests at room temperature are performed using molecular dynamics on all configurations of single-walled carbon nanotubes up to 4 nm in diameter. Distributions of the Young's modulus, Poisson's ratio, ultimate tensile strength and fracture strain are determined and reported. The results show that the chirality of the nanotube has the greatest influence on the properties. An artificial neural network is developed for the dataset obtained by molecular dynamics and used to predict the mechanical properties. It is clearly shown that Deep Learning provides accurate predictions, with the further advantage that thermal fluctuations are smoothed out. In addition, a through analysis of the effect of dataset size on prediction quality is performed, providing modeling strategies for further researchers.
\end{abstract}

\end{frontmatter}

\section{Introduction, motivation and outline}
Molecular dynamics (MD) simulations of the mechanical behavior of carbon nanotubes (CNTs) were at the peak of interest in the research community about 10-15 years ago \cite{yakobson1996nanomechanics, Belytschko2002, liew2004, Chowdhury2012, mylvaganam2004important, Ma2010, Kok2016, Chowdhury2010, Zhao2010, Agrawal2008, dereli2007temperature, Fu2007, gupta2005elastic, Zhang2004, Mielke2004, Troya2003, Zhou2002, Yu2000a, lu1997elastic}. Nevertheless, there are some recent contributions that give an indication of still unresolved issues on carbon nanotubes, both in MD and experiments. First, although significant progress has been made in the experimental analysis of CNTs, there is still a lack of benchmarks that can be used for MD verification. The majority of the MD works refer to the classical work \cite{Yu2000a} with experimental tensile tests on multi-walled carbon nanotube (MWCNT) ropes. For instance, \cite{Jhon2016} recently attempted to bring MD results closer to these experiments by introducing defects into the nanotube structure. The other problem is that most of the existing studies focus solely on MD modeling aspects. The available models are typically tested on a small selection of single-walled carbon nanotube (SWCNT) configurations, usually of the armchair or zigzag type. Chiral SWCNTs are less commonly analyzed. This problem is pointed out in another recent paper \cite{Yazdani2017} - a thorough investigation of the mechanical behavior of CNTs is still missing. In this direction, \cite{Yazdani2017} chooses a wider range of SWCNTs than usual and tests them under tension and torsion at different temperatures. Temperature in MD simulations should be treated very carefully due to the involved thermal fluctuations of the carbon atoms, which complicates efforts to provide a single value as a stochastic representative of a property at a given temperature. All this leads toward the starting point for the present research: to obtain a set of comprehensive results on the mechanical properties of SWCNTs at room temperature.

One possible way to obtain such an estimate at a given temperature is to use a simple averaging procedure. A given model can be run several times, and the results obtained are then averaged, and the standard deviation is provided. However, as the number of different SWCNT configurations increases, this procedure becomes somewhat impractical due to the associated computational costs. At this point, we strongly believe that deep learning methods can be helpful in reducing such computational costs.

Deep Learning (DL), applied to artificial neural networks, is nowadays a cornerstone of machine learning (ML). The paper \cite{hinton2006fast} is generally considered as the beginning of the modern DL approach. Since then, DL has revolutionized ML and subsequently computer science, making its way into general public and high-tech industry under the broader and popular term of artificial intelligence. However, when it comes to the subject of MD simulations of carbon nanotubes, it is still used rather sparingly. Among those, \cite{forster2020deep} shows that chiral indices can be successfully determined by applying DL to electron microscopy images of CNTs. To obtain a realistic reference dataset, a series of 261 CNTs with different chiralities were simulated using MD. Each individual configuration is equilibrated at 300 K and 1000 independent runs were performed. To allow fast calculations, the Tersoff potential was used. Then, convolutional neural networks (CNN) were employed for image analysis and CNT classification. An overall accuracy of 90.5\% was achieved. Although not related to CNTs but to graphene, \cite{dewapriya2020characterizing} demonstrates the ability of Deep CNN to predict the fracture stress in defective graphene samples over a wide temperature range. Datasets from MD of different sizes (250, 500 and 1000 samples) are used to train the network. It is shown that even the shallow neural network can predict the fracture stress of graphene. More complex behavior involving distribution of defects required deep neural networks (DNN). A further contribution to carbon materials-carbon fibers uses ML to predict the ultimate tensile strength and Young's modulus \cite{shirolkar2021investigating} and achieves $R^2$ values of 0.85 and 0.67 for the latter properties, respectively. MD simulations on MWCNT were performed in \cite{xiang2020machine}. Number of walls, chirality, crosslink density and diameter were varied. The effect of thermal fluctuations was avoided after equilibration by reducing the temperature to 1 K. An iterative procedure using MD and genetic algorithm was applied to 1000 randomly selected samples to find the configuration with the highest tensile strength. The obtained errors in the predicted ultimate tensile strengths were up to 5\%. The dimensionality of the obtained results was then reduced by a self-organizing map. Finally, it is also found that the armchair CNTs can withstand higher tensile stresses and strains than zigzag CNTs. Further benefits of dimensionality reduction are demonstrated in \cite{pathrudkar2020interpretable}, again on MWCNTs. Developed DNN predicts the behavior of an unknown MWCNTs configuration with high accuracy. Multi-gene genetic programming can also be used to model the compression behavior of CNTs \cite{vijayaraghavan2014estimation}. Further illustrations of ML applications involving carbon include estimation of shear strength of carbon nanotube-polymer interfaces \cite{rahman2021} based on MD simulations, macroscopic delamination of carbon fibre based composites \cite{zhang2021} or prediction of adsorption energies of methane species on Cu-based alloys as an alternative to DFT calculations \cite{zhang2021b}.

The above brief review of the state of the art shows that there is little DL research on the mechanical behavior of carbon nanostructures. The present research work aims to contribute to this field in two ways. Firstly, a complete set of MD simulations of tensile tests on SWCNTs is presented, covering all possible SWCNT configurations for diameters up to 4 nm. Special care is taken to document all details of the simulations. As a result, distributions of the Young's modulus, Poisson's ratio, ultimate tensile strength and fracture strain with respect to the chirality indices are provided. Such a detailed overview of the mechanical properties is presented for the first time in the literature. Based on these results, a comprehensive set of DL simulations was carried out with the aim of establishing a realistic predictive model of the mechanical properties of SWCNT at room temperature. As further research of this nature can be envisaged, and in order to avoid excessive computational costs associated with accompanying MD calculations, guidance is provided on the selection of smaller, yet representative, datasets. Finally, the Closing remarks section summarizes the main findings of the research.

\section{Molecular Dynamics Model}
\label{sec_MD_model}

\subsection{Choice of potential}
\label{subsec:potential}
The choice of the potential used to describe intermolecular interactions has a significant impact on the calculated properties. In the present case, the selection was motivated by a recent work \cite{qian2021comprehensive} in which eight widely used potentials were analyzed in detail and compared with density functional theory (DFT). The main findings favor machine learning interatomic potentials over carbon bond order potentials, with the recent GAP -20 potential \cite{rowe2020accurate} considered the best candidate. Nevertheless, to compare the present results to existing ones, a carbon bond order potential is chosen, in particular the modified AIREBO \cite{Shenderova2000}. All potentials have some kind of drawback and the published results are contradictory in some aspects \cite{lebedeva2019elastic, qian2021comprehensive, jensen2015simulation}, but the modified AIREBO is the only potential among the considered carbon bond order potentials that does not exhibit unphysical stress increase  at fracture \cite{qian2021comprehensive}. Although it is known \cite{lebedeva2019elastic} that modified AIREBO, as well as some other potentials, exhibits more nonlinear stress-strain behavior than that observed by DFT, the overall prediction results of mechanical properties for CNTs \cite{qian2021comprehensive} (Tab. IX), ranks it second to the GAP-20 potential, with DFT results serving as a reference.

Thus, in the present case the modified AIREBO potential is used to capture the physically realistic behavior at fracture. As noted in \cite{Shenderova2000} by the authors of the original REBO potential, the cutoff function introduces a sharp increase in the interatomic forces as the fracture point is approached \cite{mylvaganam2004important}. In this way, the breaking force is significantly overestimated. Although the exact details of the implementation are not given in \cite{mylvaganam2004important}, the modification relies on moving the cutoff distance far from the inflection point. To circumvent this issue in the present research, the cutoff radius used in the interatomic potential was set to 2 $\AA$. A similar approach was also used in \cite{Chowdhury2010, Chowdhury2012,Mielke2004,Belytschko02,Khare2007,Troya2003,Zhang2005,Zhang2004,Zhao2010,Jhon2016}, among others. For instance, in the case of zigzag CNTs  at room temperature, this reduces the strain at fracture to about 0.10-0.15 and the ultimate tensile strength to 80-120 GPa. When the original potential was used \cite{Yakobson1997,Zhou2002,Fu2007}, the strain at fracture is within the range of 0.35-0.45 and the tensile strength is up to 400 GPa for a (10,10) SWCNT \cite{Hirai2003}. This is also more consistent with experiments on MWCNT \cite{Yu2000}, that also show lower values: maximal strain at fracture 0.13 and tensile stresses up to 40 GPa. Since the choice of CNT thickness remains to be somewhat arbitrary among various research groups, it should be noted that all stresses reported above were calculated with the CNT thickness 0.34 nm.

\subsection{MD - structures and boundary conditions}
\label{subsec:MD_model}
The preprocessing step involved the generation of the respective SWCNT configurations by specifying positions of the atoms \cite{scikit}. Only pristine SWCNTs with open ends were considered. In all cases, the SWCNT atoms were divided into three groups. Each SWCNT end formed one group and was used to enforce the boundary conditions. All atoms positioned within 5 $\AA$ of the end were considered part of these groups. Displacements of atoms belonging to one of these groups were constrained from moving, which provided a support for the SWCNT during the tensile test. The velocity of the atoms of the other group was prescribed and served as a source of the tensile load on SWCNT. In all cases considered, the prescribed velocity was adjusted to match the strain rate 0.001 ps$^{-1}$ throughout the simulation. Atoms that did not belong to either of the above two groups were located in the middle part of the SWCNT and formed the largest third group. The middle part was able to stretch, and its ratio of length to initial diameter was approximately equal to 5. To obtain a complete picture of the nanotube behavior, SWCNTs with theoretical diameters $D_\mathrm{th}=\frac{b\sqrt{3}}{\pi}\sqrt{n^2+nm+m^2}$ ranging from 0.36 nm for the smallest SWCNT (4,1) to 3.916 nm for the largest SWCNTs (50,2) were considered. The lengths of the SWCNTs ranged from 1.899 to 19.836 nm. A pair of natural numbers $(n,m)\in\mathbb{N}$ defining the chirality of the CNTs were in the range of $n\in\lbrace 3,..,51 \rbrace$, $m\in\lbrace 0,..,35 \rbrace$, $n \ge m$ and $0.36 \text{ nm} \le D_\mathrm{th}\le 4 \text{ nm}$, so a total of 818 different SWCNT configurations were analyzed. This also means that chiral angles are $0\degree \le \Theta \le 30\degree$. Since the diameter of a SWCNT is typically 1.0-2.0 nm \cite{Ma2010}, this should cover all nanotubes with regular dimensions, as well as all smaller and some larger ones. The number of atoms involved in the calculations ranged from 126 for (4,1) to 10097 for (50,2) nanotubes.

\subsection{MD - analysis and setup}
\label{subsec:MD_analysis}
The present simulations were carried out in LAMMPS \cite{Plimpton1995}. The first step of the simulation was the energy minimization of the carbon nanotube, which was performed on the provided initial configuration. The tolerance criteria used were $10^{-10}$ eV for the energy and $10^{-10}$ eV/$\AA$ for the force. The simulation box boundaries were considered to be non-periodic and fixed. When the minimum configuration was reached, the equilibration of the structure was performed. During this phase, the temperature of the system was increased from 0 K to 300 K. The equilibration consisted of 50000 time steps, where the duration of each step was 0.5 fs. The total duration of the equilibration was 25 ps. The duration of the time step was chosen following the recommendation in \cite{mylvaganam2004important} that the time step should be less than 10\% of the vibration period of an atom. This is also consistent with the choice of time step by other authors \cite{Fu2007,Jhon2016}. The Nose-Hoover thermostat was used. During the equilibration phase, both ends of the CNT were held fixed. This effectively ensures that support forces are co-linear and not repositioned due to self-excitation by thermal vibrations of the atoms. After stabilizing the temperature, tensile stretching of the specimen was enforced, also using the same thermostat. 

Finally, although the stochastic nature of the thermal fluctuations of the atoms cannot be avoided at nonzero temperatures, the influence can be minimized. In the present study, each configuration was tested for three different random initial states and the obtained results were averaged. Thus, a total of 2454 MD simulations were performed. 

\subsection{Postprocessing}
\label{subsec:MD_Post}
The change in diameter was monitored at the ring of atoms closest to the middle cross-section of the SWCNT throughout the complete simulation. The distance from the centroid of the ring was averaged to obtain the current radius. Elongation was calculated with respect to the rings of the group of atoms used for support and loading, located adjacent to the central and deformable part of the SWCNT.

The virial stresses of the central part were homogenized and used as the average virial stress in an axially loaded SWCNT. These stresses are then multiplied by the current volume of the specimen. The current volume was approximated as a hollow cylinder with the current length of the central part of the SWCNT and the current diameter at the middle of the specimen. The thickness was assumed to be $t=0.34$ nm, which seems to be the dominant choice in the literature. In the line with this, the stresses calculated in this way are the true stresses. Such an approach effectively cancels bending stresses occurring in the bending mode self-excited by thermal vibrations, as observed in early papers \cite{yakobson1996nanomechanics,overney1993structural}. Remaining stresses can then be attributed to axial loading. Note that this is different from most approaches by other authors who use the theoretical diameter $D_\mathrm{th}$, and therefore report engineering stresses. Changes in the diameter due to minimization and equilibration at a given temperature are not included in approaches based on $D_\mathrm{th}$. Therefore, $D_\mathrm{th}$ differs from the actual initial diameter that CNT acquires after the minimization and equilibration procedures, even before further variations due to the contractions caused by tensile loading take place, Fig.~\ref{fig:app_D_vs_Dth}. The strain was calculated as usual, from the current elongation divided by the initial length of the central part.

The Young's modulus is usually calculated in two ways. One approach is based on the potential energy \cite{gupta2005elastic} and the other on the linear regression analysis of the initial part of the stress-strain curve \cite{liew2004}. In the vast majority of cases, the exact description of the range used in the calculation is not given. In the present case, the standard linear regression procedure was used, using the stresses and strains described above. The proportionality limit is assumed to be $\varepsilon=0.01$. Finally, the ultimate tensile strength (UTS, $\sigma_\mathrm{max}$) and fracture strain $\epsilon_\mathrm{max}$ are determined and reported.

\subsection{MD results and discussion}
\label{subsec:MD_results}
\begin{figure}
	\centering
	\includegraphics[scale=0.6]{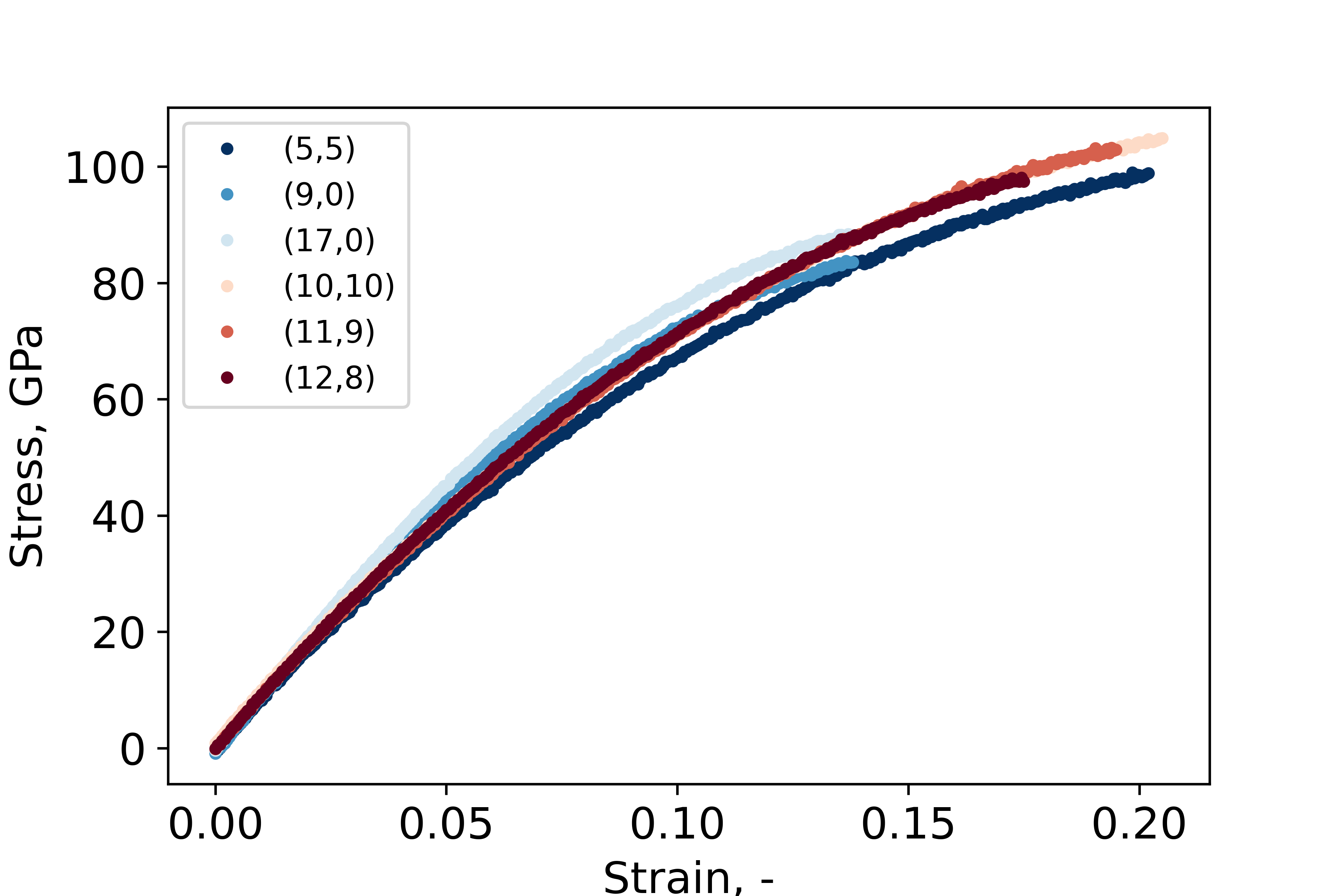}
	\caption{A selection of tensile stress-strain curves for configurations listed in Tab.~\ref{tab:MD_comparison}.} 
	\label{fig:StressVerif}
\end{figure}

First, and to ensure that the results obtained are consistent with those already published by other authors, several popular benchmark SWCNT configurations are selected and results are compared, Tab.~\ref{tab:MD_comparison}. The selection includes armchair, zigzag and chiral SWCNTs with small, medium and large diameters. It is difficult to make the comparison rigorously, since the models used in the literature differ in many details. This starts from the choice of potential, length, loading rates, nanotube thickness to computational details. Moreover, most results reported in the literature use the theoretical diameter $D_\mathrm{th}$ of the SWCNT in the stress calculations. The latter assumption particularly affects differences observed in the Young's modulus and ultimate tensile strength. Consequently, the comparison can only be an approximate one. Nevertheless, the present results can be judged to be of similar quality to those available in the literature. Some of the stress-strain curves of the configurations listed in Tab.~\ref{tab:MD_comparison} are shown in Fig.~\ref{fig:StressVerif}.

\begin{table}
\centering
\begin{tabular}{|c|c|c|c|c|}
\hline
\rule[-1ex]{0pt}{2.5ex} Conf. & Ref. &$E$/$E_\mathrm{ref}$ (GPa) & UTS/UTS$_\text{ref}$ (GPa) & $\epsilon_\text{max}$/$\epsilon_\text{max,ref}$  \\
\hline
\rule[-1ex]{0pt}{2.5ex}  (5,5)	  	  &  \cite{Meo06}		  & 916.9/894.7	 & 100.7/123.0 & 0.209/0.216  \\
\hline
\rule[-1ex]{0pt}{2.5ex}  (5,5)	  	  &  \cite{Canadija17}	  & 916.9/1039	 &             &              \\
\hline
\rule[-1ex]{0pt}{2.5ex}  (9,0)	  	  &  \cite{Meo06}		  & 982.4/939.1	 &  84.5/94.0  & 0.139/0.164  \\ 
\hline
\rule[-1ex]{0pt}{2.5ex}  (10,10)	  &  \cite{Chowdhury2012} & 909.0/666	 & 105.5/115.3 & 0.207/0.294  \\
\hline
\rule[-1ex]{0pt}{2.5ex}  (10,10)	  &  \cite{Yazdani2017}	  & 909.0/910	 & 105.5/120.5 & 0.207/0.196 \\
\hline
\rule[-1ex]{0pt}{2.5ex}  (11,9)	 	  &  \cite{Yazdani2017}	  & 918.0/915	 & 104.1/113   & 0.196/0.182  \\
\hline
\rule[-1ex]{0pt}{2.5ex}  (12,8)	  	  &  \cite{Jhon2016}	  & 921.8/966.2	 & 98.8/117.1  & 0.177/0.176 \\
\hline
\rule[-1ex]{0pt}{2.5ex}  (12,12)	  &  \cite{Belytschko2002}&          	 & 105.8/112.0 & 0.207/0.187 \\
\hline
\rule[-1ex]{0pt}{2.5ex}  (12,12)	  &  \cite{liew2004}	  & 923.8/1043	 & 105.8/148.5 & 0.207/0.279 \\
\hline
\rule[-1ex]{0pt}{2.5ex}  (17,0)	 	  &  \cite{Yazdani2017}	  & 980.0/1045	 & 88.0/98 	   & 0.136/0.135 \\
\hline
\rule[-1ex]{0pt}{2.5ex}  (29,29)	  &  \cite{Zhang2005}	  & 			 & 106.7/105.3 &  			 \\
\hline
\rule[-1ex]{0pt}{2.5ex}  (33,24)	  &  \cite{Zhang2005}	  & 			 & 101.8/98.8  &  			 \\
\hline
\rule[-1ex]{0pt}{2.5ex}  (36,21)	  &  \cite{Zhang2005}	  & 			 & 97.5/95.7   &  			 \\
\hline
\rule[-1ex]{0pt}{2.5ex}  (50,0)  	  &  \cite{Zhang2005}	  & 			 & 89.1/88.3   &  			 \\
\hline
\end{tabular}
\caption{Comparison of part of present results ($E$, UTS, $\epsilon_\text{max}$) to existing results ($E_\mathrm{ref}$, UTS$_\text{ref}$, $\epsilon_\text{max,ref}$). In \cite{Meo06, Canadija17} finite elements were used, while all the others used MD.} 
\label{tab:MD_comparison}
\end{table}

\begin{figure}[htp]
	\centering
	\subcaptionbox {\label{fig:YMComplete}}  {\includegraphics[scale=0.43]{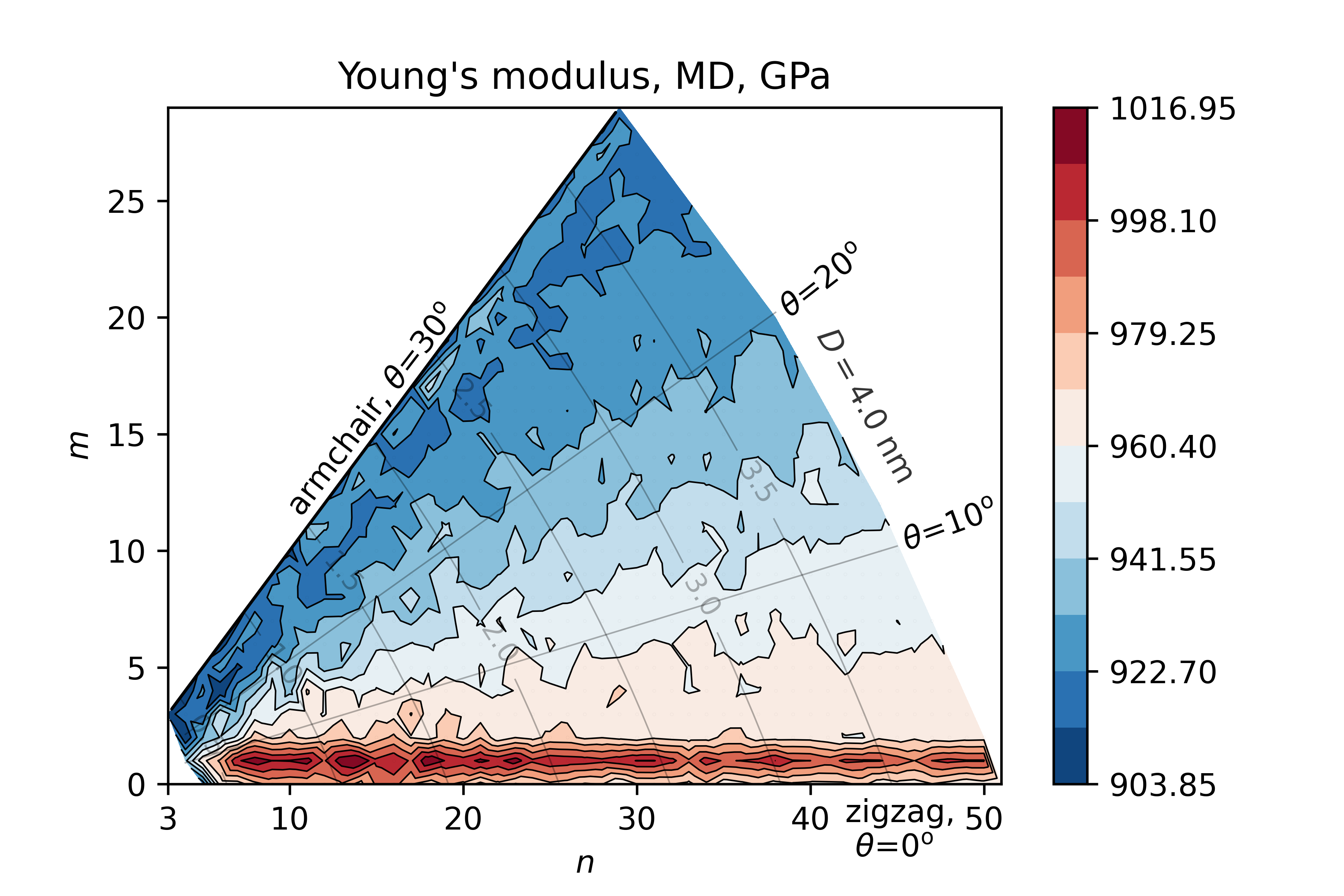}}
	\subcaptionbox {\label{fig:PoissonComplete}} {\includegraphics[scale=0.43]{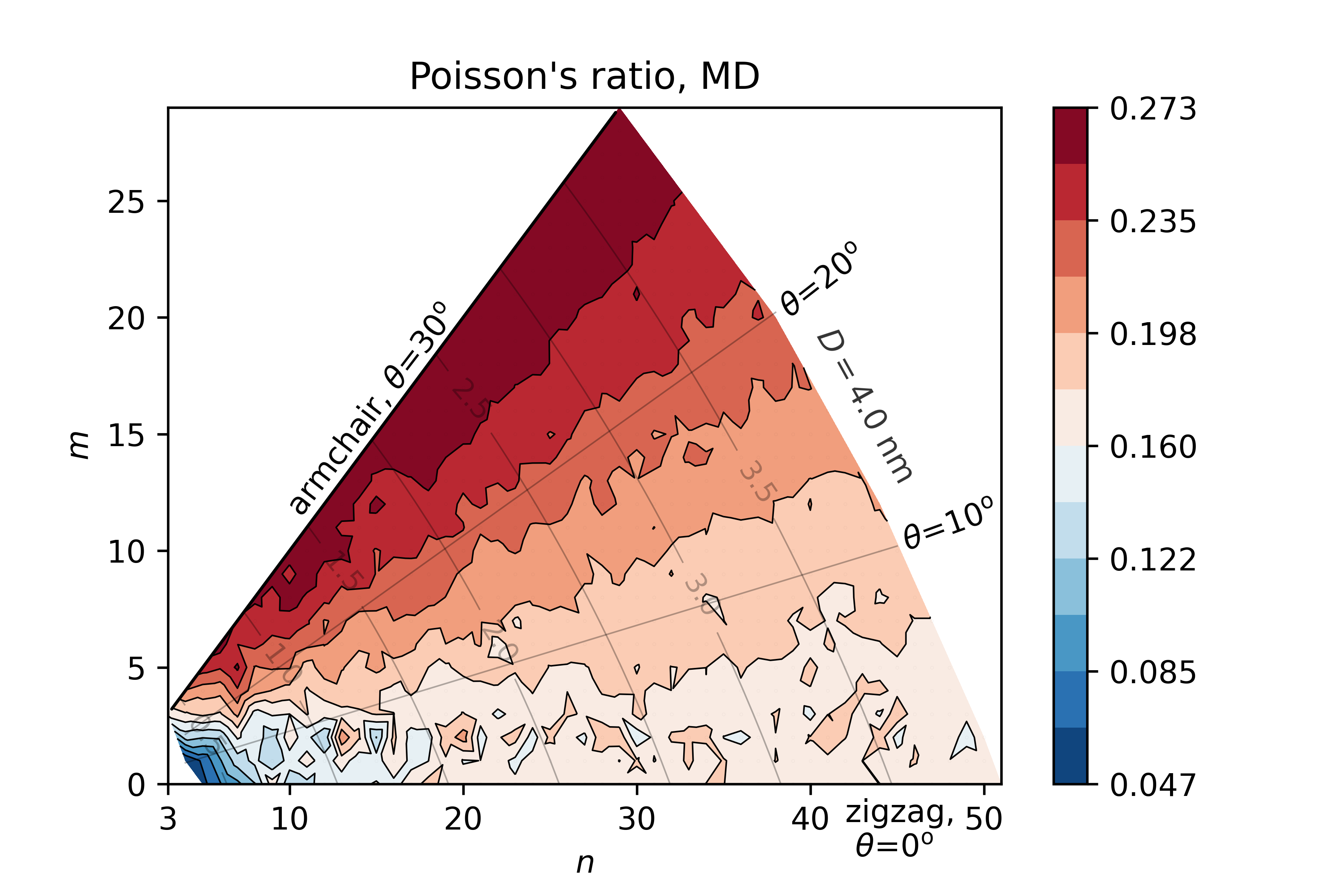}}
	\\
	\subcaptionbox {\label{fig:MaxStressComplete}} {\includegraphics[scale=0.43]{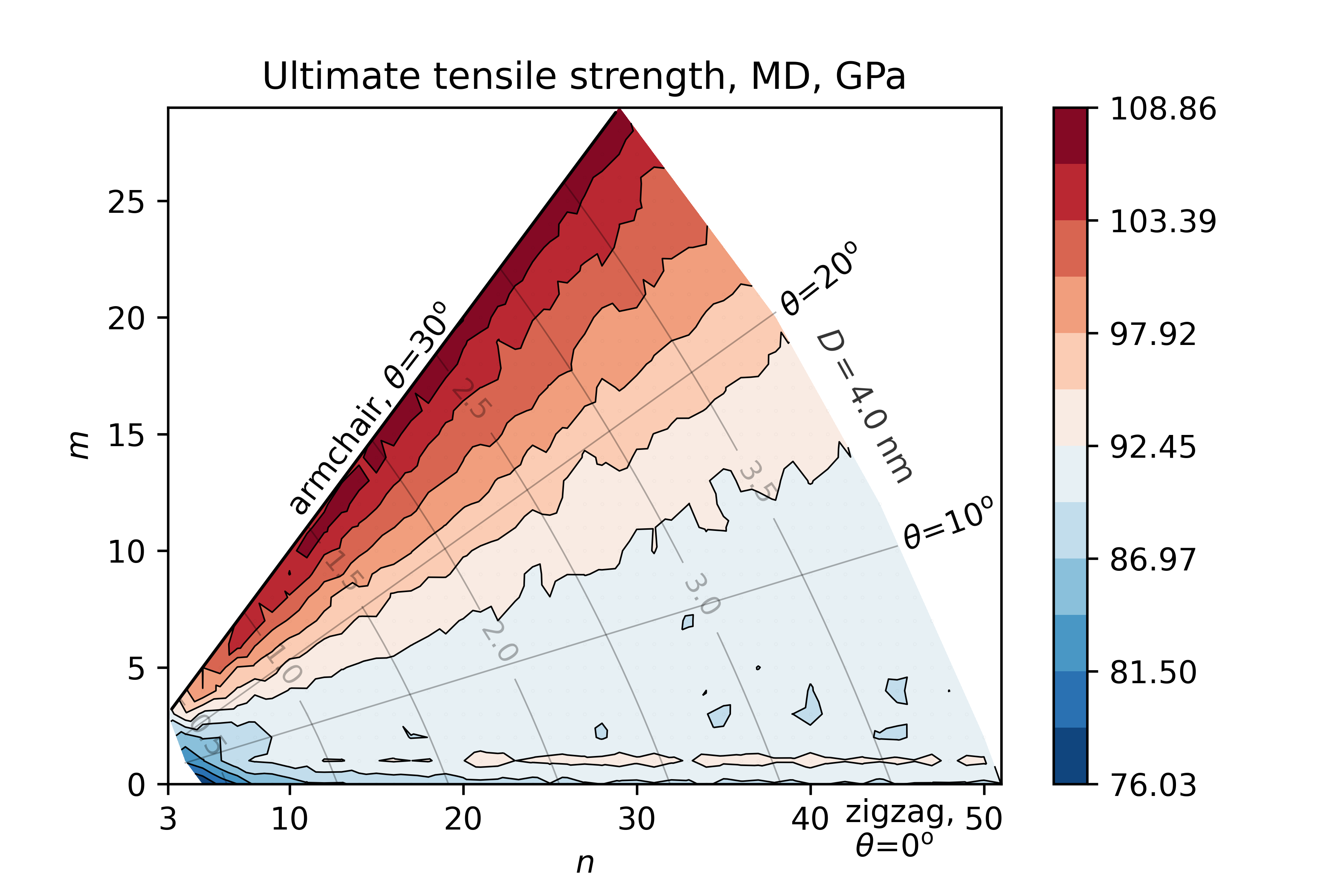}}
	\subcaptionbox {\label{fig:MaxEpsComplete}} {\includegraphics[scale=0.43]{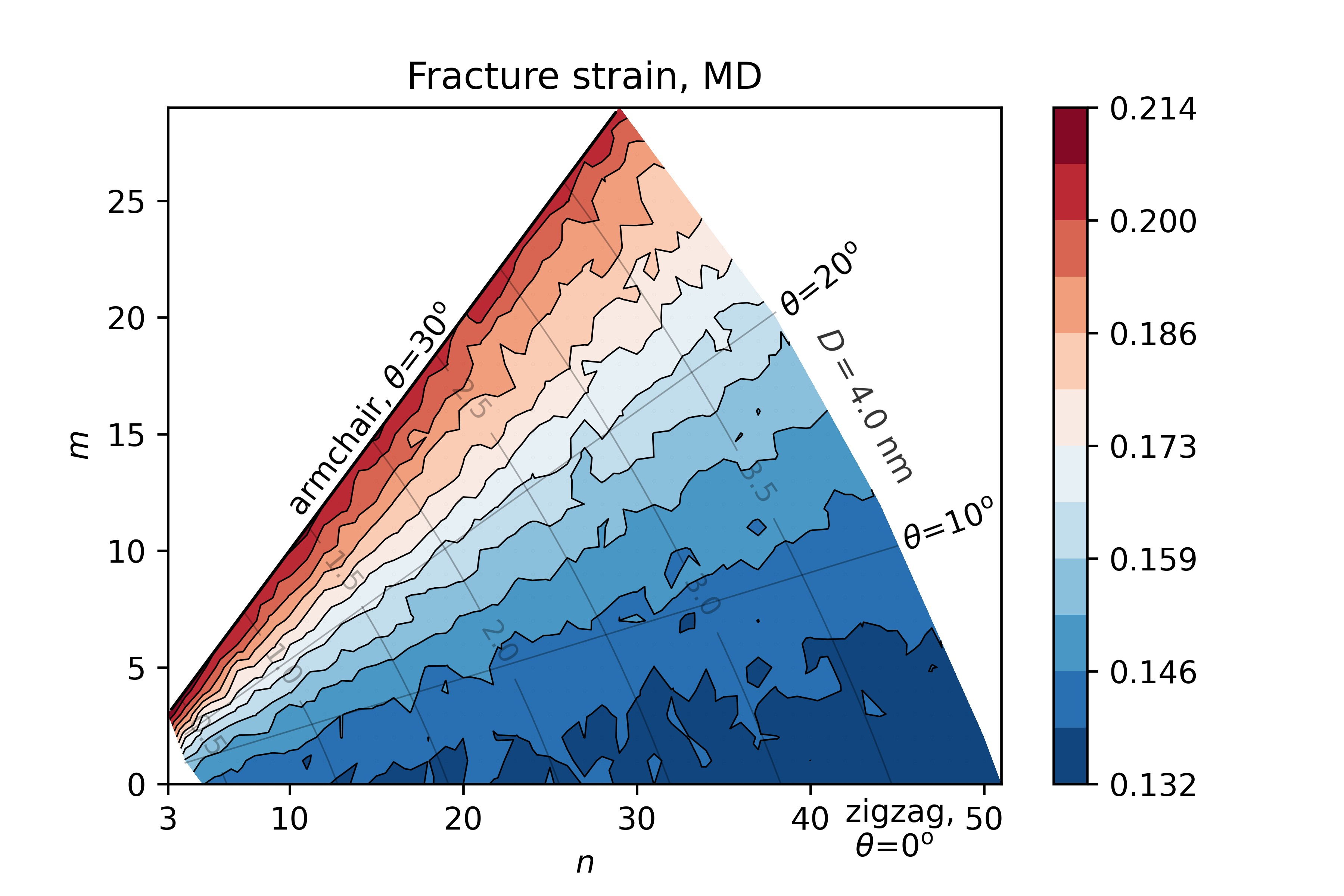}}
	\\
	\caption{Mechanical properties of SWCNT at 300 K as functions of chiral indices $n$ and $m$ obtained by MD simulations. Averaged on three sets of results. (a) Young's modulus, (b) Poisson's ratio, (c) Ultimate tensile strength, (d) Fracture strain.}
	\label{fig:MD}	
\end{figure}

The first postprocessing step involved removal of outliers, followed by averaging of the results from three simulation runs and presented for the convenience of the reader in the form of contour plots, with dependence expressed on the chiral indices $n$ and $m$, Figs.~\ref{fig:YMComplete}-\ref{fig:MaxEpsComplete}. The diameter $D_\mathrm{th}$ of the CNTs is also plotted. Fig.~\ref{fig:YMComplete} shows that, apart from a slightly disturbed Young's modulus distribution for the smallest diameter CNTs,  the chirality angle has the most significant influence, while the CNT diameter has a minor influence. A general trend of a slight decrease in Young's modulus with the increase in diameter of CNTs with the same chirality is observed. The extreme values are obtained for the armchair configuration (4,4) and the chiral nanotubes (14,1) close to zigzag configurations, Tab.~\ref{tab:MD_statistics}.

\begin{table}
	\centering
	\begin{tabular}{|>{\centering\arraybackslash}m{1.8cm}|>{\centering}m{2.8cm}|>{\centering\arraybackslash}m{2.8cm}|>{\centering\arraybackslash}m{1.8cm}|>{\centering\arraybackslash}m{2cm}|}
		\hline
		& Minimum/      & Maximum/      & Average & Standard  \\
		& configuration & configuration & 		& deviation \\
		\hline
		$E$ (GPa) 				   & 903.85/(4,4) & 1016.95/(14,1)  & 947.45  & 21.77 \\
		\hline
		$\nu$     				   &  0.047/(4,1) &   0.273/(29,27)  & 0.207   & 0.037 \\
		\hline
		$\sigma_\mathrm{max}$(GPa) &  76.03/(5,0) &  108.86/(28,27)& 94.86   & 5.529 \\
		\hline 
		$\epsilon_\mathrm{max}$    & 0.132/(48,1)&   0.214/(3,3)   & 0.159   & 0.021 \\
		\hline
	\end{tabular}
	\caption{Extreme values, averages and standard deviations of mechanical properties.} 
	\label{tab:MD_statistics}
\end{table}

The above observations are partly in contrast to what was stated in \cite{Li03}, where diameter was found to be the most influential variable regarding $E$. The latter results were obtained by finite element analysis for a small sample of zigzag and armchair CNTs in the diameter range 0.4-2.0 nm. A closer inspection of the present results in Fig. \ref{fig:YMComplete} confirms that in the case of zigzag CNTs, the Young's modulus actually increases in the indicated range, but after 2.0 nm it shows a slight downward trend. The similar slight decrease of the Young's modulus with the diameter increase is present for other chiralities as well. However, the comprehensive results reported here point to chirality as the dominant variable affecting the variations in Young's modulus. Further, in \cite{gupta2005elastic} significant increase in the Young's modulus with the increase in diameter is again observed for smaller diameter zigzag configurations, while a very small increase in the modulus is noted for other configurations. Similarly, \cite{Li03} reports that armchair CNTs exhibit a higher Young's modulus than zigzag CNTs for diameters smaller than 0.8 nm, quite in contrast to the present results. At this point, it is interesting to note that \cite{Yazdani2017} reports almost identical trends as the present ones. The decrease of the Young's modulus with the increase in chirality is also noted. However, due to the significantly smaller number of configurations considered, extreme values near the zigzag configurations are not detected. The other discrepancy is for the smallest diameter armchair CNTs, which achieve a lower Young's modulus than the rest, but thereafter a downward trend is again present. 

\begin{figure}
	\centering
	\includegraphics[scale=0.6]{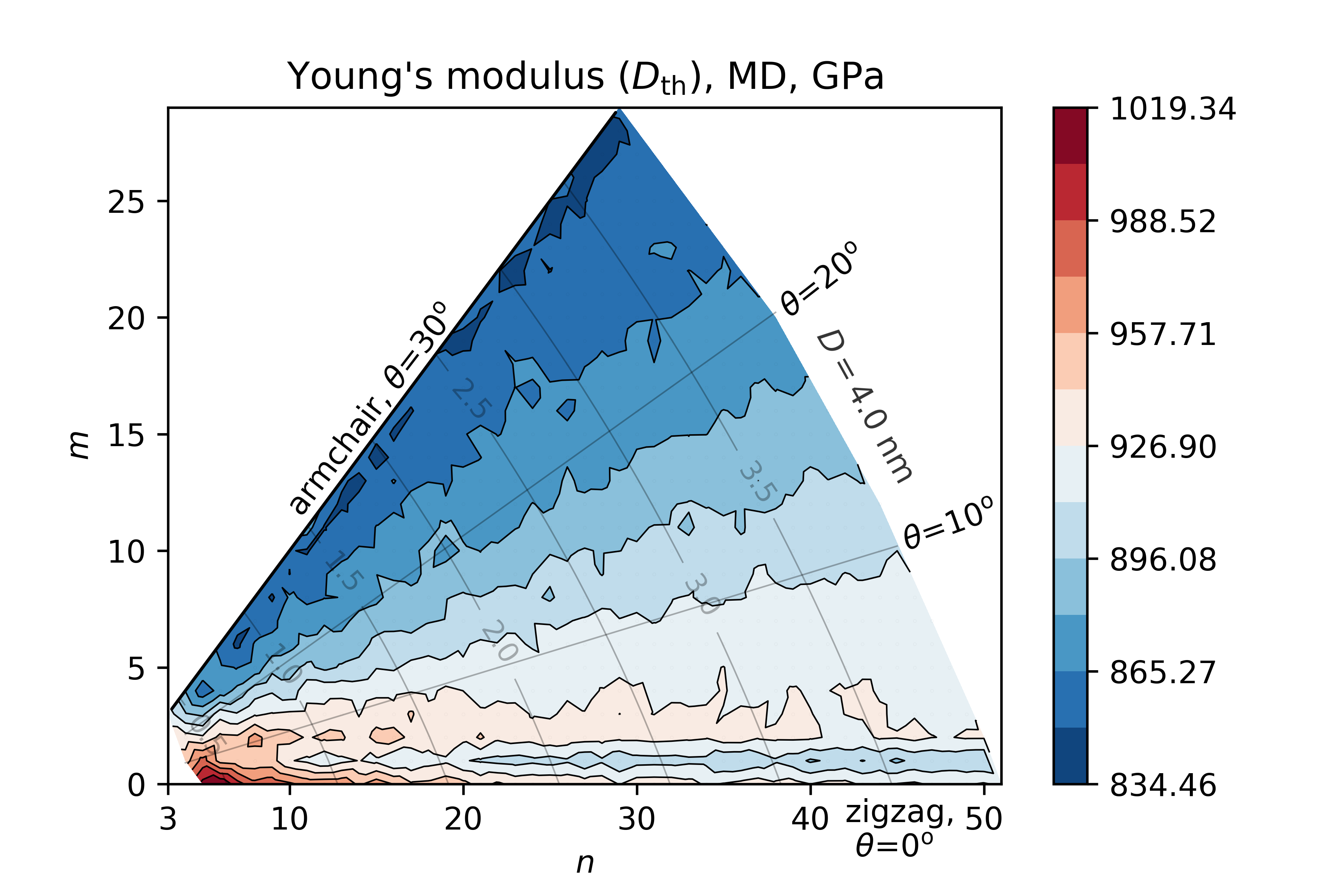}
	\caption{Graphical representation of the Young's modulus as a function of chiral indices $n$ and $m$ and calculated with $D_\mathrm{th}$. Averaged on three sets of results.} 
	\label{fig:YMDrescaled}
\end{figure}

Apart from the different diameter/length ratio and differences in MD computational details used in \cite{Yazdani2017}, at least part of the discrepancies described stem from the way the diameter used in the cross section is calculated, cf. present simulations recalculated with $D_\mathrm{th}$, Fig.~\ref{fig:YMDrescaled}. Since the actual diameter $D$ differs from the theoretical one by -4.93\% to +5.44\% (Fig.~\ref{fig:app_D_vs_Dth}), the Young's modulus calculated with $D_\mathrm{th}$ is slightly lower in most CNTs (lower $D_\mathrm{th}$ is found only in smallest CNTs thus giving higher $E$, while the rest have lower $E$). The most marked difference in Fig.~\ref{fig:YMComplete} is a barely noticeable downward trend with the increase of the diameter otherwise more pronounced in Fig.~\ref{fig:YMDrescaled}. The modulus is generally noticeably lower for zigzag nanotubes in Fig.~\ref{fig:YMDrescaled}. In summary, part of the results observed in the literature regarding the Young's modulus differs to present ones due to the conventionally used theoretical diameter.

\begin{figure}
	\centering
	\includegraphics[scale=0.5]{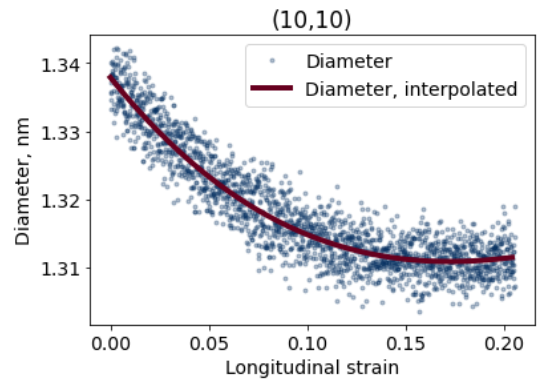}
	\caption{Variation of CNT (10,10) diameter due to thermal vibrations and smoothing.} 
	\label{fig:Diameter}
\end{figure}

\begin{figure}
	\centering
	\includegraphics[scale=0.5]{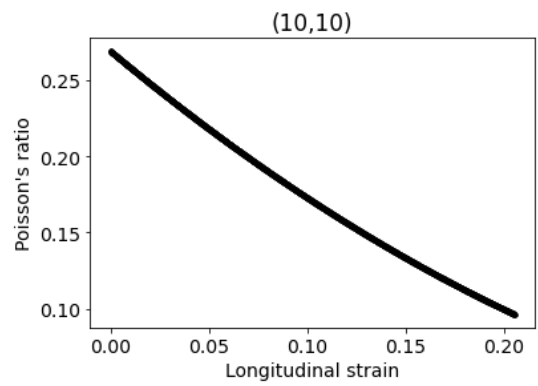}
	\caption{Change of the Poisson's ratio as a function of longitudinal strain for (10,10) CNT.} 
	\label{fig:Poisson10x10}
\end{figure}

The determination of Poisson's ratio proved to be somewhat more involved that anticipated. In particular, the amplitudes of the thermal vibrations of the atoms turned to be quite large compared to the displacements due to the cross-sectional contraction of the CNT under tensile loading, Fig.~\ref{fig:Diameter}. Therefore, as a first step, the diameter data were smoothed by applying a cubic spline. The smoothed diameter was then used to calculate the Poisson's ratio according to the standard formula throughout the complete tensile test. For illustration, the change in Poisson's ratio during the test is provided in Fig.~\ref{fig:Poisson10x10}. It can be seen that Poisson's ratio changes drastically with increase in tensile strain. 

For further analysis, the Poisson's ratio at the beginning of stretching is chosen, Fig,~\ref{fig:PoissonComplete}. Again, the dependence on diameter is markedly visible for the nanotubes with the smallest diameter. It is evident that the chiral angle is again the most influential variable. Zigzag nanotubes have the smallest Poisson's ratio and increasing the chiral angle also increases the Poisson's ratio. Thus, the highest ratios are observed in armchair nanotubes. The minimum of 0.047 is obtained in the case of (4,1) CNT and the maximum of 0.273 for (29,27). 

The present results indicate that the approach sometimes found in the literature to compare the Poisson's ratios of CNTs with different chiralities is not justified due to the pronounced dependence on chirality. A similar order of magnitude as in the present case, with increase of the ratio with increase of the diameter and strong dependence on the chirality, is obtained by ab-initio calculations in \cite{sanchez1999ab}. It is found that armchair nanotubes have slightly lower ratios than other chiralities considered. An uncertainty of 10\% is reported. In another contribution \cite{lu1997elastic}, values around 0.28 are observed for all configurations considered and obtained by a force constant model, which gives similar ratios for the armchair CNTs reported here, and somewhat higher for other chiralities. The temperature dependence is investigated in \cite{dereli2007temperature}, and $\nu=0.3$ is reported at 300 K for (10,10) CNT, which is comparable to the 0.266 obtained in this research for this particular configuration.

The behavior at fracture, Figs.~\ref{fig:MaxStressComplete}, \ref{fig:MaxEpsComplete} again shows a similar or even more pronounced influence of the chirality angle. A continuous increase in UTS and fracture strain with the increase in chirality angle is discovered, similarly to \cite{Yazdani2017,xiang2020machine}. UTS ranges from 76.03 GPa to 108.86 GPa for (5,0) and (28,27) nanotubes, respectively, Tab.~\ref{tab:MD_statistics}. The armchair configuration clearly exhibits higher UTS, while the zigzag configurations have the lowest UTS, especially those with smaller diameter. In general, increasing the diameter leads to a slight increase in UTS. The fracture strain seems to be slightly higher for the smallest diameters, but further increasing the diameter shows no significant effect. Zigzag configurations can withstand significantly lower strains than armchair CNTs before fracture occurs, with the minimum value 0.13 for the chiral nanotube (48,1). The highest strain at fracture 0.214 is obtained for the small diameter armchair nanotube (3,3).

\textbf{Remark 1.} Please bear in mind that the origin of the slight size dependence of the diameter exhibited by the Young's modulus and UTS, which is most pronounced for smaller diameters in Figs.~\ref{fig:YMComplete}, \ref{fig:MaxStressComplete}, is not entirely clear. The latter properties depend on the cross-section of the SWCNT, which is assumed to be a thin circular ring by convention. Putting aside the debated issue of the ring thickness, the cross-section is in fact an irregular polygon with vertices that does not lie in a plane orthogonal to the longitudinal axis of a chiral SWCNT. The discrepancy from a circular cross-section is largest for smaller diameter cross sections. Moreover, the difference between the theoretical diameter and the actual diameter at the simulation temperature is also important. Thus, it is not clear whether the diameter-dependent size effects are attributable to the physical effects or to the cross-section approximation. The absence of such effects in the case of fracture strain that does not depend on the cross-section assumption, Fig.~\ref{fig:MaxEpsComplete}, might indicate that the latter statement is true, but definitive proof is still not provided at this point.

\textbf{Remark 2.} A separate line of research based on size effects has been particularly active in recent years - the nonlocal mechanics of nanorods and nanobeams. Simple mechanical models and quite complex \cite{dehrouyeh2019nonlinear,Sedighi2021} nonlocal formulations can be found in the literature nowadays, making various claims concerning the nonlocality of nanoscopic structures, carbon nanotubes in particular. Static \cite{fernandez2016bending} and dynamic \cite{skoblar2021} problems in homogeneous and composite structures \cite{canadija2019nonlocal,canadija2020thermomechanics,rezaiee2020size} are addressed for both isothermal and nonisothermal environments. The nonlocal mechanics in the above works is based on the stress and strain fields in the neighborhood of a point under consideration and is also related to the size-effects in nanoscopic structures. Note that the possible absence of influence of diameter on mechanical properties, as indicated in Remark 1, should not be confused with the size-effects exploited in nonlocal mechanics that clearly stem from short- and long-ranged interactions.

\subsection{Correlation analysis}
\begin{figure}
	\centering
	\includegraphics[scale=0.75]{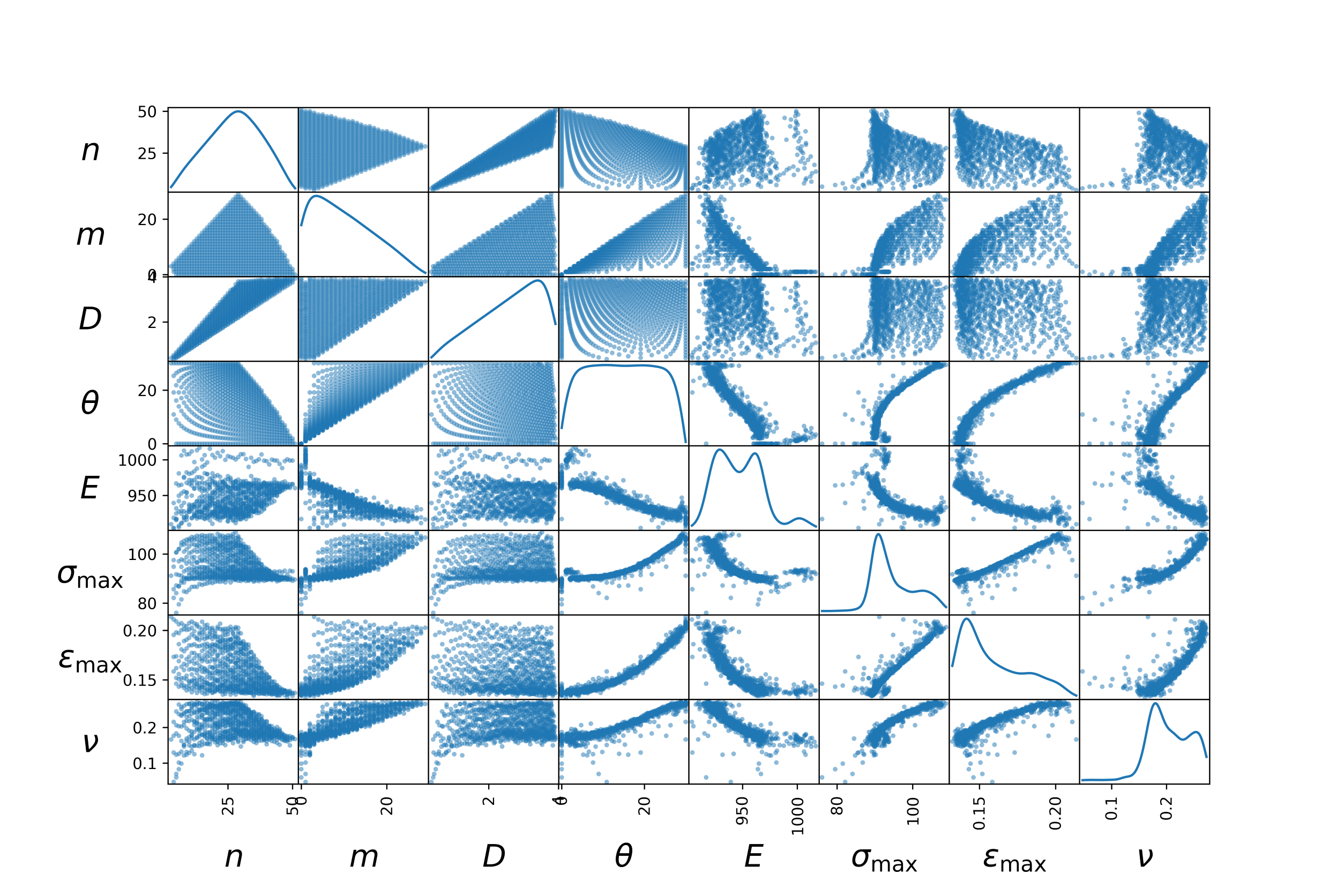}
	\caption{The scatter matrix plot of selected attributes with a kernel density estimation at the diagonal.} 
	\label{fig:Correlation}
\end{figure}

To gain further insight into the relationship between mechanical properties, a correlation analysis is performed. Pearson's correlation coefficients were calculated for selected properties, Tab. \ref{tab:Correlation}. The corresponding scatter matrix is shown in Fig.~\ref{fig:Correlation}. The weak dependence of Young's modulus, Poisson's ratio, UTS, and strain at fracture on diameter has already been shown in Figs.~\ref{fig:MD}. The values of the Pearson coefficient related to diameter are quite close to zero, especially for the UTS and fracture strain, while the points in the graphs are well scattered, indicating the same conclusion. The deviation from zero can be attributed to the fact that the SWCNTs with the smallest diameter show some dependence on the diameter. In strong contrast to the latter observation, the chiral angle plots show marked band structures with high absolute values of the Pearson coefficients. A strong negative correlation of the Young's modulus and strong positive correlations of other properties with respect to the chiral angles are clearly evident. This is consistent with the discussion in Sec.~\ref{subsec:MD_results} related to Figs.~\ref{fig:MD}, which highlights chirality as the most influential variable. The same is also reflected in the mutual correlations between Young's modulus, Poisson's ratio, UTS and fracture strain. Apart from a rather strong negative correlation between Young's modulus and other properties, pronounced positive correlations are visible in all diagrams.
'
\begin{table}
	\centering
	\begin{tabular}{|r|c|c|c|c|c|c|}
		\hline
		& $D$   & $\theta$ & $E$  & $\nu$  & $\sigma_\mathrm{max}$ & $\epsilon_\mathrm{max}$    \\
		\hline
		$D$ 			 & 1.000	& -0.011 &	-0.111 &	0.118 &	0.041 &	-0.124 \\
		\hline
		$\theta$ 		 &-0.011	&  1.000 &	-0.942 &	0.935 &	0.918 &	0.979 \\
		\hline
		$E$				 &-0.111  & -0.942 &	1.000  &-0.898	& -0.848 &	-0.914 \\
		\hline
		$\nu$			 & 0.118	& 0.935	 &-0.898   &	1.000	& 0.918 &	0.897 \\
		\hline
		$\sigma_\mathrm{max}$	 &0.041	& 0.918	 &-0.848   &	0.918	& 1.000	& 0.909 \\
		\hline
		$\epsilon_\mathrm{max}$  &-0.124	& 0.979	 &-0.914   &	0.897	& 0.909 &	1.000 \\
		\hline
	\end{tabular}
	\caption{The Pearson's correlation coefficients for selected attributes.} 
	\label{tab:Correlation}
\end{table}

\section{Deep learning framework for SWCNT}

\subsection{Datasets}
\label{subsec:datasets}
Data obtained by MD simulations is exploited in two ways. The first dataset is generated as follows. The results of tensile tests of 818 SWCNT configurations are averaged over three sets of simulations and used as the dataset. The dataset if freely available at \cite{dataset2021}. This preprocessing is carried out on MD results and described in Sec.~\ref{subsec:MD_results}. If a single result from three sets for $E$, UTS or $\varepsilon_\mathrm{max}$ differed by more than 10\% from other two results, it was judged as an outlier and excluded from simulation. Due to larger fluctuations observed, threshold of 50\% with the above procedure was used for Poissons' ratios. In total, four Young's moduli, ten $\varepsilon_\mathrm{max}$ and seven Poisson's ratio were classified as outliers across all three sets with results. To avoid any ambiguity in terms of the theoretical and actual SWCNT diameter, the feature set consisted of chirality indices $n$ and $m$. The labels are represented by the initial Young's modulus, the initial Poisson's ratio, UTS and the strain at fracture. Note that the relationship between indices, diameters and chiralities will be simply found out by the DNN and included in the model.

The dataset is randomly split into the training, validation, and test sets in a 70:15:15 ratio. As usual, the test set was not seen during training and used to access the performance of DNN. Note that in this research, all possible combinations of SWCNT within the given diameter range (Sec.~\ref{sec_MD_model}) were generated and included in the dataset. Consequently, a sufficiently complex neural network could be found to provide a near-ideal fit, and can thus be reduced to a kind of lookup table. However, in the present case, the neural network could be exploited to provide an additional benefit - the smoothing of thermal fluctuations in the MD results. Admittedly, the degree of smoothing is somewhat subjective and remains an arbitrary choice.

The second group of datasets starts from the above dataset. To investigate the influence of the size of the dataset, smaller datasets were generated as different fractions of the complete dataset. Dataset sizes ranged from 5\%-95\% of the full dataset with a step of 5\%. For each fraction, ten different random selections were used in calculations. To account for more pronounced changes of properties with respect to the diameterin the case of smaller diameter SWCNTs, the full dataset was split into two smaller sets before applying the aforementioned fractions: one with SWCNTs up to 1 nm in diameter and one with larger SWCNTs. Then, the prescribed fraction was applied to both subsets to select the required number of SWCNTs. Instead of taking only a certain fraction of the complete set of configurations, this approach ensures that a reasonable number of smaller SWCNTs are included in the dataset. Otherwise, since there are many more SWCNTs with larger diameters, they would dominate and the changes observed in the smaller diameter nanotubes would not be adequately detected by DNN due to the small number of points. Third, in addition to the averaged MD data obtained by three different sets of MD simulations, each of the three contributing sets was used individually as a source for the dataset, again applying the fraction methodology described. This provided information on predictions based on a single set of results in which the averaging is not performed. This kind of analysis is of interest since computational time required for MD simulations to prepare the dataset is reduced by the same fraction. On the other side, this has a negative effect on the accuracy of predicted properties, so a guidance on the fraction value is required. 

\subsection{Deep neural network architecture and parameters}
DNN was developed in Keras. Several different architectures and parameter choices were considered and the following one is selected in all cases. The neural network was sequential, with dense layers consisting of 128-64-64-32-32-32-32-32-8-8-4 neurons. Apart from the first and last layer, which used linear activation, all the others employed ReLU activation functions. Input and output data were normalized between $-1$ and $+1$. The loss function used was the mean squared error (MSE) in conjunction with the Adam optimizer with a default learning rate of 0.001. Early stopping was triggered if the validation loss did not improve for 500 epochs, taking the best result as the final model. The batch size of 64 was used. To increase the DL model stability, accuracy and robustness 10-fold cross-validation was employed. The DNN that resulted in the minimal loss on the test set was selected as the reference one.

\subsection{Results}
\subsubsection{DNN model of SWCNT}
\label{subsubsec:DL_model}
\begin{figure}
	\centering
	\includegraphics[scale=0.5]{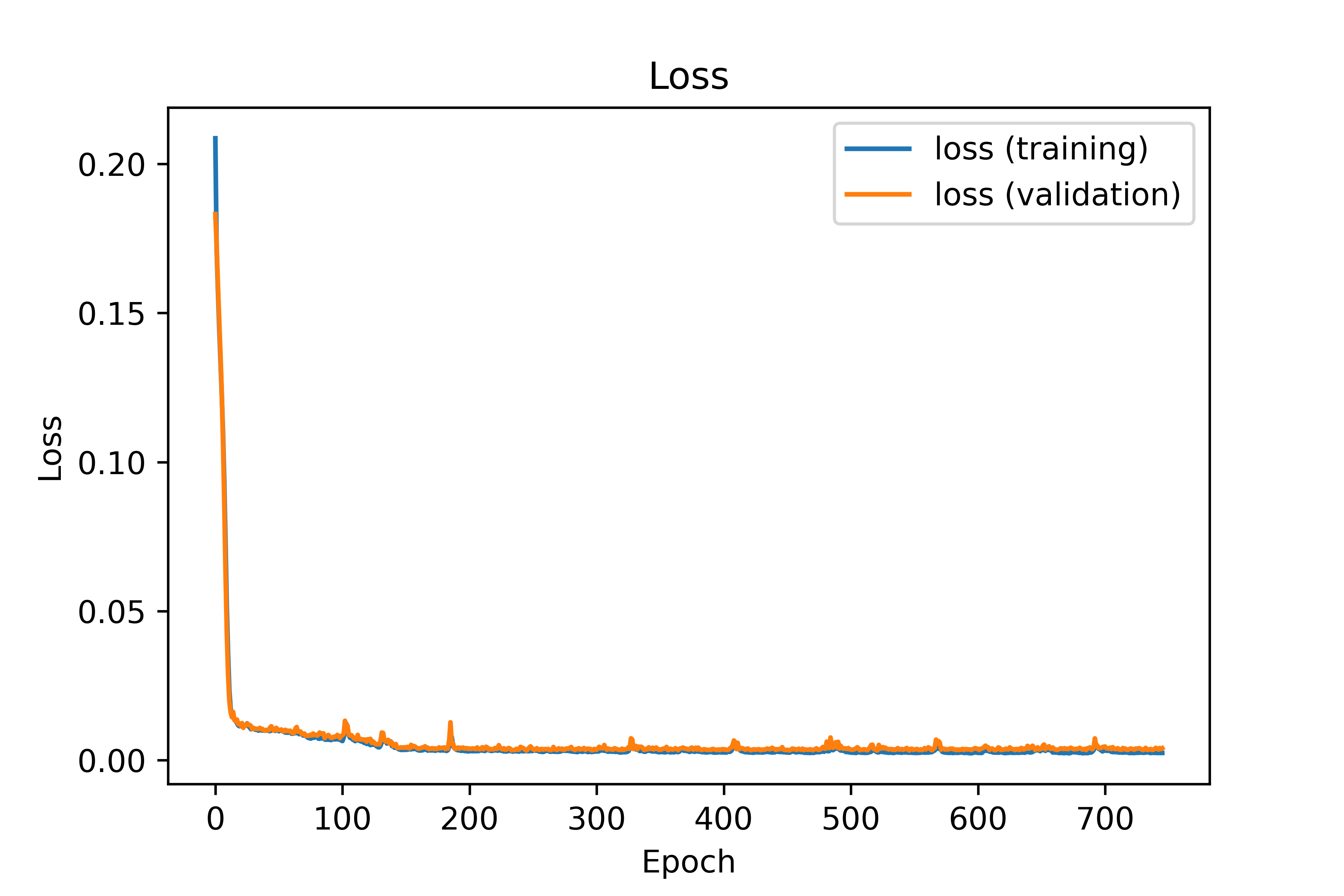}
	\caption{Loss vs. Epoch for predictions of averaged MD results.} 
	\label{fig:loss}
\end{figure}

Fig.~\ref{fig:loss} shows the history of the loss. A rapid drop in loss is obtained at the beginning of the fitting procedure, while the subsequent part of the curve corresponds to finding a better local optimum. The simulations generally lasted 1000-2000 epochs, with an average 1247. Early stopping with respect to the validation loss was activated. For a random split into training/validation/test set, the final losses (MSE) of the reference DNN were 0.00277/0.00323/0.00484, respectively.

As already pointed out, Young's modulus and Poisson's ratio distributions obtained using MD still contain some degree of noise, Figs.~\ref{fig:YMComplete}, \ref{fig:PoissonComplete}. Even at the macroscopic scale, the measurement of these properties is prone to error and should be performed conscientiously. Nanoscopic structures are strongly affected by the thermal vibrations of carbon atoms, which further complicates the determination of $E$ and $\nu$. As a result, $R^2$ is slightly lower than for the UTS and fracture strain, but still very high, Tab.~\ref{tab:R2}. Although such a pathway is not investigated in the present study, a more complex neural network with a higher $R^2$ score could be found out. As already explained, the present goal is not to fully reproduce MD simulations, which still contain noise and do not represent a high quality stochastic average. Arguably, a slightly lower $R^2$ value with smoother distributions actually represents a kind of smoothing filter of the results otherwise strongly influenced by thermal fluctuations, Figs.~\ref{fig:ML_E}, \ref{fig:ML_P}. In this way, a slightly reduced $R^2$ should actually mean a better representation of the stochastically averaged properties. Such behavior of DNN can be easily explained. In averaging, only three different MD simulation results for a given SWCNT configuration are considered, while in DNN training, the neighboring configurations are also accounted for. This does not mean that the averaging procedure can be abandoned in favor of DNN. It still remains an important part of the preprocessing step.

The obtained coefficients of determination $R^2$ for the UTS and fracture strain, Tab.~\ref{tab:R2}, are quite high. Clearly, more uniform distributions obtained by MD simulations, Figs.~\ref{fig:MaxStressComplete}, \ref{fig:MaxEpsComplete}, also allow a better approximation by the DNN, as indicated by the rather high $R^2=0.99$ for the whole dataset. Similar $R^2$ were obtained for the validation and test datasets, indicating that no overfitting occurs. The graphical representations of the distributions obtained by the DNN, Figs.~\ref{fig:ML_UTS}, \ref{fig:ML_FS} appears to be reasonable and smoother than a statistical average obtained solely by averaging the MD results.
 
\begin{figure}[htp]
	\centering
	\subcaptionbox {\label{fig:App_Diff_E}}  {\includegraphics[scale=0.43]{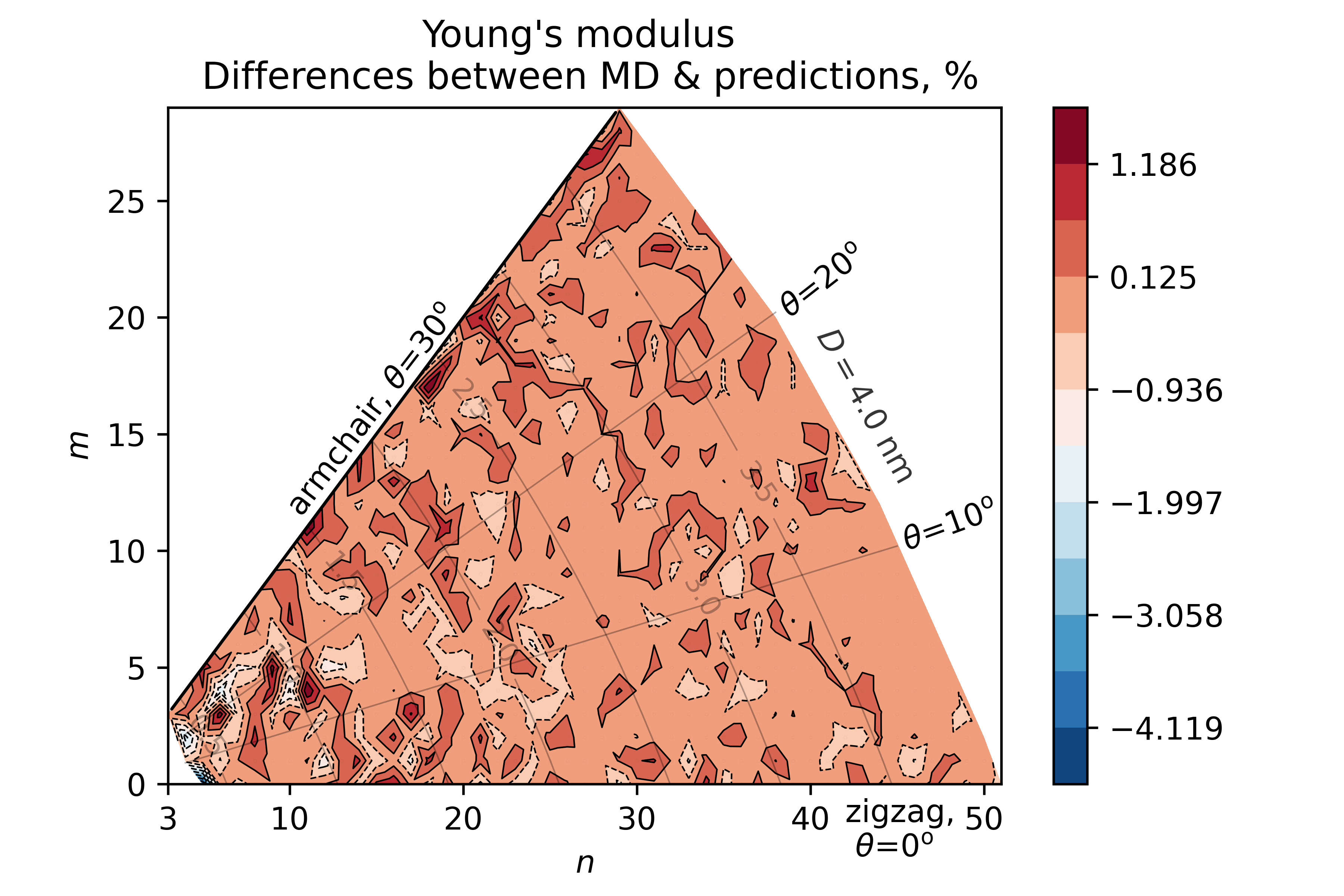}}
	\subcaptionbox {\label{fig:App_Diff_P}}  {\includegraphics[scale=0.43]{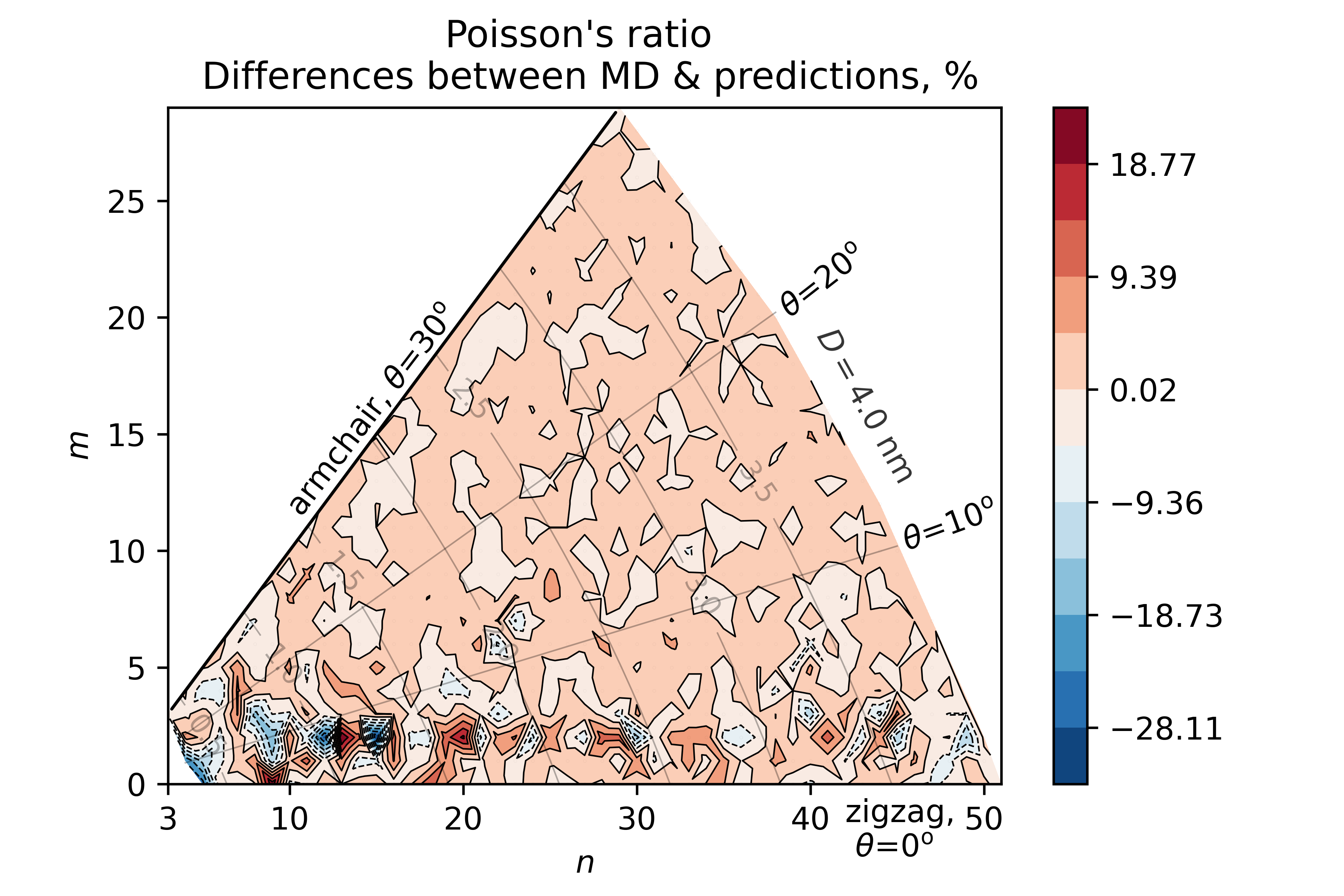}}
	\\
	\subcaptionbox {\label{fig:App_Diff_UTS}}  {\includegraphics[scale=0.43]{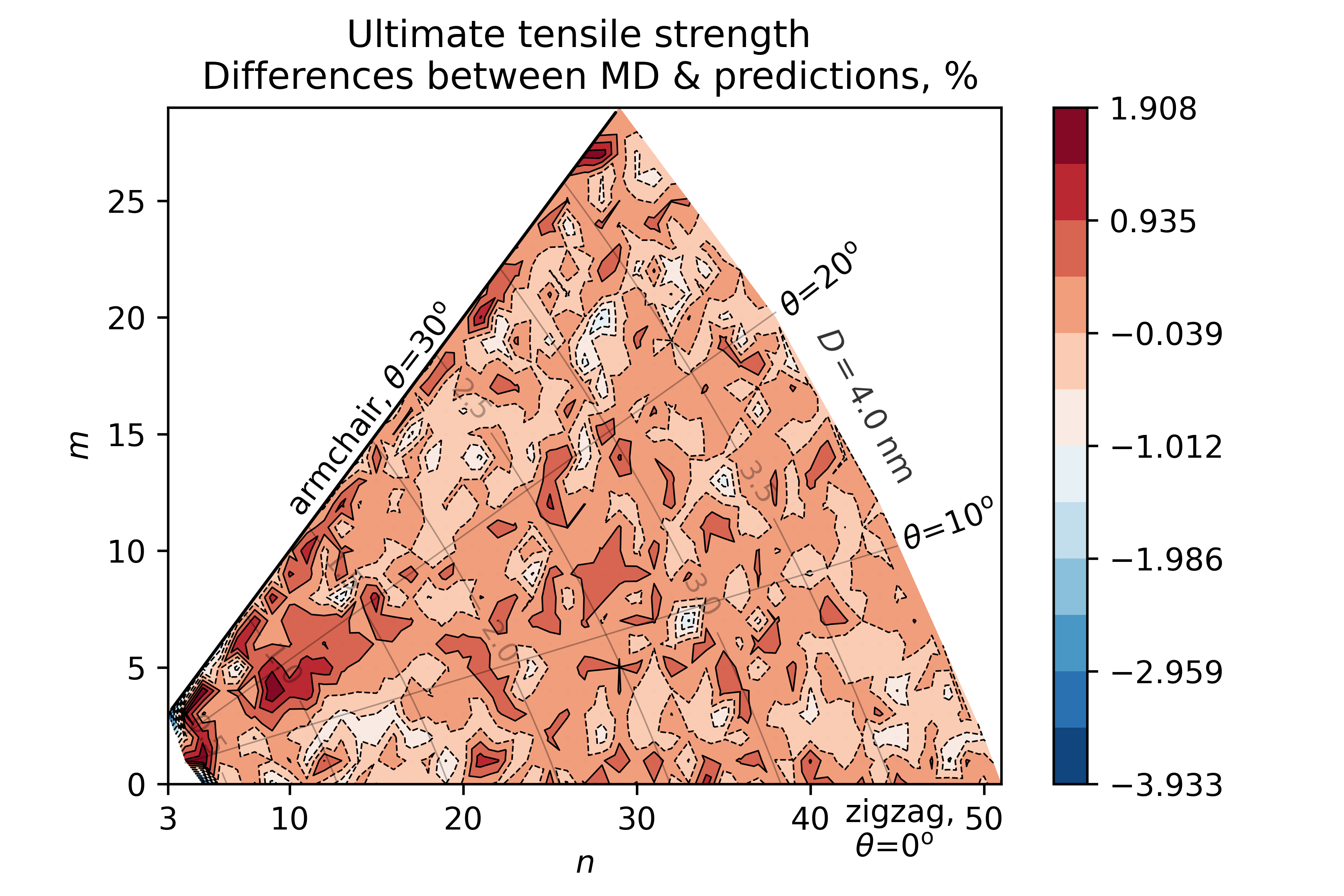}}
	\subcaptionbox {\label{fig:App_Diff_FS}}  {\includegraphics[scale=0.43]{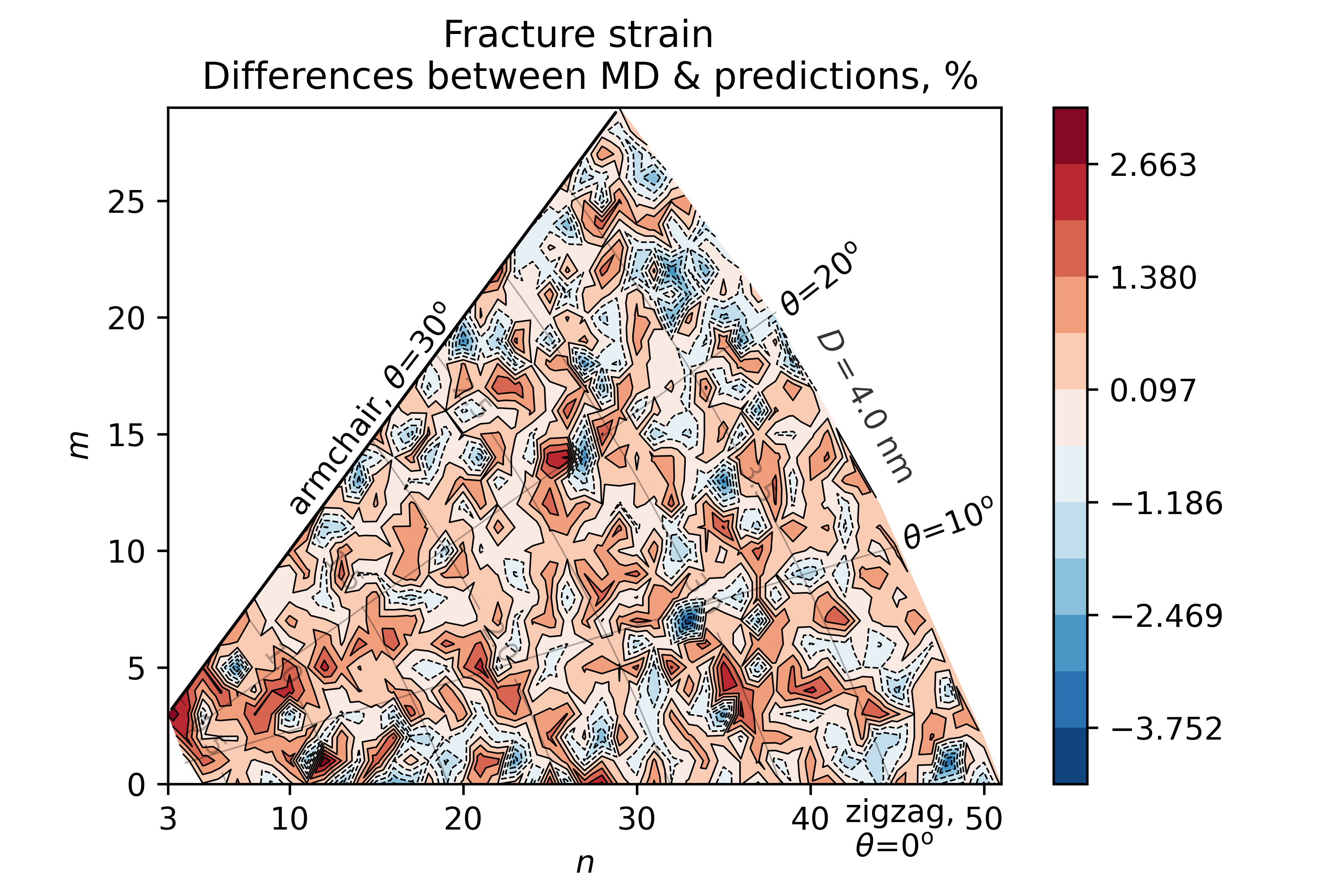}}
	\\
	\caption{Differences (\%) between mechanical properties of SWCNT calculated by MD and predicted by DNN. (a) Young's modulus, (b) Poisson's ratio, (c) Ultimate tensile strength, (d) Fracture strain.}	
	\label{fig:DiffMDML}
\end{figure}

The differences between the predictions and the MD results for the individual results are provided in Fig.~\ref{fig:DiffMDML}. With the exception of Poisson's ratio, the errors for all SWCNT configurations range from $+2.66$\% to $-4.12$\%. In the case of the Young's modulus and ultimate tensile strength, the largest errors are noticed for the nanotubes with smaller diameters, while the vast majority of errors in the other cases are below 1\%. The errors in the fracture strain are well distributed over all diameter ranges. As for Poisson's ratio, errors up to $+18.77$\% and $-28.11$\% are observed, but again mostly for the smaller diameter nanotubes and here for just a few nanotubes with chiralities close to the zigzag SWCNTs. The remaining differences in the predictions remain quite small.
 
\subsubsection{Influence of the dataset size}
One of the outcomes of the present research is to provide some kind of guidance on the choice of dataset size for further computations. In particular, investigations of SWCNT behavior using a "brute force" approach represented by the complete set of possible SWCNT configurations performed in this work might not be always feasible due to the MD computational costs. For instance, such simulations could include, longer SWCNTs, different loading rates, multi-walled nanotubes, etc. To this end, as described in Sec.~\ref{subsec:datasets} several fractions of the present dataset were considered as a training basis for a DNN with the same architecture as before. Both averaged data and three individual contributing sets of MD results were used. For each fraction, ten different sets were generated and used in training. The mean result obtained from these ten calculations per fraction is reported below.

\begin{figure}[htp]
	\centering
	\subcaptionbox {\label{fig:R2_E_comparison}}  {\includegraphics[scale=0.43]{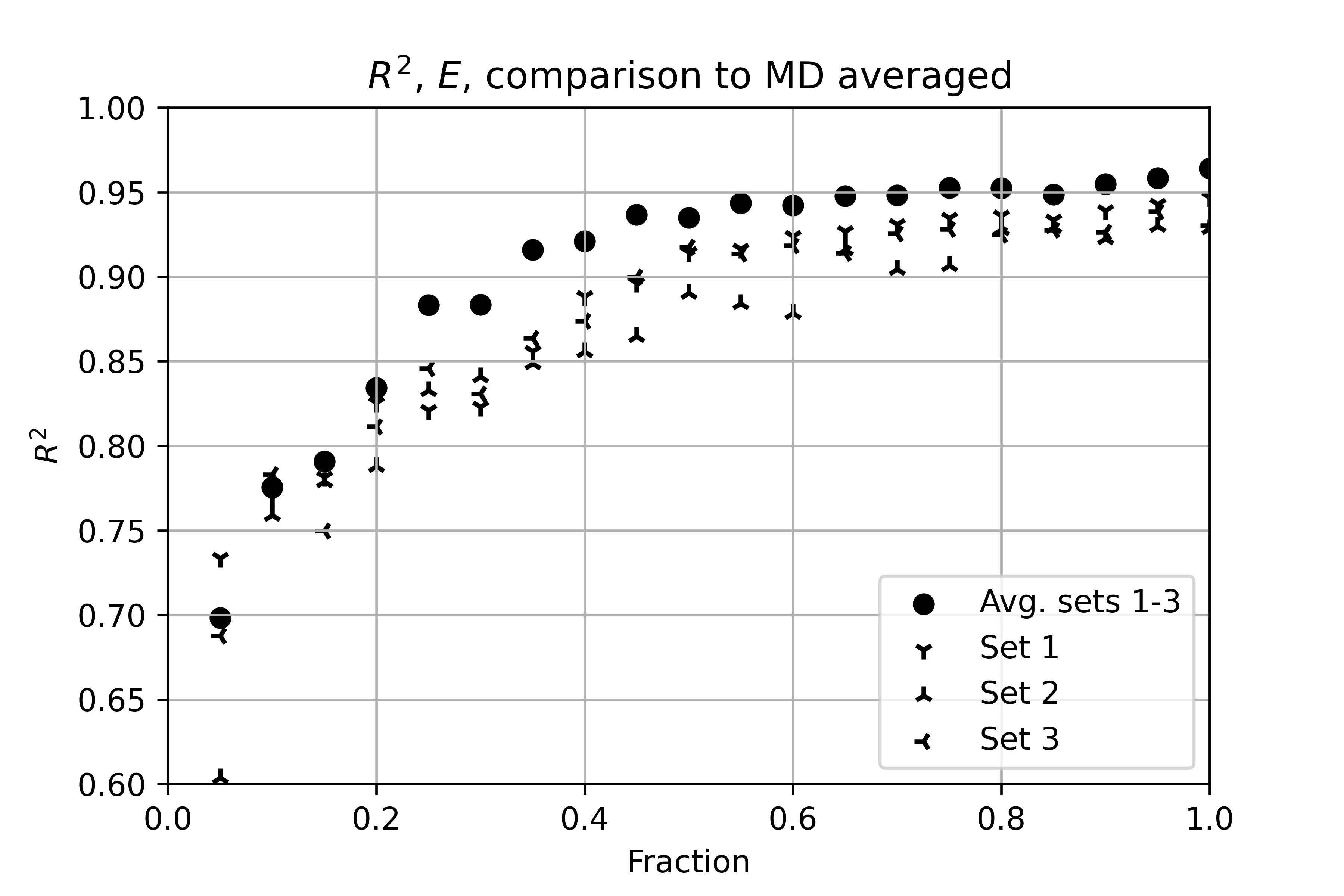}}
	\subcaptionbox {\label{fig:R2_P_comparison}}  {\includegraphics[scale=0.43]{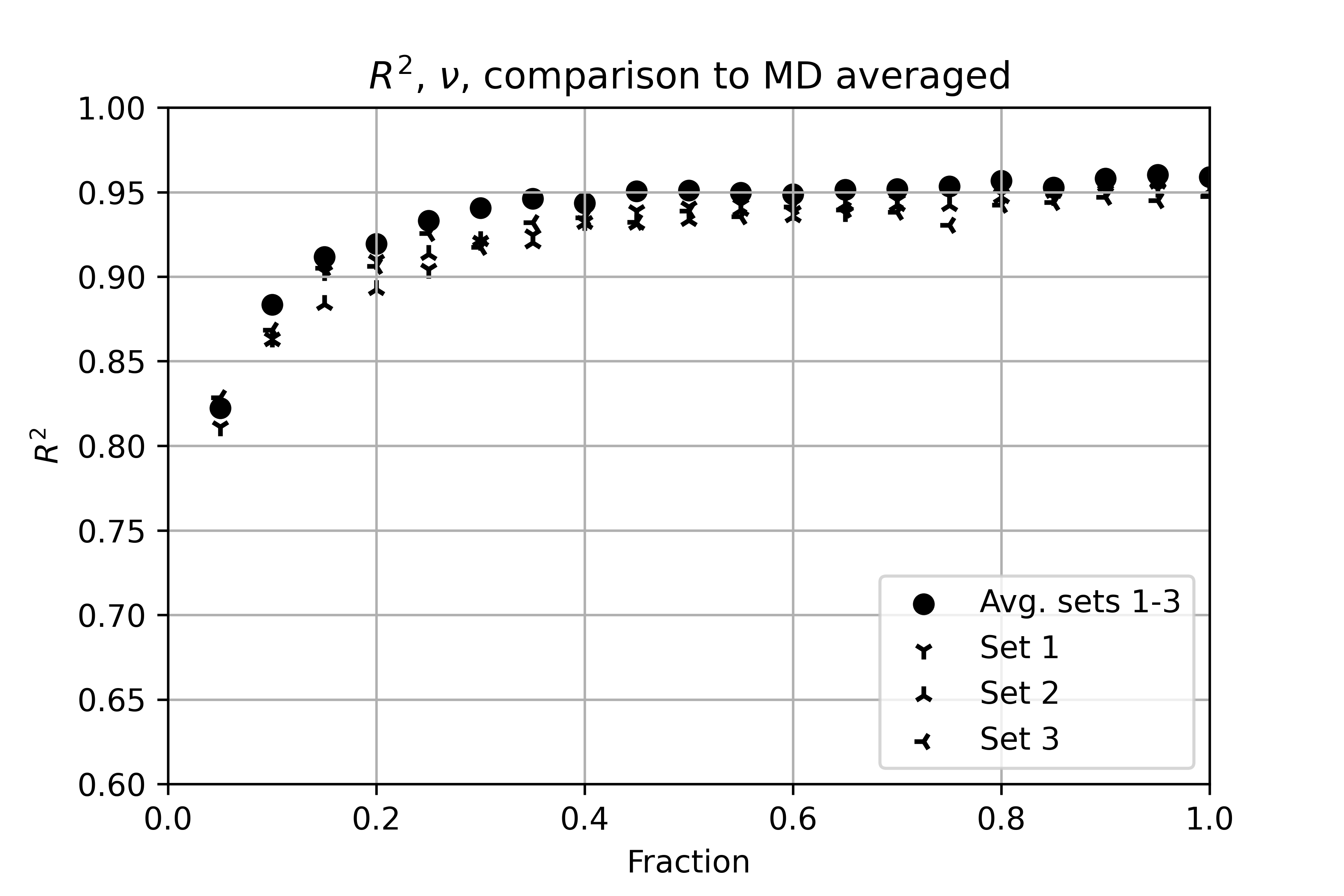}}
	\\
	\subcaptionbox {\label{fig:R2_UTS_comparison}}  {\includegraphics[scale=0.43]{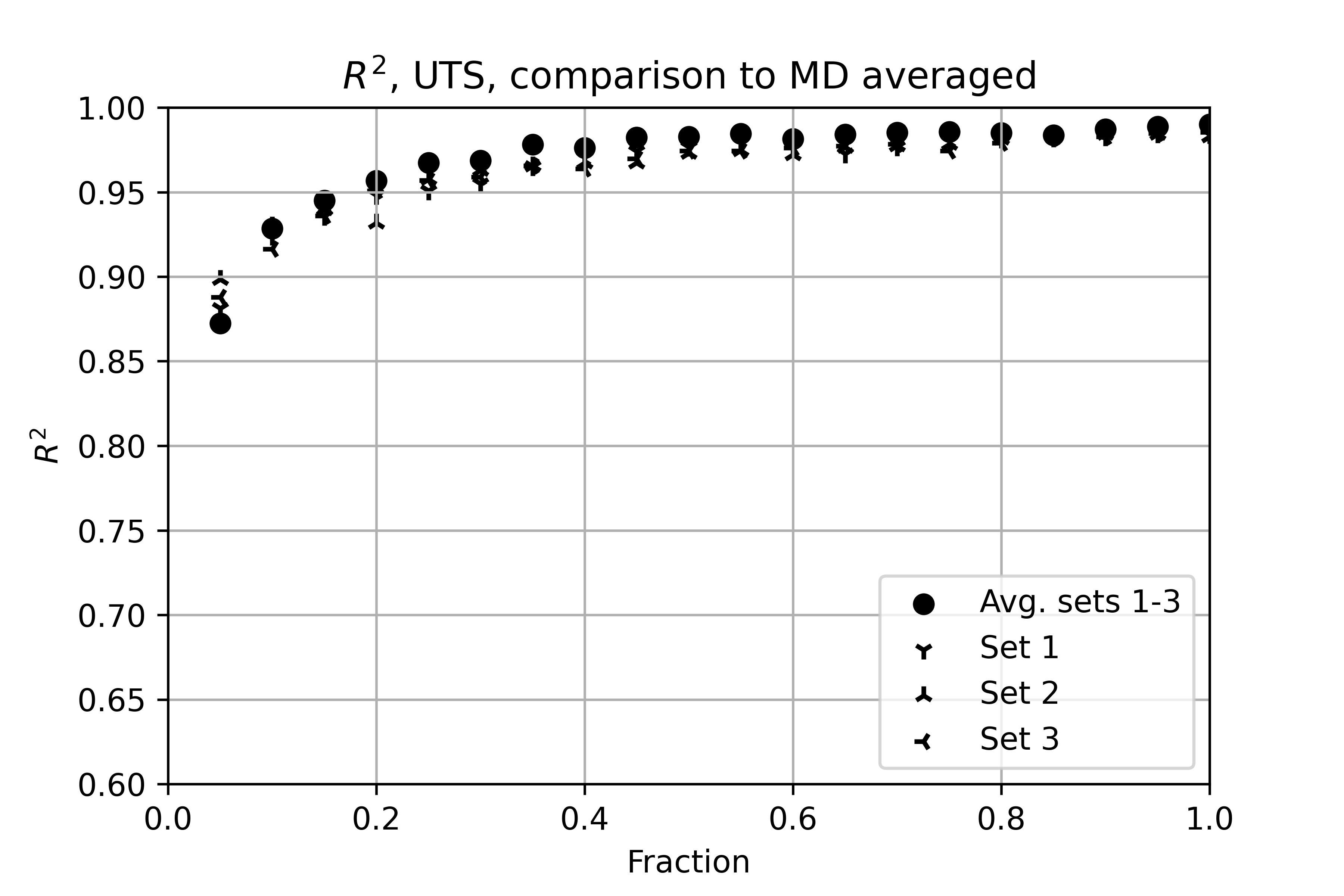}}
	\subcaptionbox {\label{fig:R2_FS_comparison}}  {\includegraphics[scale=0.43]{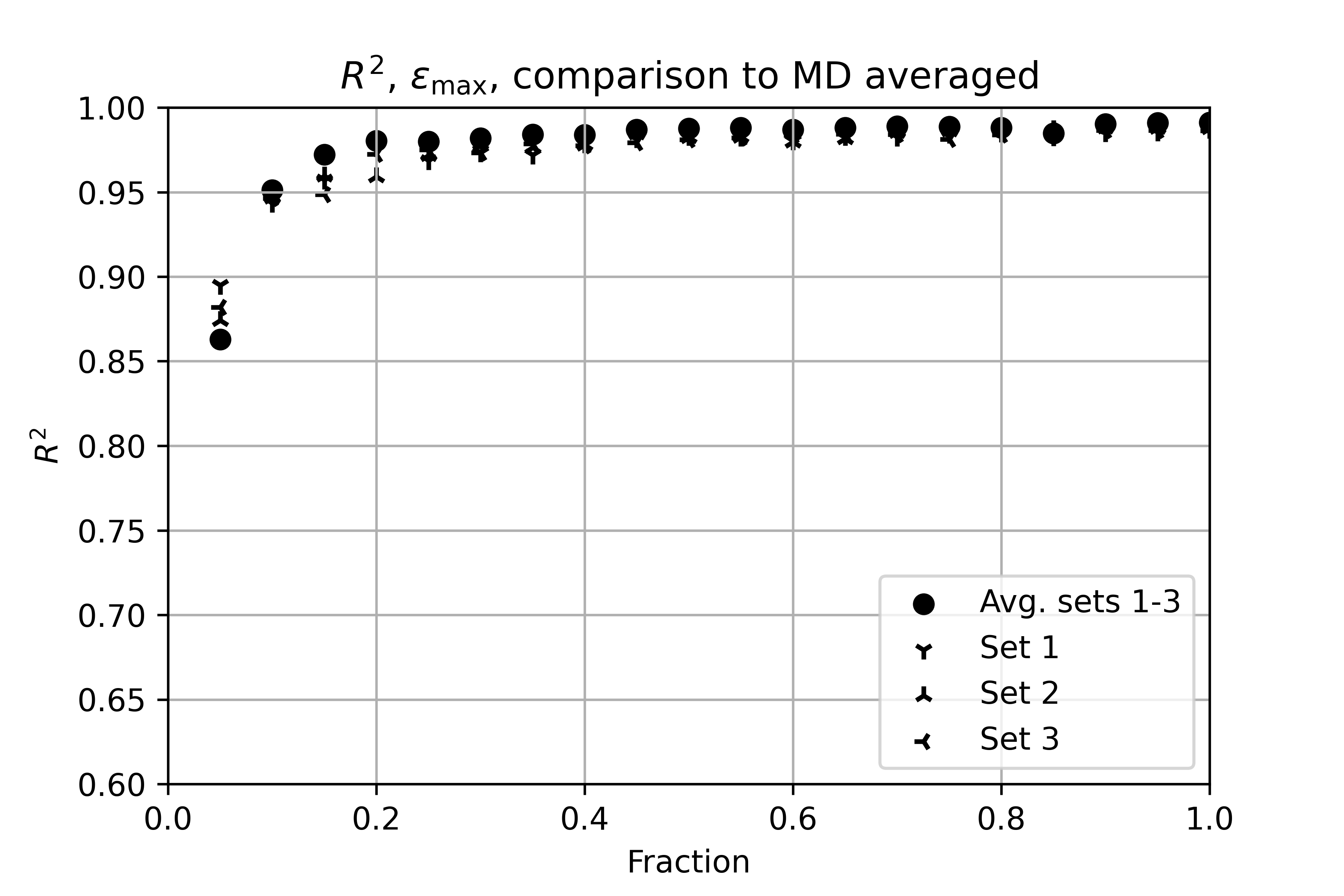}}
	\\
	\caption{Comparison of $R^2$ for different analysis set sizes to MD results obtained with averaged set and individual sets 1-3. Each point is obtained as a mean value of 10 different fraction sets. (a) Young's modulus, (b) Poisson's ratio, (c) Ultimate tensile strength, (d) Fracture strain.}	
	\label{fig:R2}
\end{figure}

The obtained dependence on the fractional size is illustrated in Fig.~\ref{fig:R2}. Almost the same order of approximation is obtained for UTS and the fracture strain, even when the dataset is only one third of the averaged dataset. Single MD datasets show only slightly lower performance than the averaged one (with the exception of the lowest fraction). This is quite convenient since the computation of these properties is expensive due to the requirement of a full MD simulation. A similar conclusion can be drawn for Poisson's ratio, although the difference between averaged data and single sets is more pronounced. Either the averaged approach with 30\% of the averaged set (or 245 out of 818 possible configurations) or a single set of MD simulations but on 75\% dataset size (or 614 configurations) should be chosen to achieve approximately the same level of prediction as the full set of averaged simulations. For the Young's modulus differences between averaged/single sets are even more pronounced, indicating the averaged approach as the recommended one. Thus, a larger data set is preferred especially for the determination of Young's modulus, but also Poisson's ratio. The computational effort compared to the determination of UTS and fracture strain should not be higher, since the determination of the former properties does not require a complete MD simulation, but only the initial part of the tensile test. It should also be noted that the time required to train the present DNN is only a few minutes on an average PC for the complete dataset, compared to more than two weeks required to run three full sets of MD simulations. No effort has been made to document the time required to train the DNN for different fractions that are even less demanding, but there is a general trend toward shorter training times for smaller fractions. 

\begin{figure}[htp]
	\centering
	\subcaptionbox {\label{fig:ML_E}}  {\includegraphics[scale=0.43]{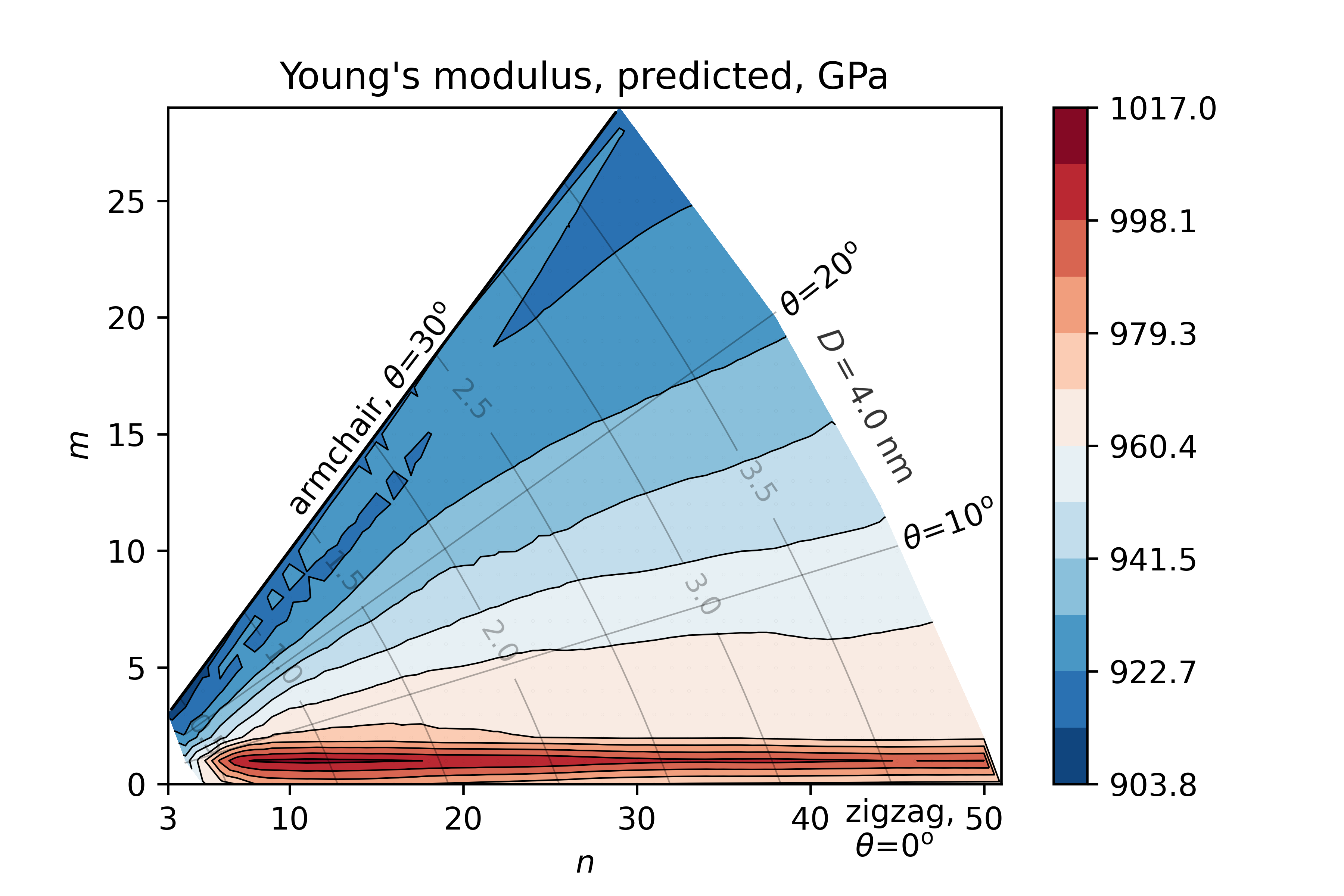}}
	\subcaptionbox {\label{fig:ML_P}}  {\includegraphics[scale=0.43]{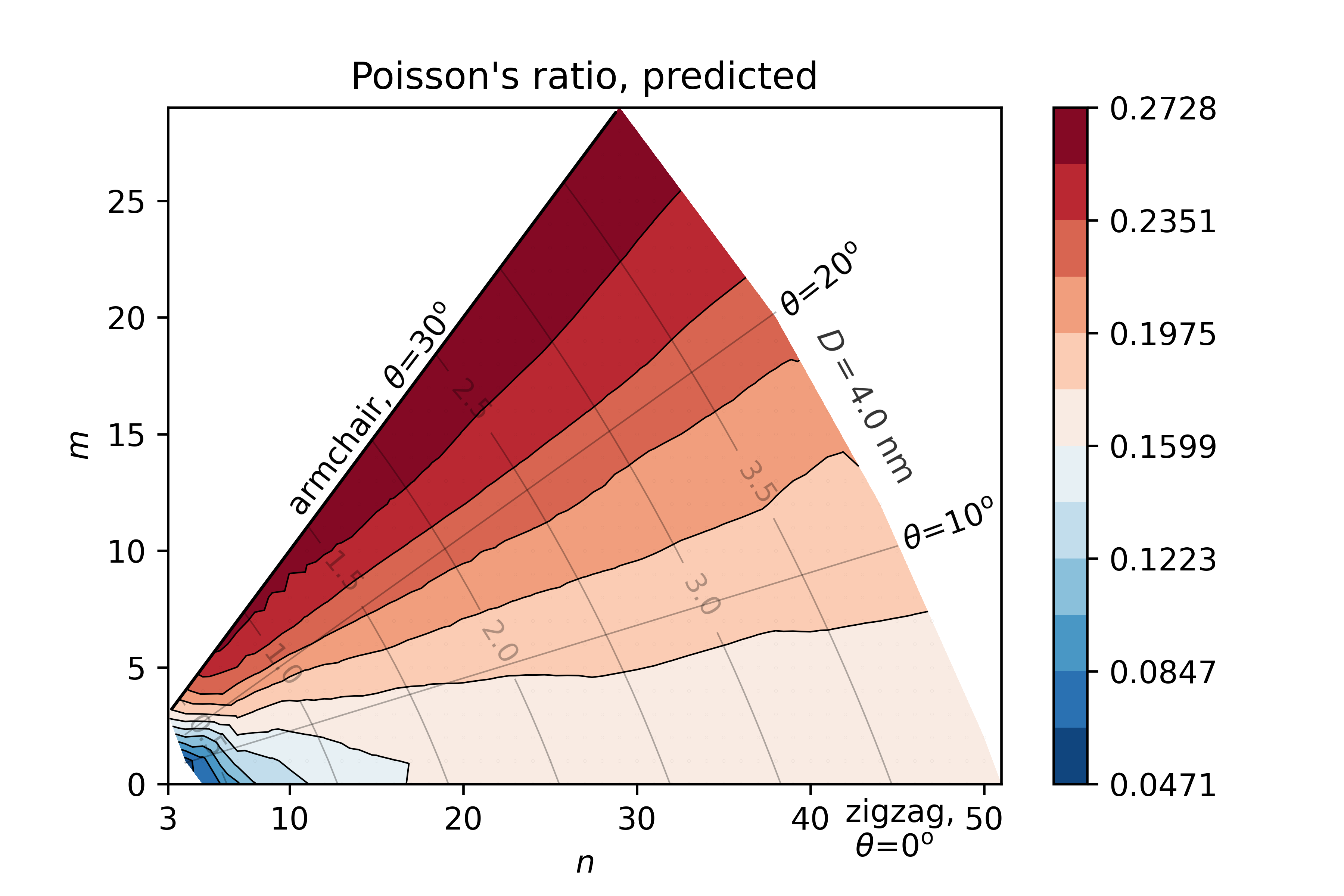}}
	\\
	\subcaptionbox {\label{fig:ML_UTS}}{\includegraphics[scale=0.43]{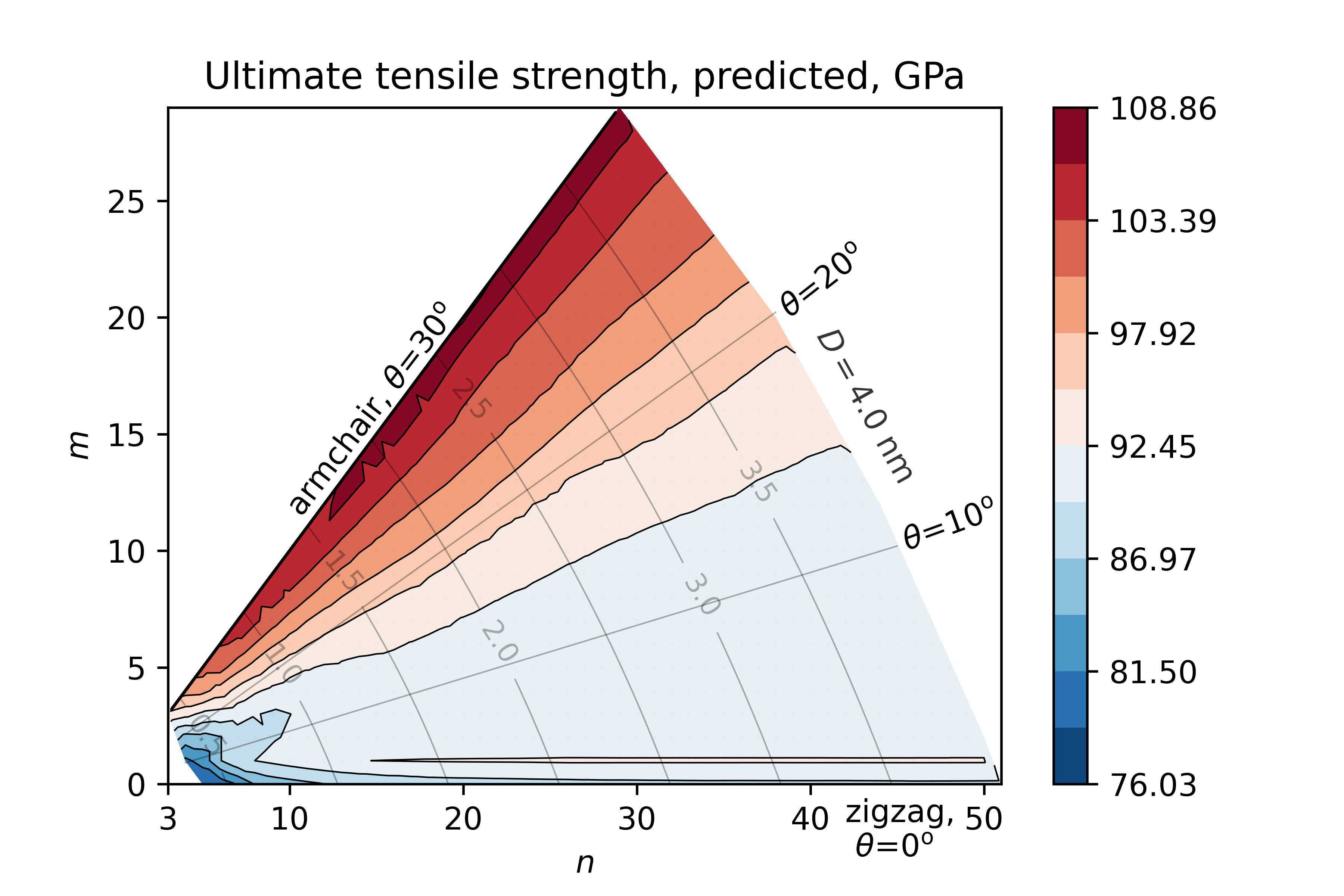}}
	\subcaptionbox {\label{fig:ML_FS}} {\includegraphics[scale=0.43]{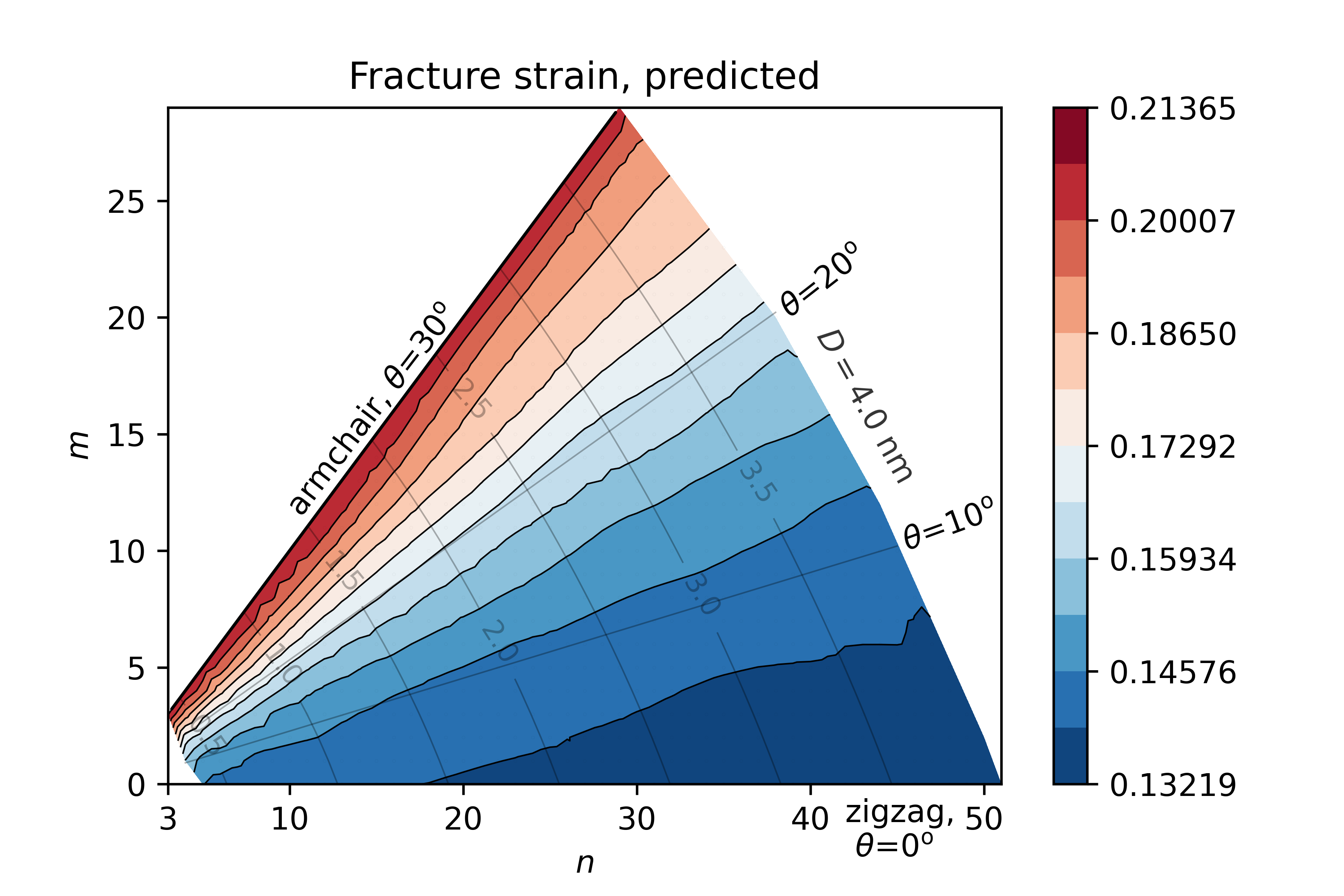}}
	\\
	\caption{Mechanical properties of SWCNT as predicted by DNN from the averaged MD dataset. (a) Young's modulus, (b) Poisson's ratio, (c) Ultimate tensile strength, (d) Fracture strain.}
	\label{fig:ML}	
\end{figure}

For the convenience of the reader, other results of interest are graphically documented in the Appendix. The comparison of the obtained distributions for each property, but in the case of a complete individual set (Set 1), is provided in Fig.~\ref{fig:AppSingleML}, cf. for the averaged set of results Figs.~\ref{fig:MD},\ref{fig:ML}. Both MD simulation and prediction results are given. Although the MD results of a single set clearly show more non-uniform distributions than the averaged ones, this is not the case for the DL predictions, confirming the hypothesis of the beneficial influence of the deep learning methodology on the determination of the properties of nanoscopic structures.

The effect of the size of the dataset on the distribution of a particular property, as computed for a single set of MD results, is given in Figs.~\ref{fig:AppfracML_YM}-\ref{fig:AppfracML_FS}. The similar results, but for the averaged set, are shown in Figs.~\ref{fig:AppfracML_YMavg}, \ref{fig:AppfracML_FSavg}. Again, the beneficial effect of DNN averaging is clearly noticed, especially in the case of Young's modulus and Poisson's ratio, while the favorable distributions of the ultimate tensile strength and fracture strain is obtained even for the smaller sizes of the dataset. These observations can be further supported by analyzing $R^2$ values, if instead of MD results as the reference DL predictions Fig.~\ref{fig:ML} are used, Fig.~\ref{fig:AppR2_ML}. It is clear that the same prediction quality is obtained for the dataset obtained by a single set of simulations as small as 35\% and 20\% (or 286 and 164 configurations) for UTS and fracture strain, respectively. As with the Young's modulus and Poisson's ratio, the beneficial effect of the dataset size and DNN averaging is again noticed. 

\begin{table}
	\centering
	\begin{tabular}{|c|c|c|c|c|}
		\hline
		& Complete  & Training & Validation & Test   \\
		\hline
		$R^2(E)$		 		 & 0.963	 & 0.974	& 0.966		 & 0.908  \\
		\hline
		$R^2(\nu)$		 		 & 0.964	 & 0.963	& 0.959		 & 0.969  \\
		\hline
		$R^2(\sigma_\mathrm{max})$	 & 0.992	 & 0.993	& 0.991	 	 & 0.986  \\
		\hline
		$R^2(\epsilon_\mathrm{max}$)  & 0.993	 & 0.994	& 0.991	 	 & 0.989  \\
		\hline \hline
		Loss					 & 0.00373	 & 0.00249	& 0.00341	 & 0.00531 \\
		\hline
	\end{tabular}
	\caption{Averaged $R^2$ and loss on ten random splits into training/validation/test sets.} 
	\label{tab:R2}
\end{table}

\section{Closing remarks}
The main findings of the present research can be summarized as follows. A comprehensive series of MD simulations were carried out, and Young's moduli, Poisson's ratios, ultimate tensile strengths and fracture strains for all SWCNTs with diameters less than 4.0 nm were determined. The tensile tests were carried out at 300 K. The simulation results reveal that the armchair configuration has the lowest Young's modulus but the highest Poisson's ratios, ultimate tensile strengths and fracture strains. Zigzag configurations show the opposite behavior - the highest Young's modulus, but lower other mechanical properties. To be more precise, the maximum Young's modulus is not found in the zigzag configuration, but in chiral SWCNTs close to zigzag configurations ($n$,1). Otherwise, the distribution is rather uniform, with a strong correlation of the mechanical properties with the chiral angle. The diameter seems to be important only for the SWCNTs with the smallest diameter, although such a conclusion might follow from the convention that the cross section is a thin circular ring, rather than an irregular polygon.

Based on the MD results, a deep learning neural network was developed. The obtained results clearly indicate that the neural network can predict mechanical properties with high accuracy. It is also shown that such an approach has a beneficial effect on smoothing noise originating from thermal fluctuations of carbon atoms. Although standard averaging and outlier removal in the preprocessing stage also filters out a certain amount of noise, DNN results are significantly smoother than averaged MD results. Nevertheless, averaging still represents an important part of the procedure. In addition, a series of tests on datasets of different sizes shows that the most computationally intensive properties - the ultimate tensile strength and fracture strain can be obtained fairly accurately from a single set of simulations representing about 1/3 of the possible configurations. As for the Poisson's ratio, but especially for the Young's modulus, a more extensive set of simulations should be used, preferably averaging results from multiple MD simulations. However, this can be realized with much shorter simulations than for the former properties, which only concern the initial part of the tensile test. 

Although the applicability of the present results may be quite large, the reader is cautioned that the DL model was not tested on SWCNTs with diameters larger than 4.0 nm. Although the predictions are easy to obtain, their accuracy is unknown at the moment, which limits the application only to diameters smaller than 4.0 nm. Another limitation is that known inconsistencies between the AIREBO potential and experiments, as discussed in Sec.~\ref{subsec:potential}, are inherently included in the DL model. Although the possible availability of a better potential will certainly affect the predicted properties, it is not expected to significantly change the modeling strategy used in the present machine learning framework. 

Finally, the results of the research at hand provide hints on modeling strategies for investigations on further influencing variables on the mechanical behavior of SWCNT - like length or loading rate effects, at room but also at other temperatures. Such investigations should be based on MD simulations performed with datasets of the size described above and complemented by a carefully developed deep learned neural network for a complete representation of the sought properties.
 
\section*{Acknowledgments}
This work has been supported in by Croatian Science Foundation under the project IP-2019-04-4703. This support is gratefully acknowledged.

\section*{Research data}
\setcounter{figure}{0}
The input dataset file with data about all SWCNT configurations is available at Čanađija, Marko (2021), “SWCNT Dataset”, Mendeley Data, V1, doi: 10.17632/vwd34rvphw.1,  \url{http://dx.doi.org/10.17632/vwd34rvphw.1#file-d10cae96-ca17-40c4-8db8-b6c964ff6c74}.

\appendix

\section{Supplementary figures}
\setcounter{figure}{0}
The present investigation yielded a variety of results that are hopefully interesting for the reader. For the sake of clarity of presentation, only the most important illustrations are included in the main body of the manuscript, the others being presented in the Appendix. Comments are provided in the main text.

\begin{figure}
	\centering
	\includegraphics[scale=0.6]{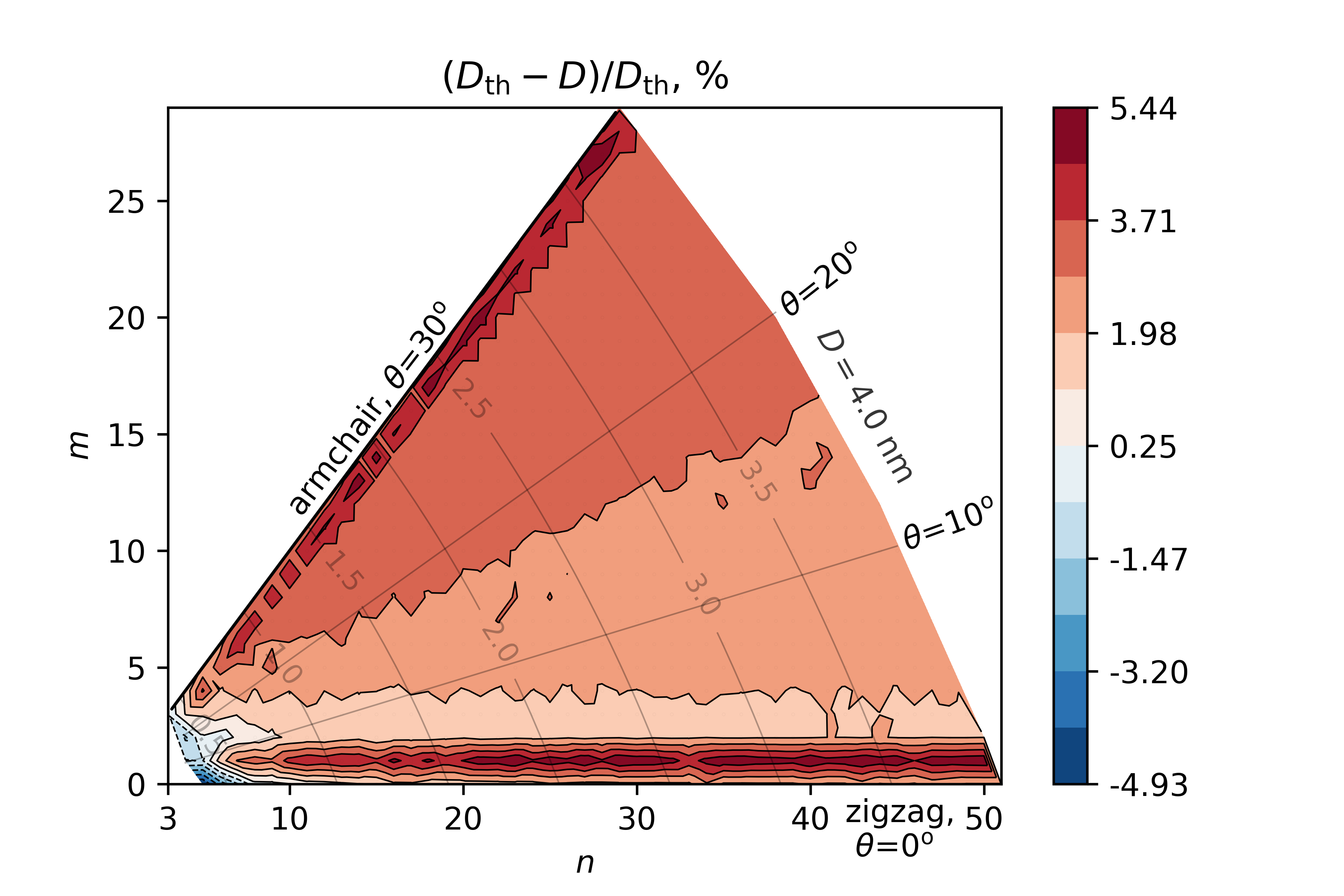}
	\caption{Differences (\%) in the theoretical diameter $D_\mathrm{th}$ and actual initial diameter $D$.} 
	\label{fig:app_D_vs_Dth}
\end{figure}

\begin{figure}[htp]
	\centering
	\subcaptionbox {\label{fig:App_IF2_MD_E}}  {\includegraphics[scale=0.43]{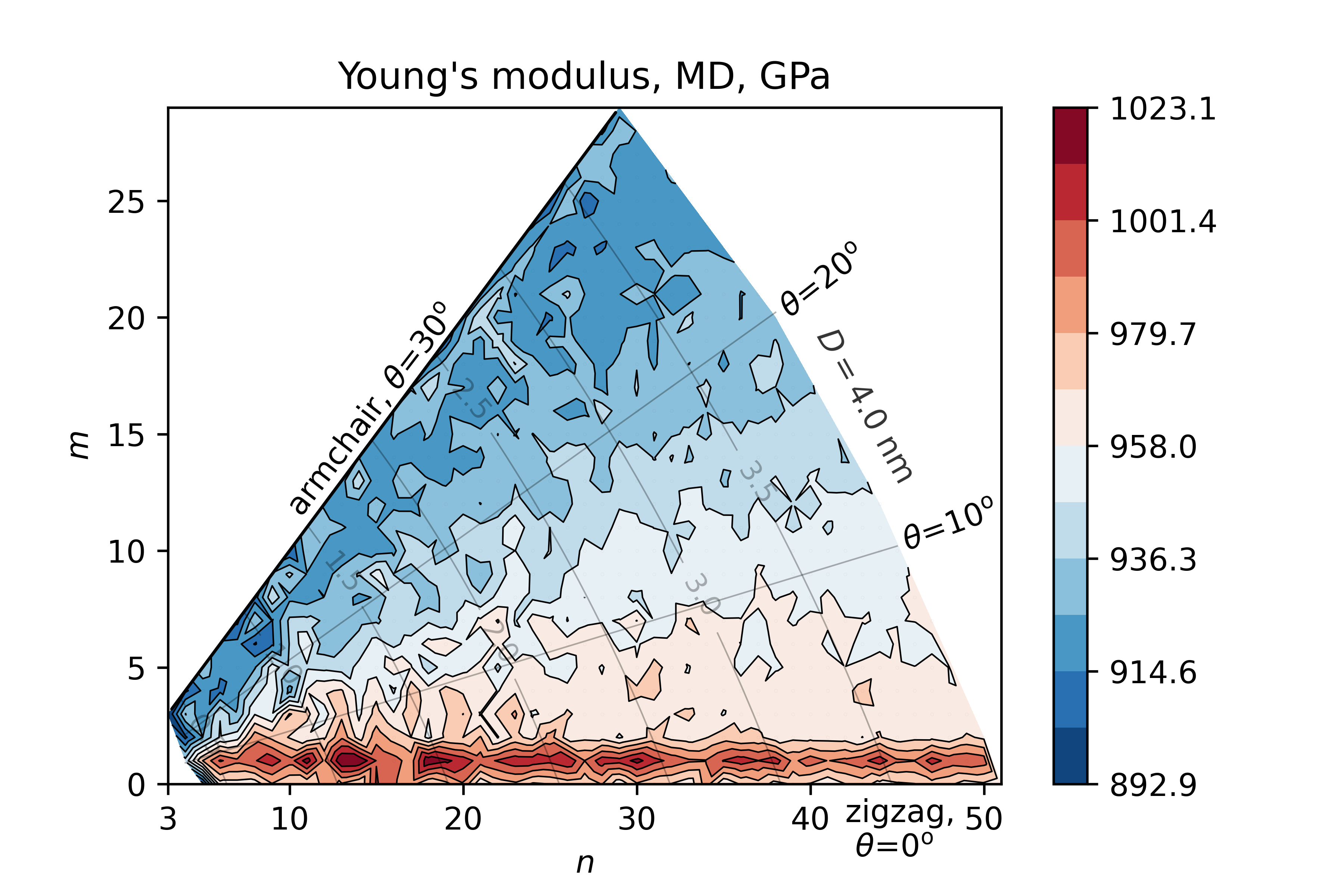}}
	\subcaptionbox {\label{fig:App_IF2_ML_E}}  {\includegraphics[scale=0.43]{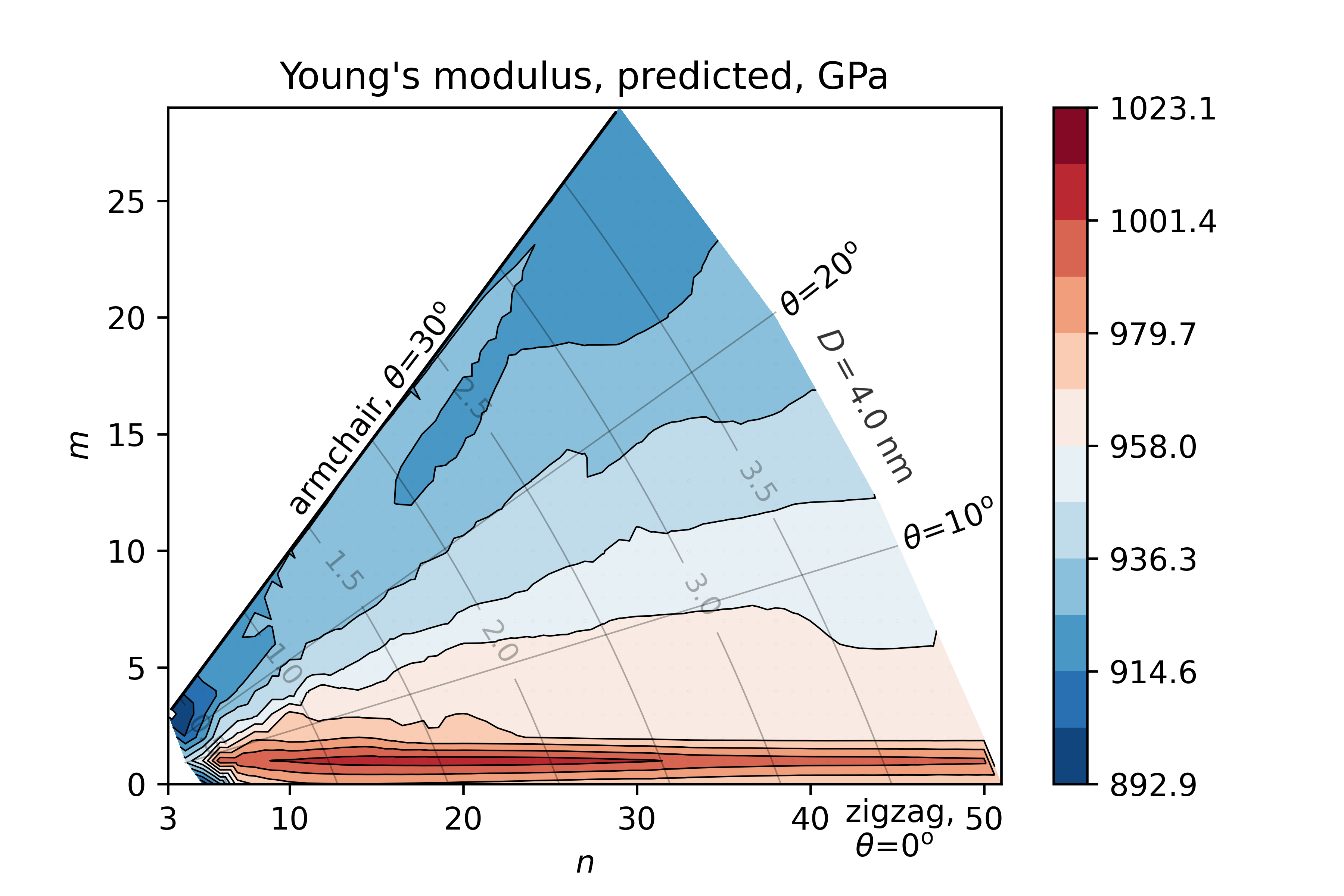}}
	\\
	\subcaptionbox {\label{fig:App_IF2_MD_P}}  {\includegraphics[scale=0.43]{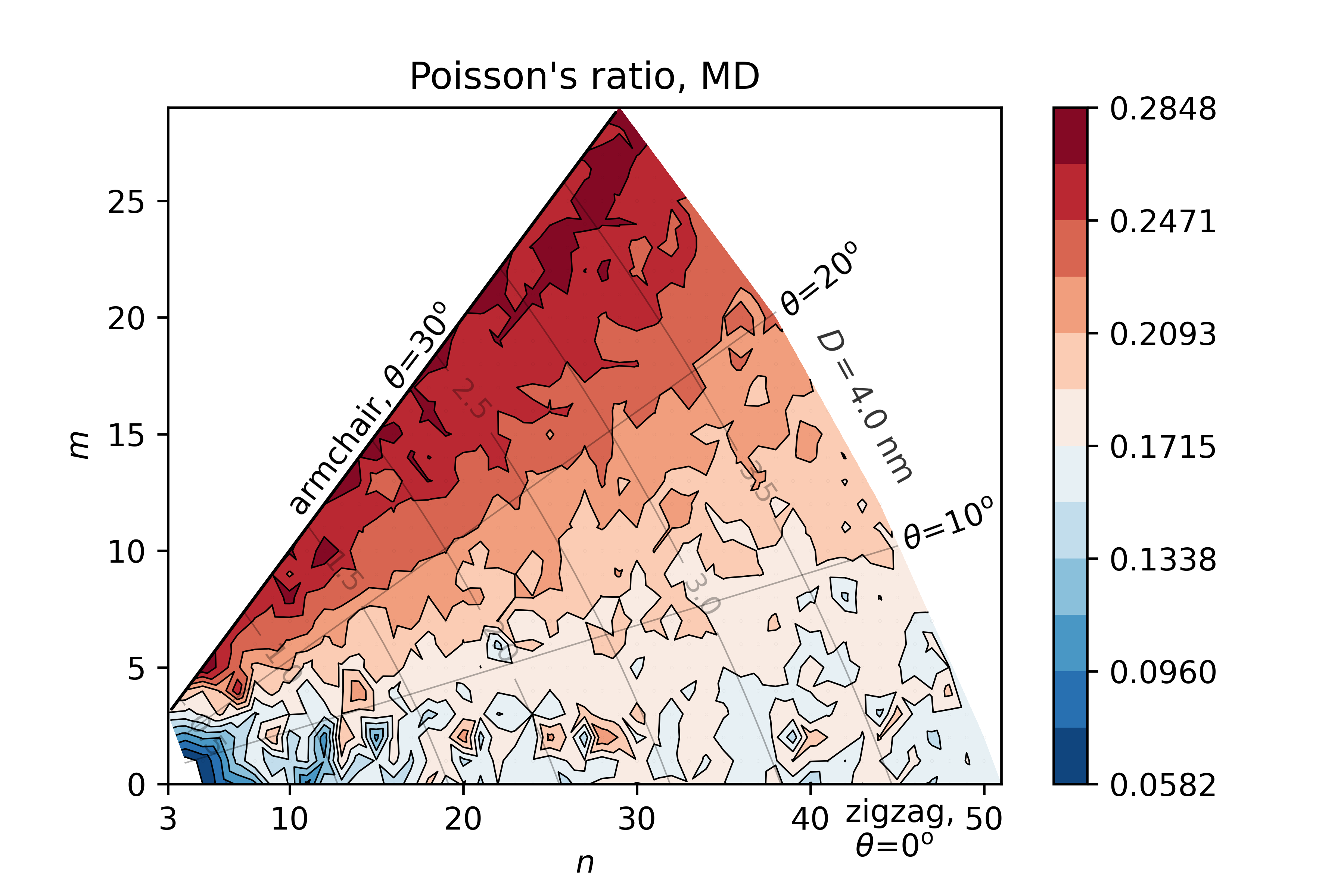}}
	\subcaptionbox {\label{fig:App_IF2_ML_P}}  {\includegraphics[scale=0.43]{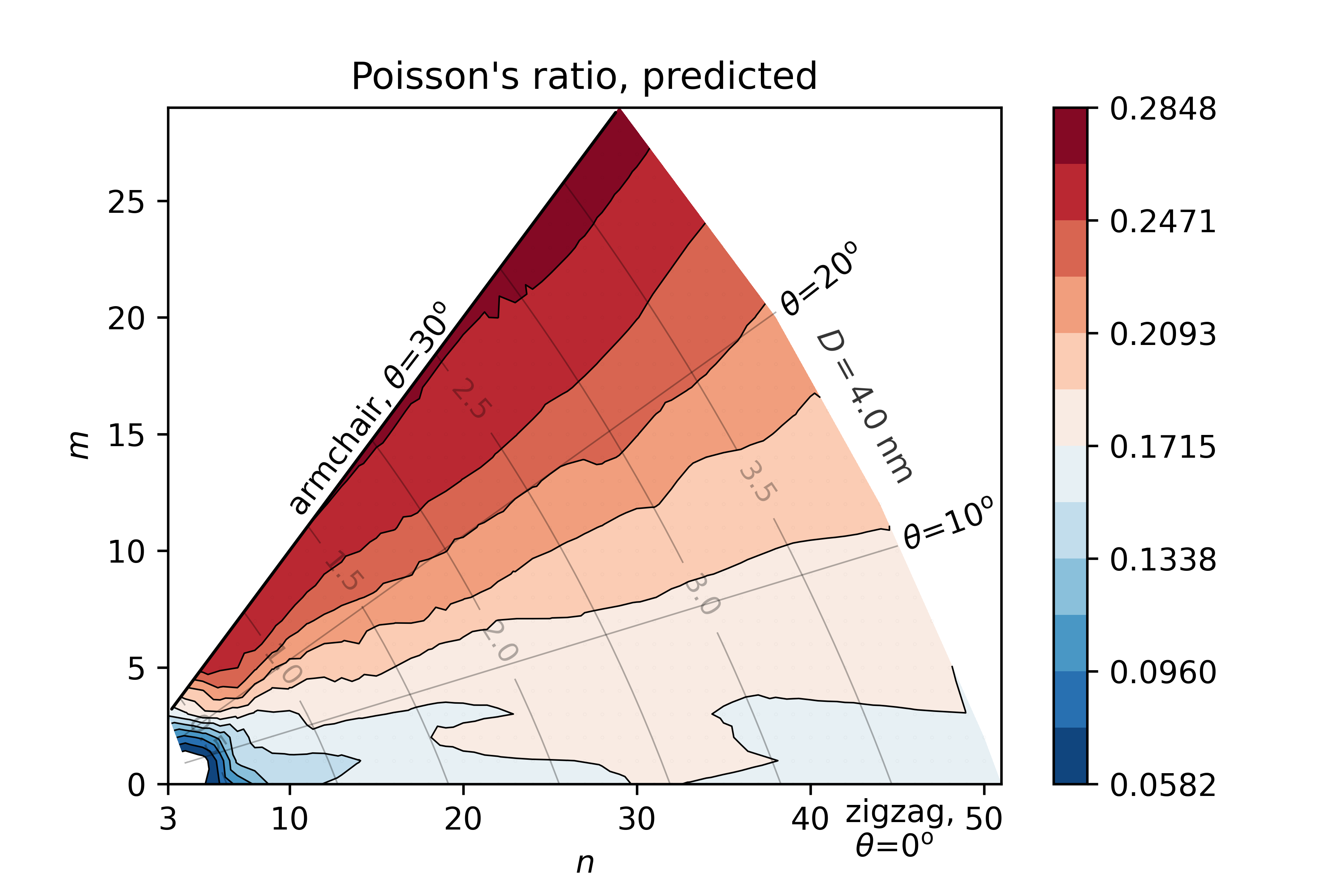}}
	\\
	\subcaptionbox {\label{fig:App_IF2_MD_UTS}}  {\includegraphics[scale=0.43]{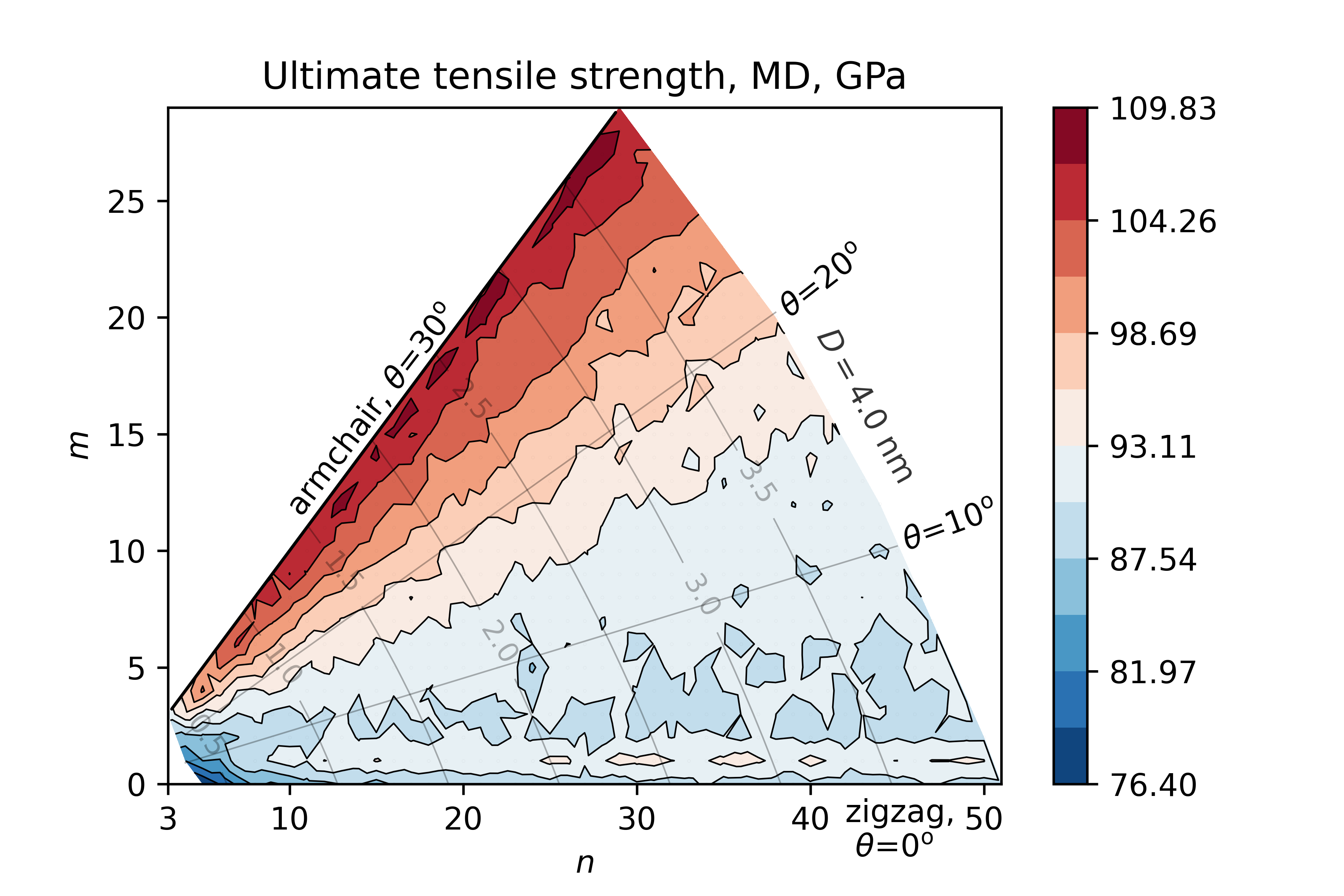}}
	\subcaptionbox {\label{fig:App_IF2_ML_UTS}}  {\includegraphics[scale=0.43]{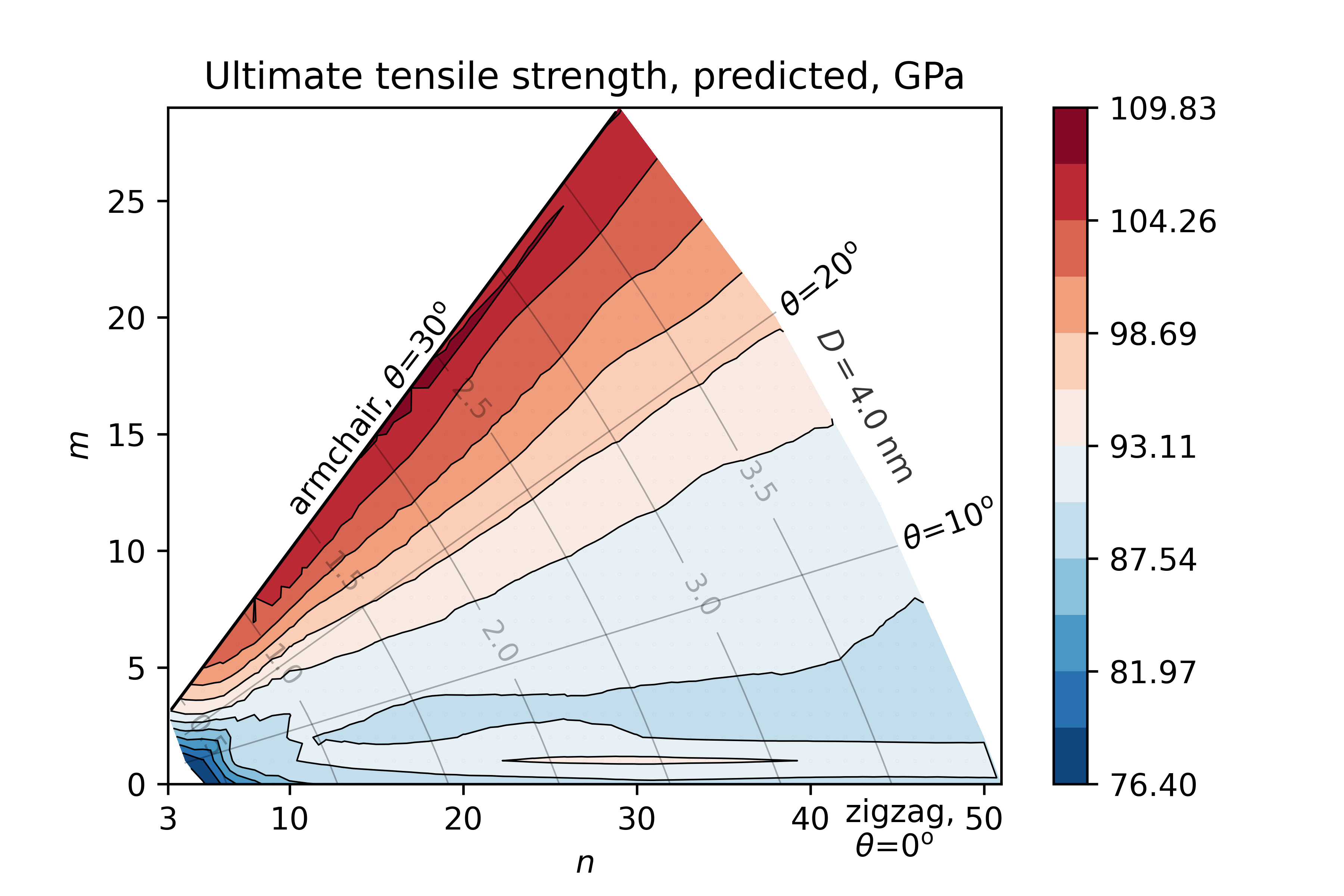}}
	\\
	\subcaptionbox {\label{fig:App_IF2_MD_FS}}  {\includegraphics[scale=0.43]{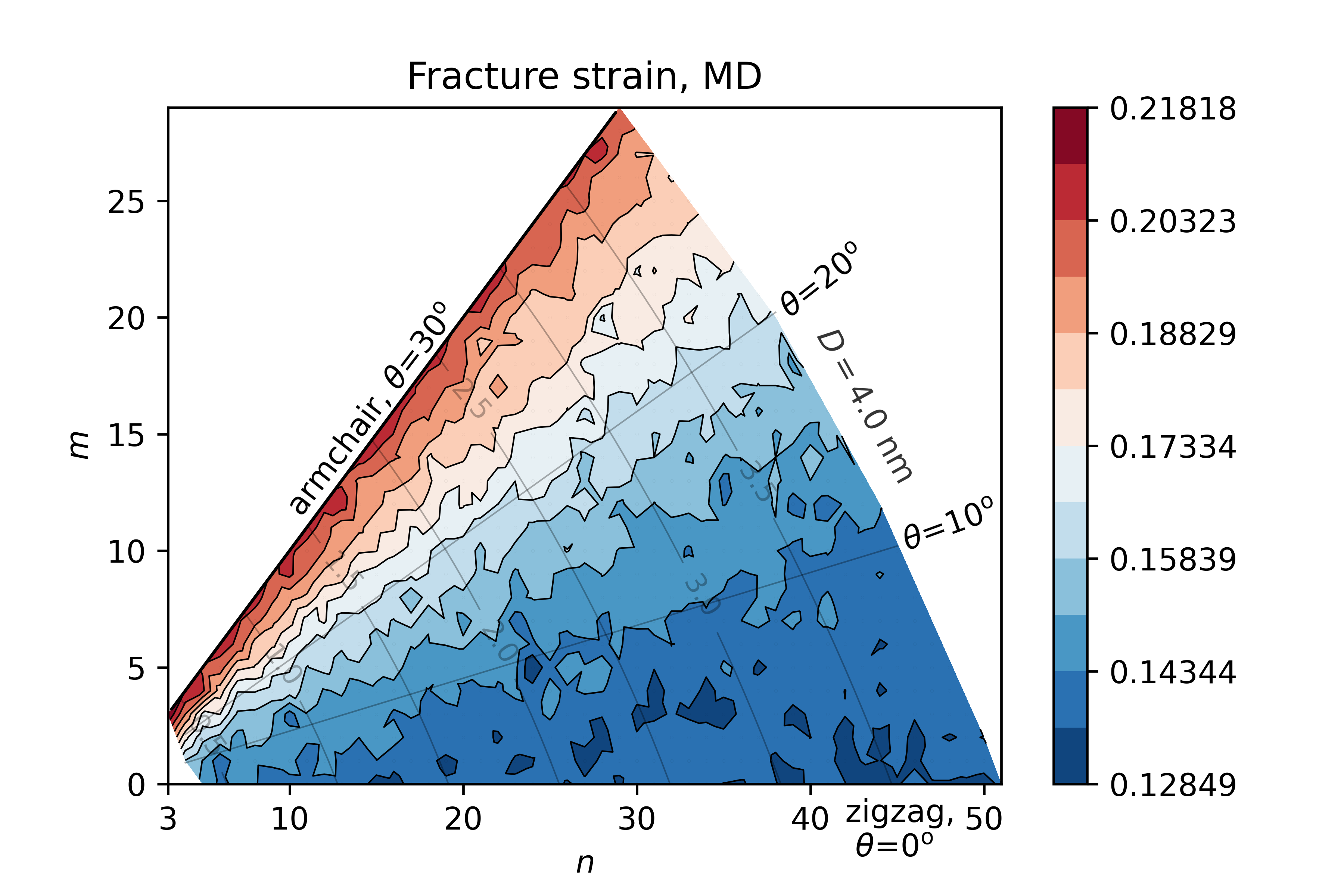}}
	\subcaptionbox {\label{fig:App_IF2_ML_FS}}  {\includegraphics[scale=0.43]{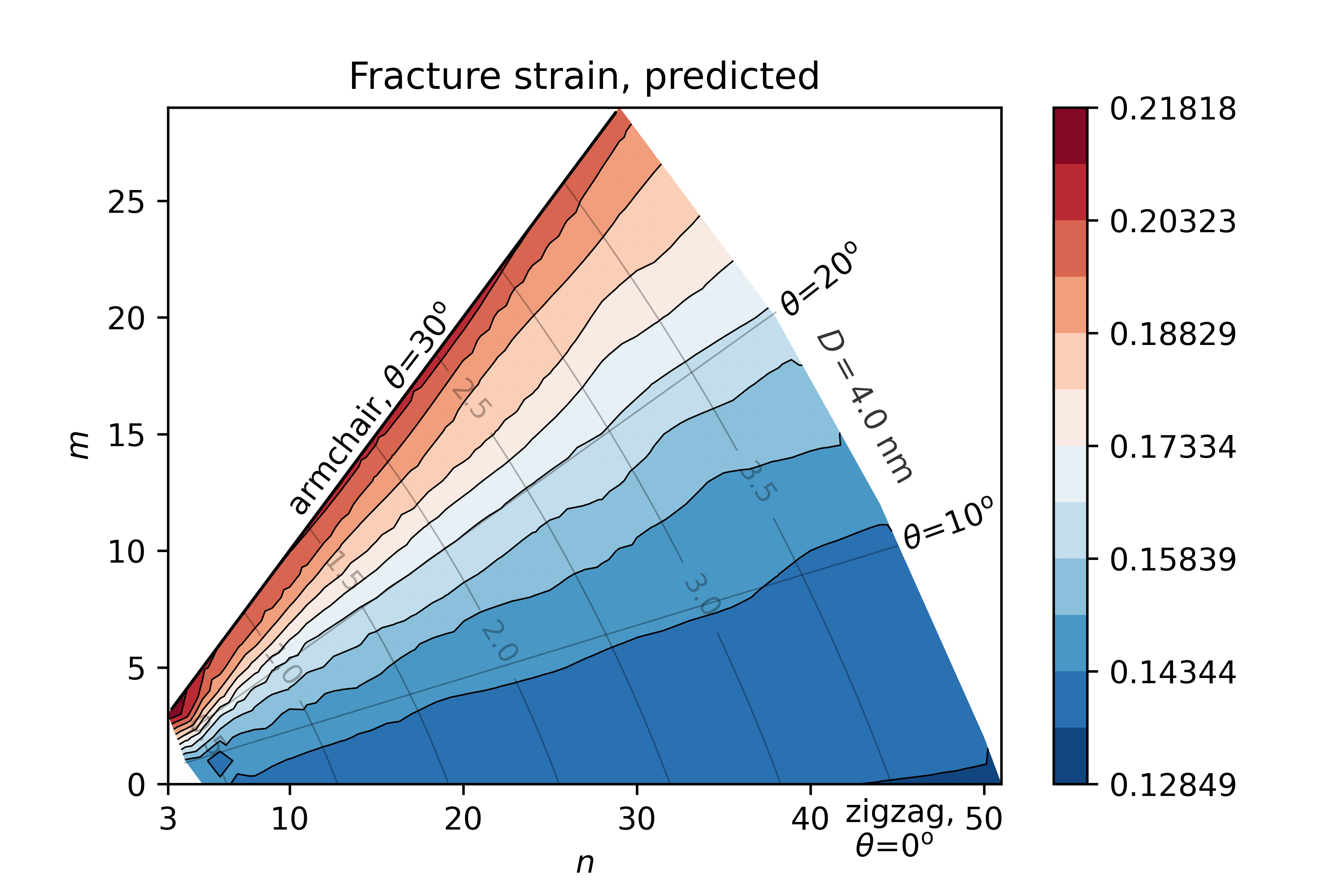}}
	\\
	\caption{Mechanical properties of SWCNT as calculated by MD on a single set of results (Set 1) and predicted by DNN. (a, b) Young's modulus (MD, DL), (c, d) Poisson's ratio (MD, DL), (e, f) Ultimate tensile strength (MD, DL), (g, h) Fracture strain (MD, DL).}	
	\label{fig:AppSingleML}
\end{figure}

\begin{figure}[htp]
	\centering
	\subcaptionbox {\label{fig:App_f10IF2_ML_E}}  {\includegraphics[scale=0.43]{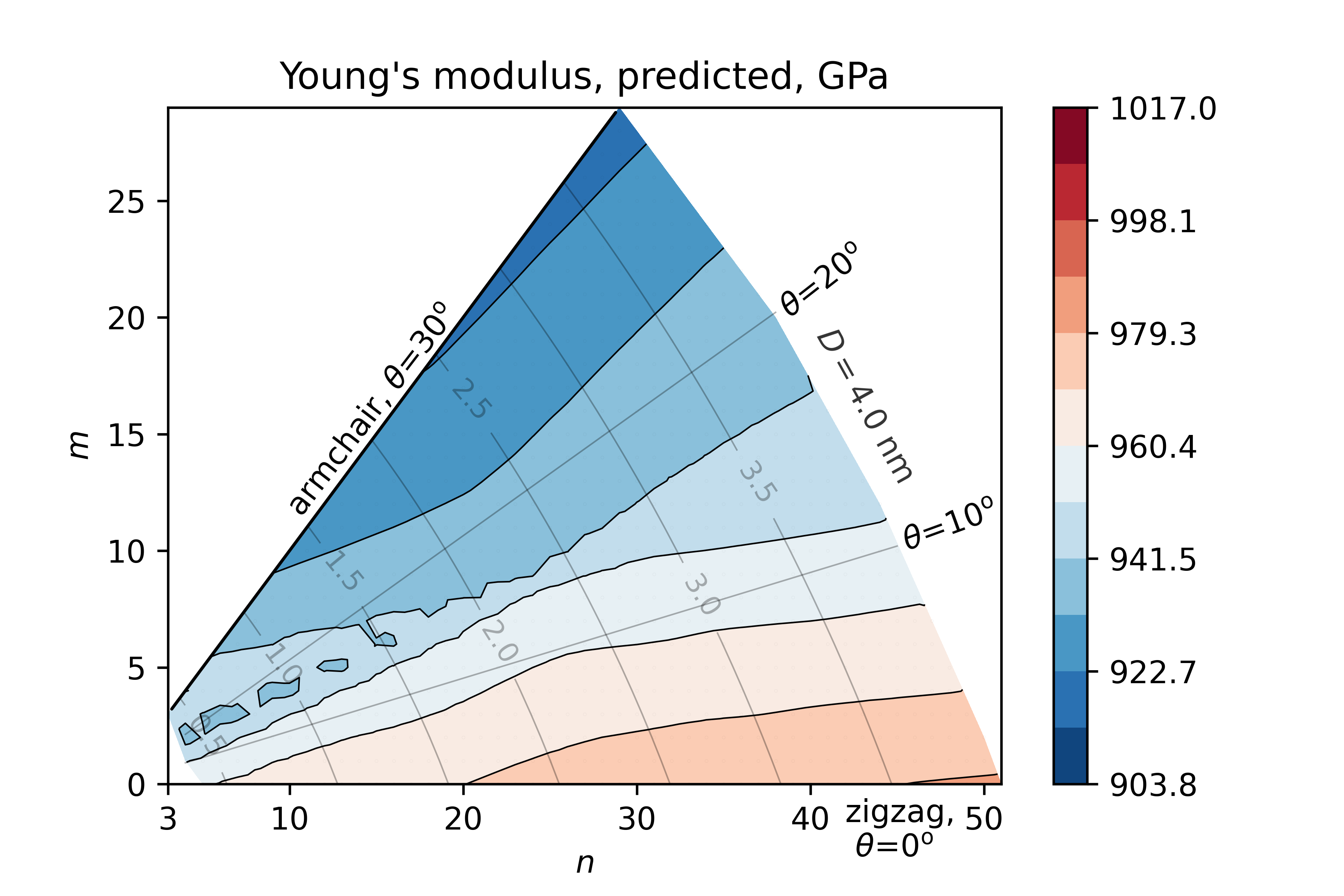}}	
	\subcaptionbox {\label{fig:App_f25IF2_ML_E}}  {\includegraphics[scale=0.43]{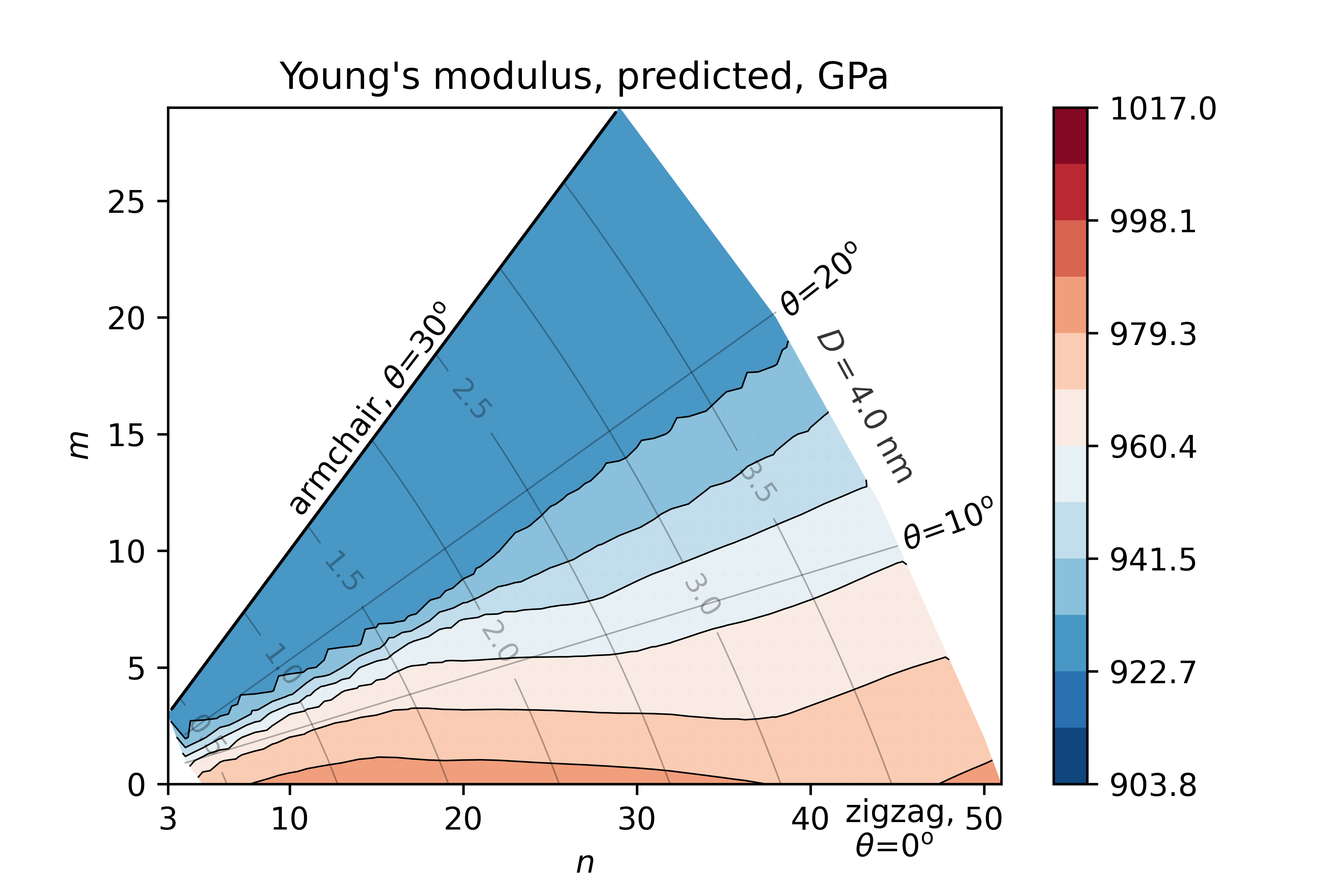}}
	\\
	\subcaptionbox {\label{fig:App_f50IF2_ML_E}}  {\includegraphics[scale=0.43]{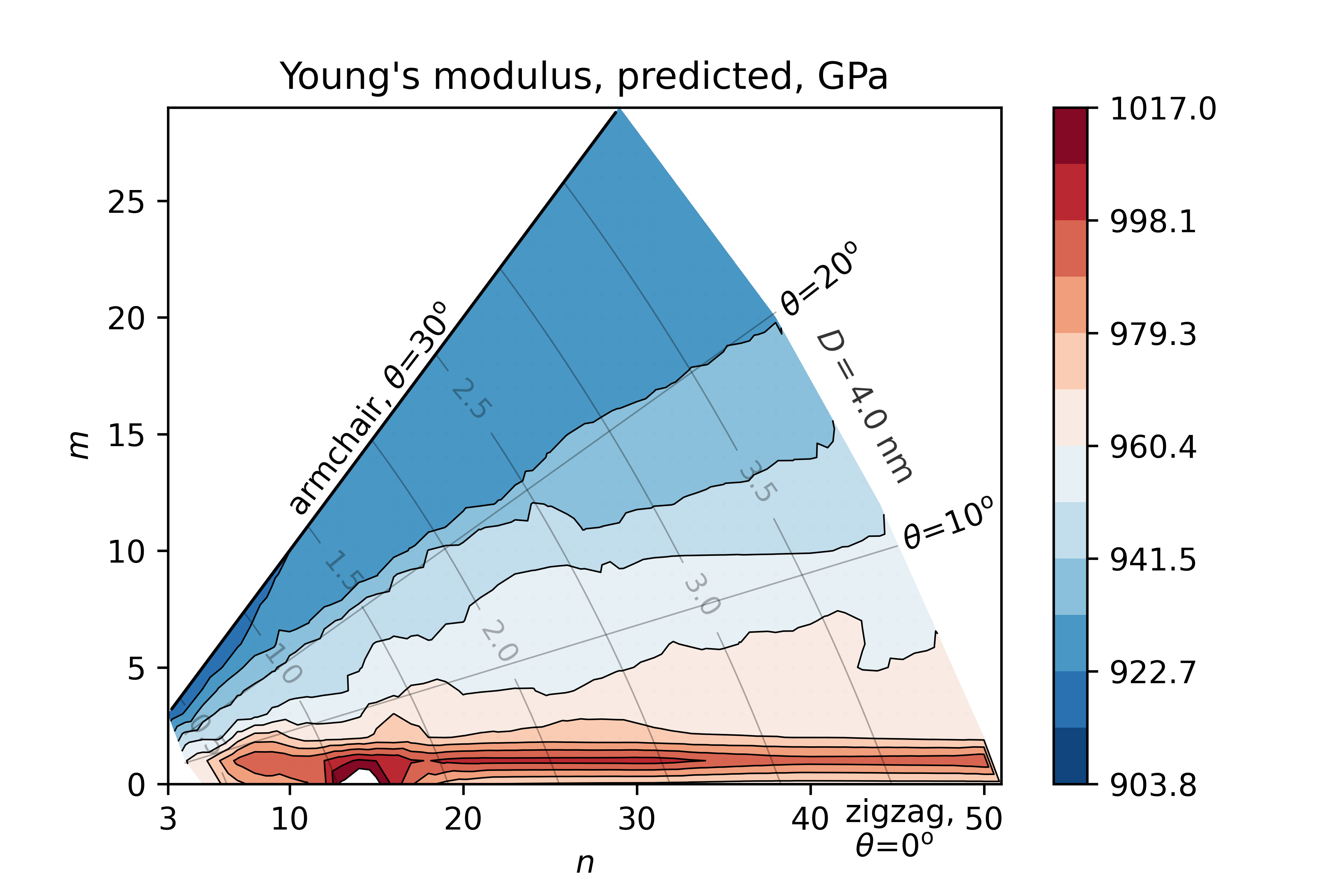}}
	\subcaptionbox {\label{fig:App_f75IF2_ML_E}}  {\includegraphics[scale=0.43]{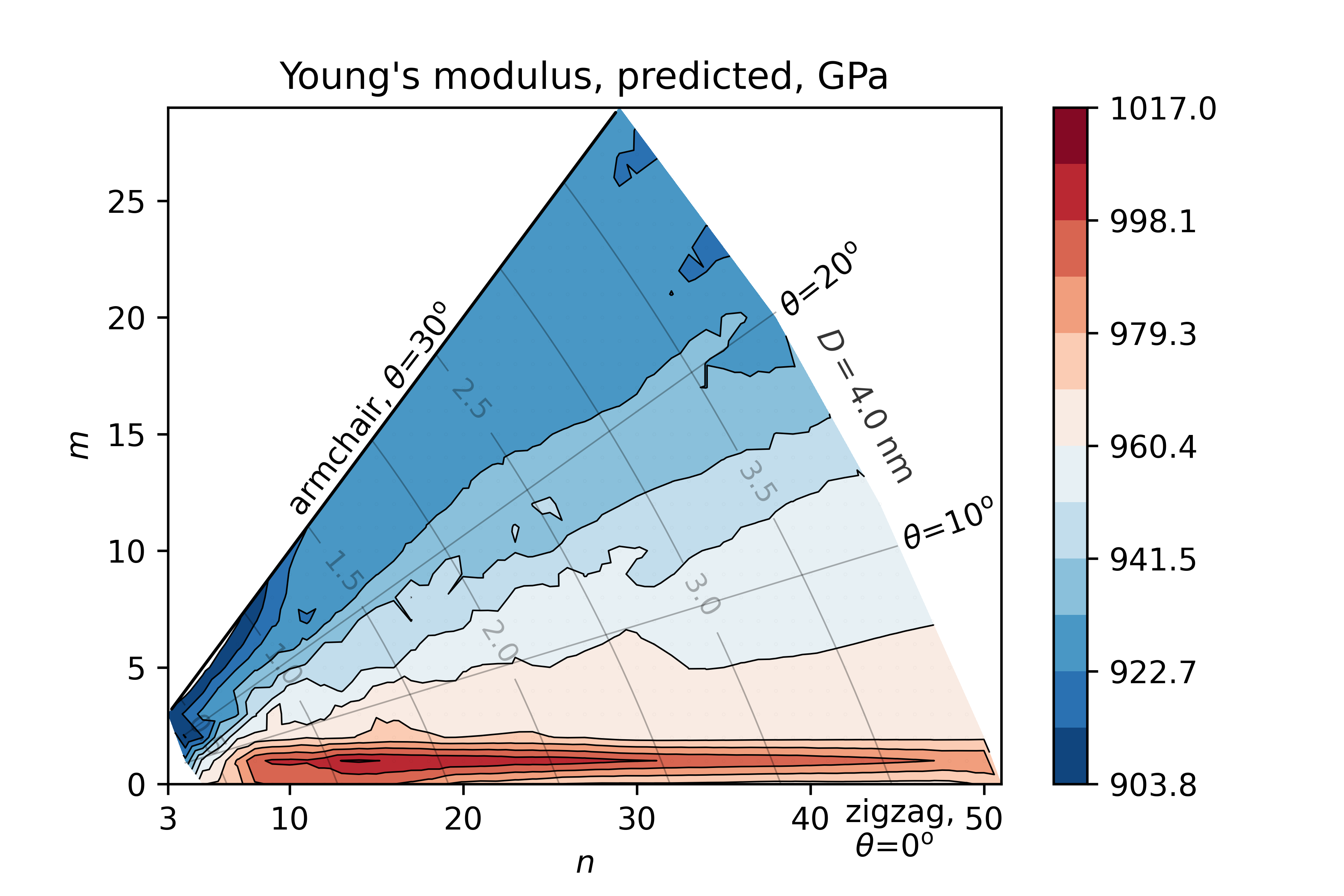}}
	\\
	\caption{Young's modulus of SWCNTs as predicted by DNN on a single set of results (Set 2) for different set sizes/fractions.  Contour limits are adjusted to MD averaged results, Fig.~\ref{fig:MD}. White areas indicate results outside of contour limits. (a, b, c, d) Set fractions/Number of configurations/$R^2$: 10\%/82/0.323, 25\%/205/0.449, 50\%/409/0.813, 75\%/614/0.699.}	
	\label{fig:AppfracML_YM}
\end{figure}

\begin{figure}[htp]
	\centering
	\subcaptionbox {\label{fig:App_f10IF2_ML_P}}  {\includegraphics[scale=0.43]{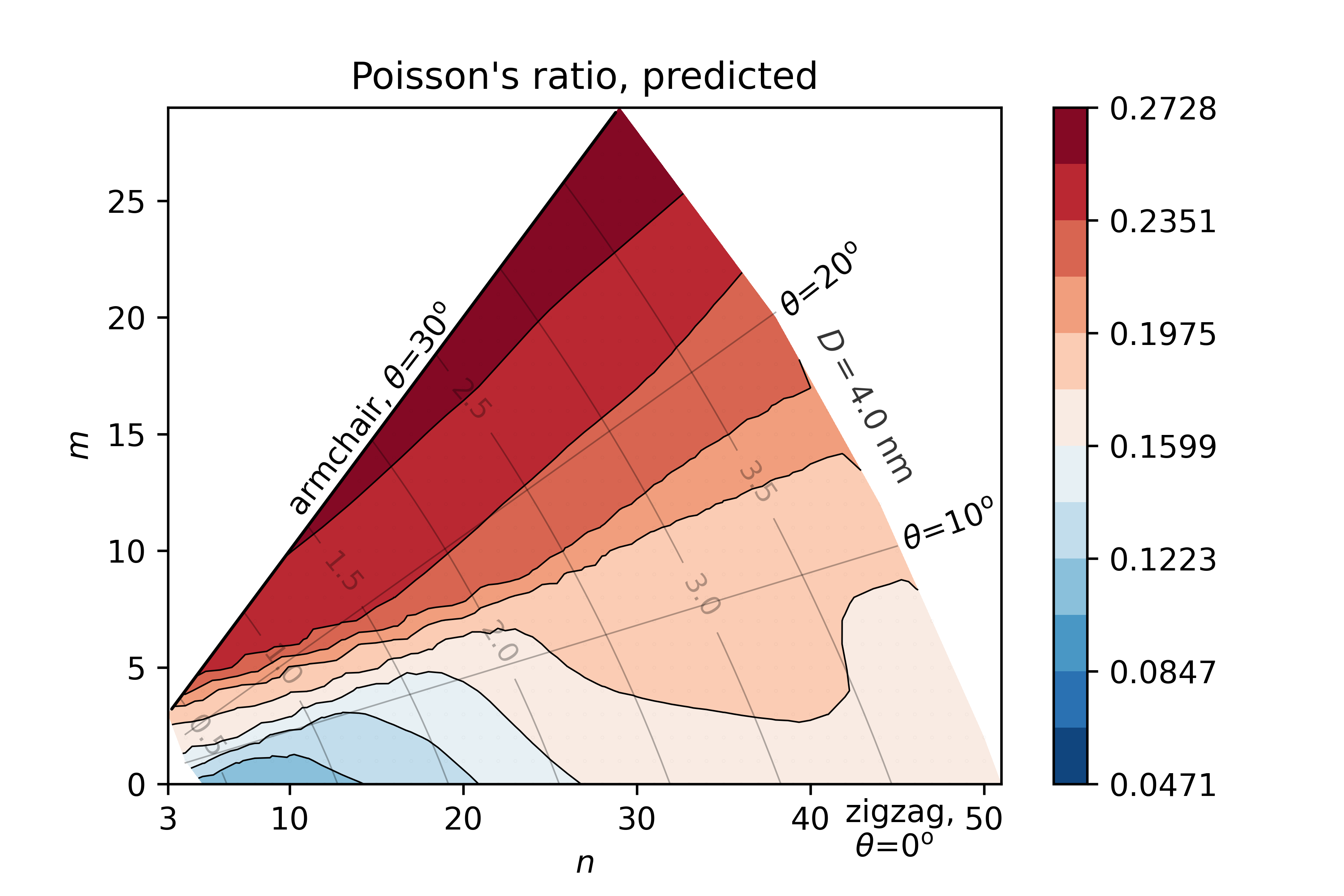}}
	\subcaptionbox {\label{fig:App_f25IF2_ML_P}}  {\includegraphics[scale=0.43]{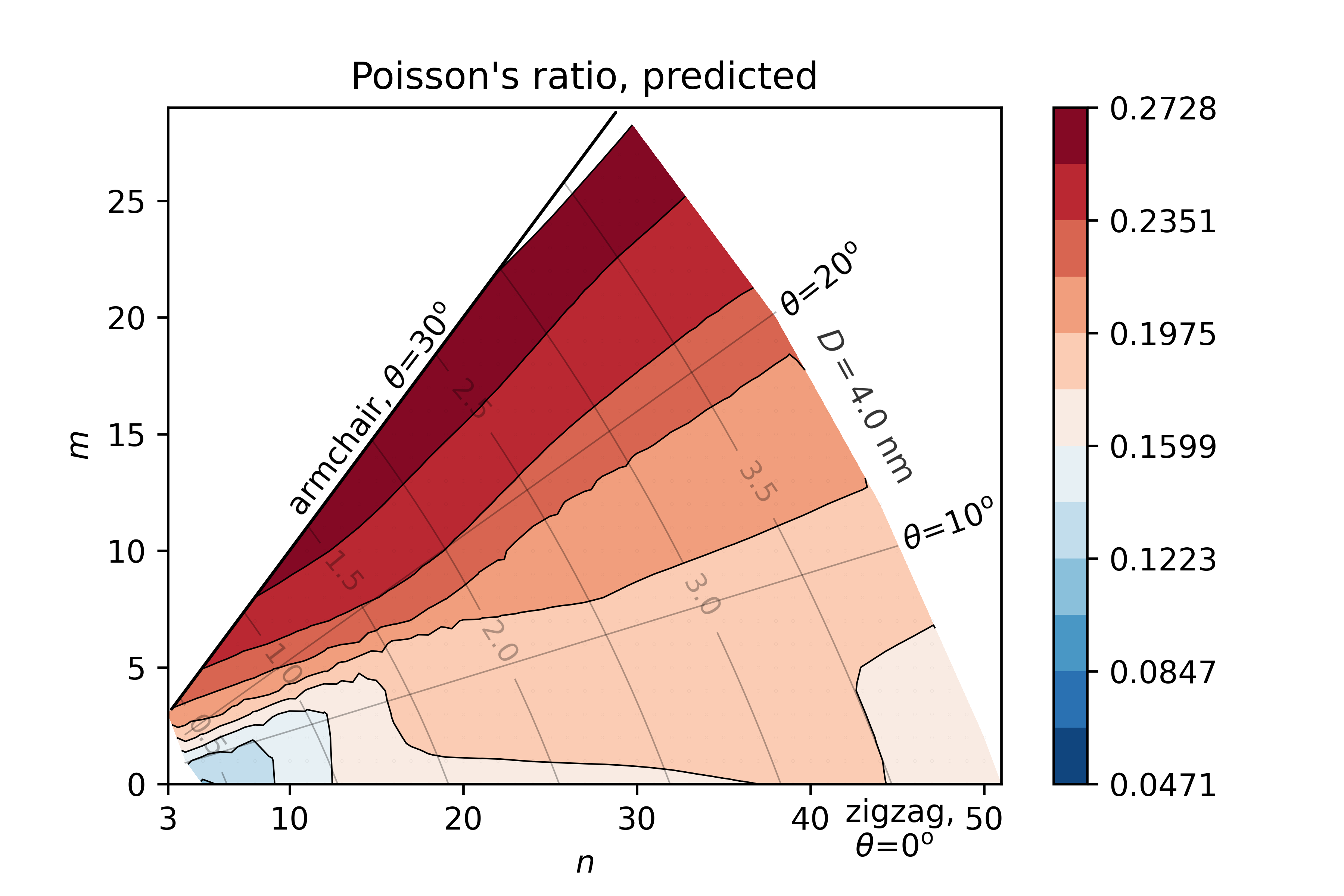}}
	\\	
	\subcaptionbox {\label{fig:App_f50IF2_ML_P}}  {\includegraphics[scale=0.43]{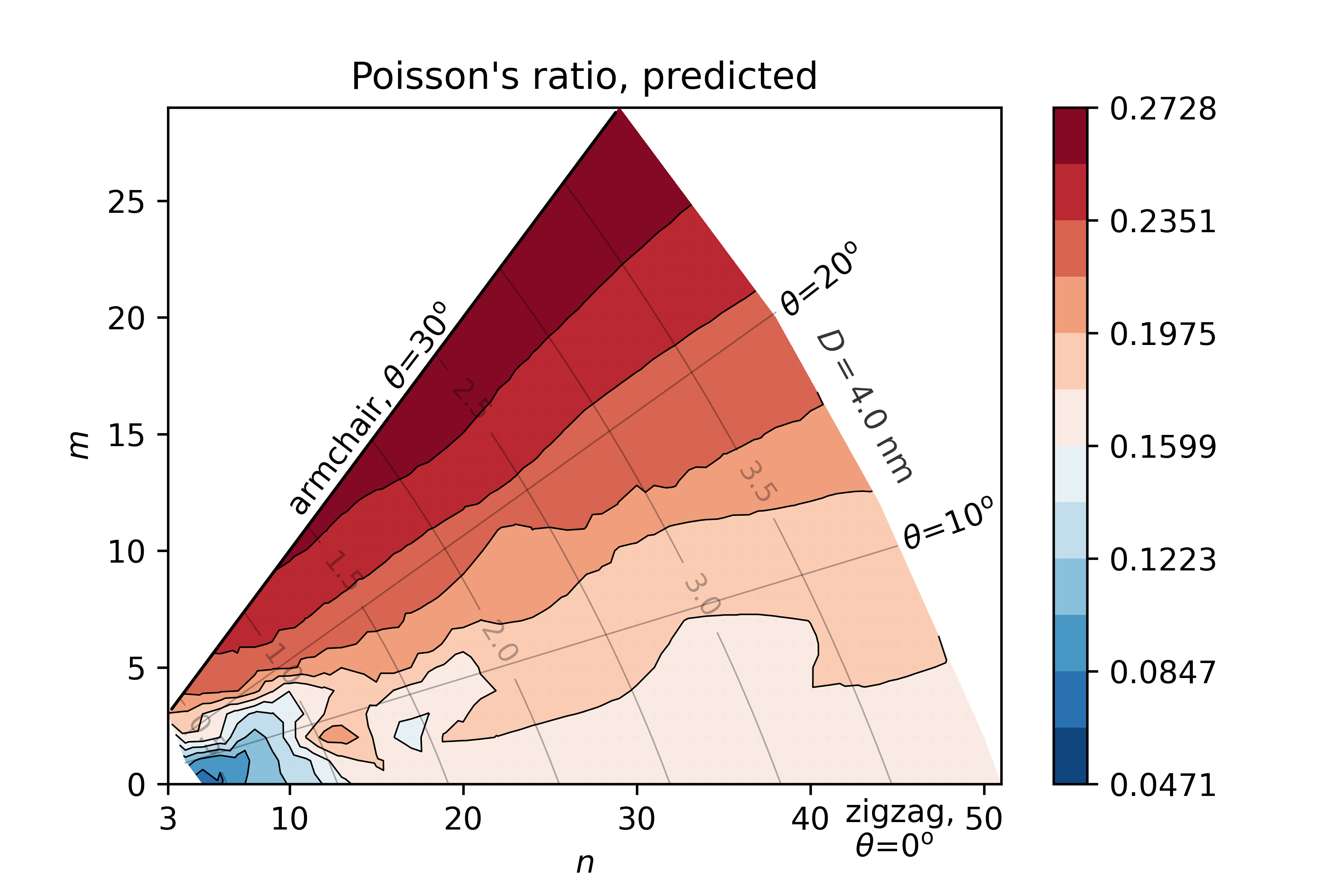}}
	\subcaptionbox {\label{fig:App_f75IF2_ML_P}}  {\includegraphics[scale=0.43]{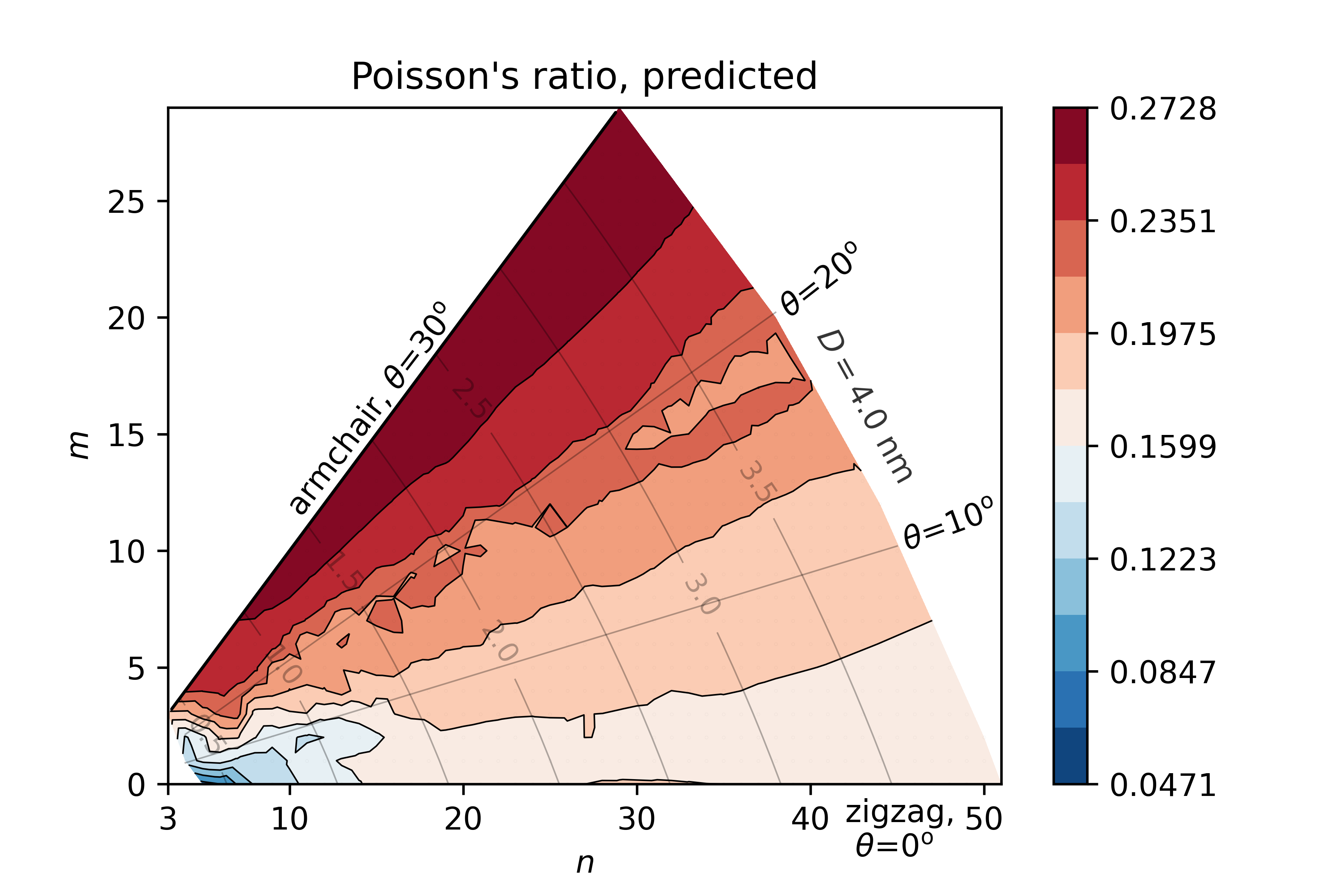}}
	\\
	\label{fig:AppfracML_P}\caption{Poisson's ratio of SWCNTs as predicted by DNN on a single set of results (Set 2) for different set sizes/fractions.  Contour limits are adjusted to MD averaged results, Fig.~\ref{fig:MD}. (a, b, c, d) Set fractions/$R^2$: 10\%/0.883, 25\%/0.908, 50\%/0.944, 75\%/0.925.}	
\end{figure}

\begin{figure}[htp]
	\centering
	\subcaptionbox {\label{fig:App_f10IF2_MD_UTS}}  {\includegraphics[scale=0.43]{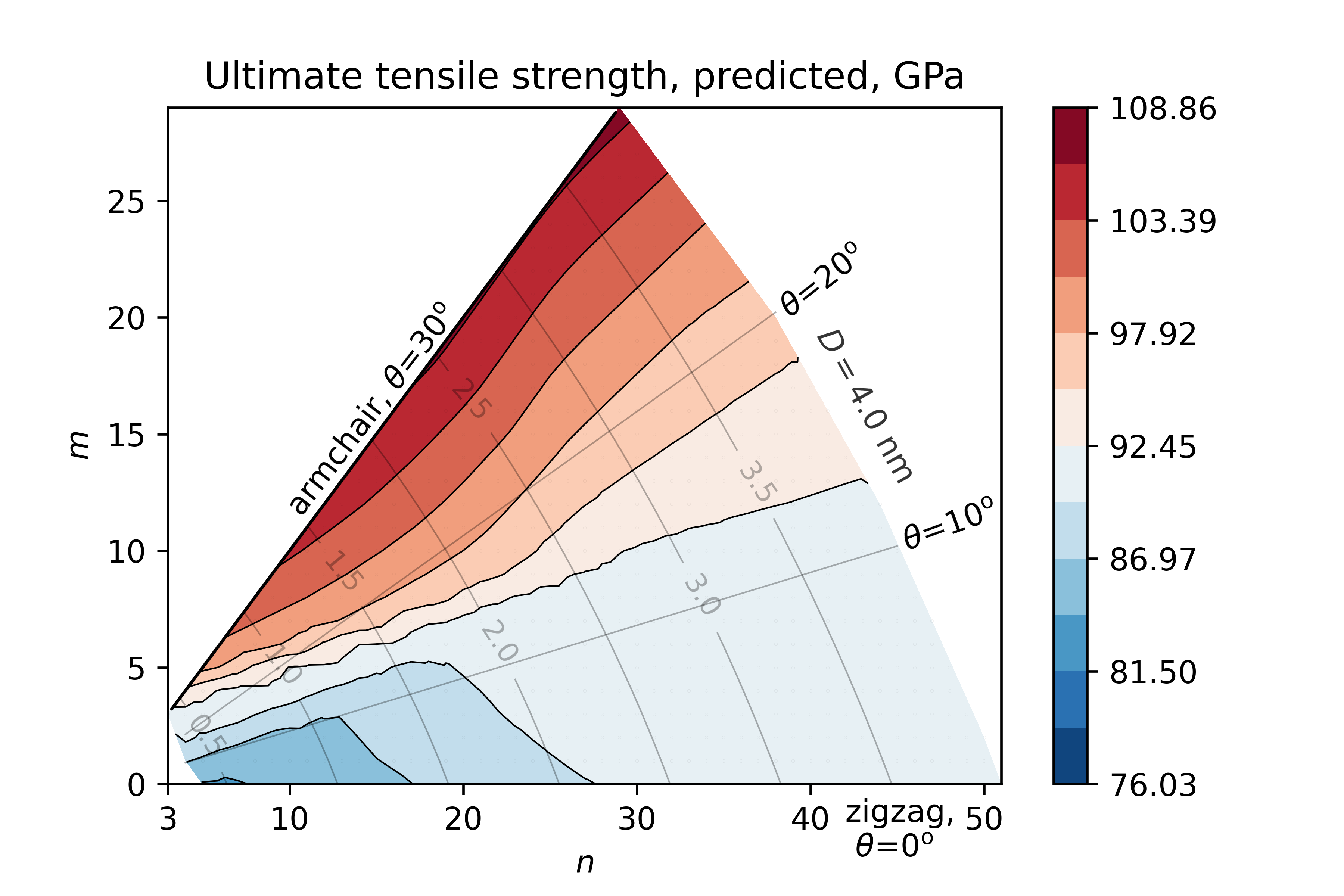}}
	\subcaptionbox {\label{fig:App_f25IF2_MD_UTS}}  {\includegraphics[scale=0.43]{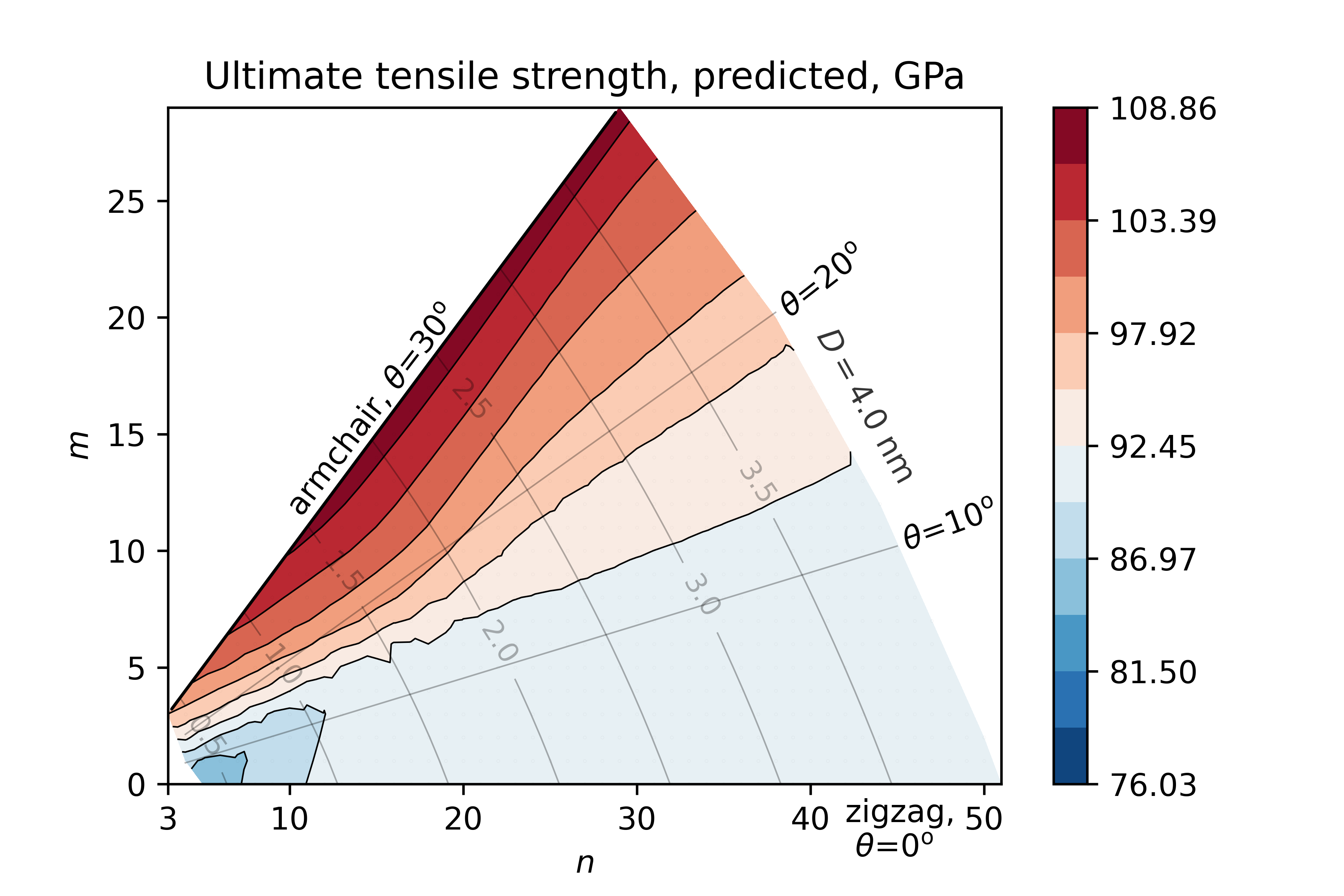}}
	\\	
	\subcaptionbox {\label{fig:App_f50IF2_ML_UTS}}  {\includegraphics[scale=0.43]{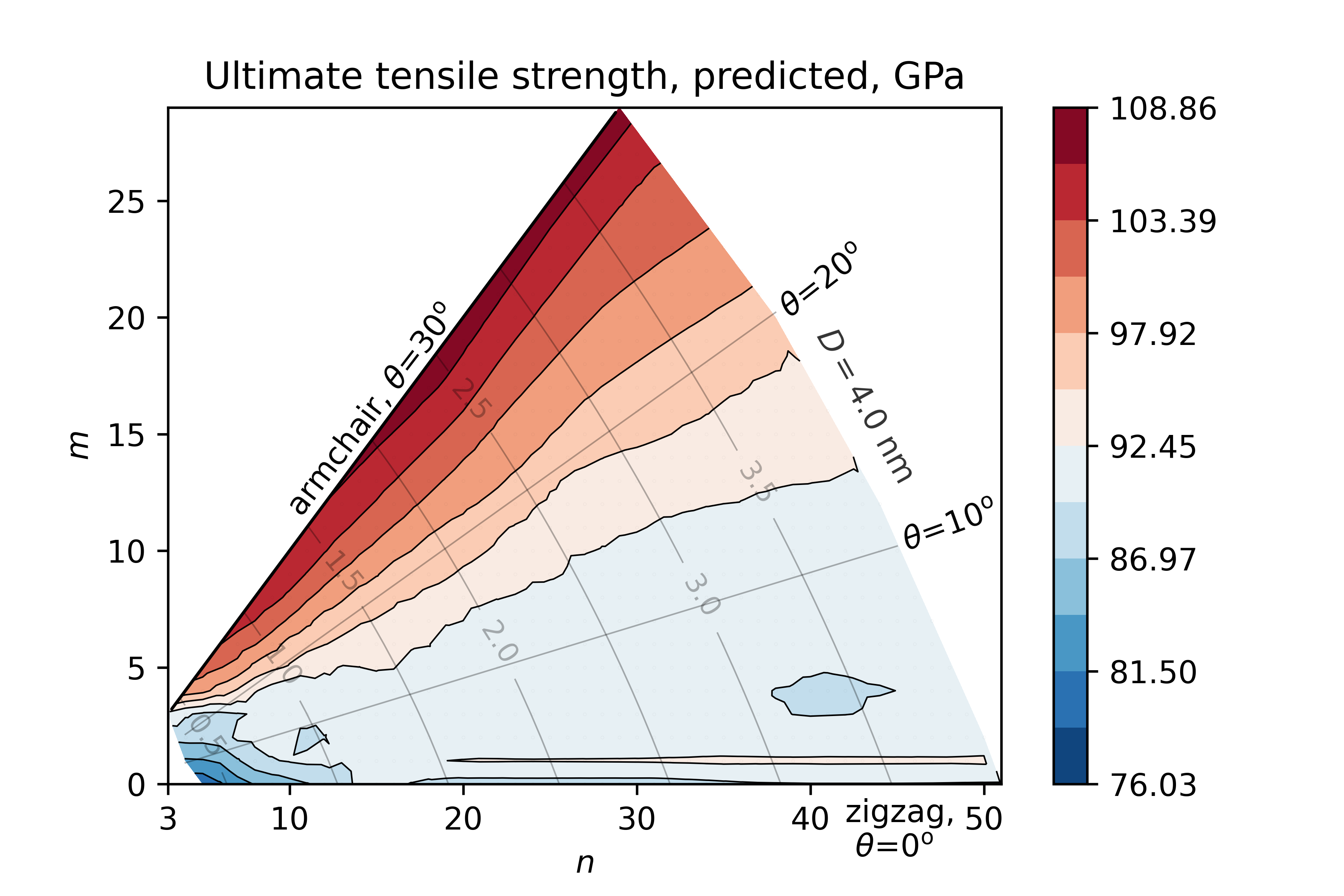}}
	\subcaptionbox {\label{fig:App_f75IF2_ML_UTS}}  {\includegraphics[scale=0.43]{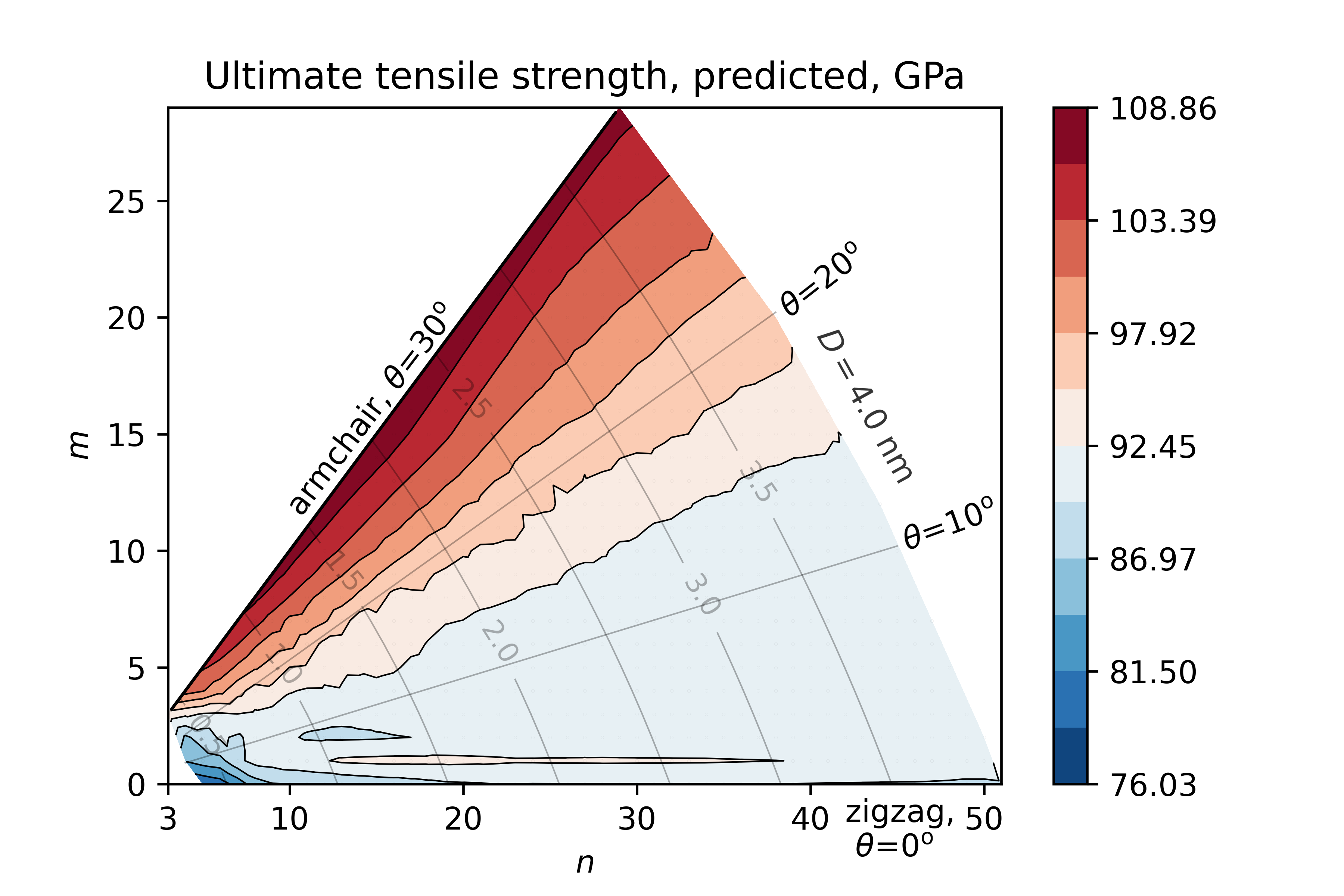}}
	\\
	\caption{Ultimate tensile strength of SWCNTs as predicted by DNN on a single set of results (Set 2) for different set sizes/fractions.  Contour limits are adjusted to MD averaged results, Fig.~\ref{fig:MD}. (a, b, c, d) Set fractions/$R^2$:10\%/0.946, 25\%/0.948, 50\%/0.970, 75\%/0.963.}	
	\label{fig:AppfracML_UTS}
\end{figure}

\begin{figure}[htp]
	\centering
    \subcaptionbox {\label{fig:App_f10IF2_ML_FS}}  {\includegraphics[scale=0.43]{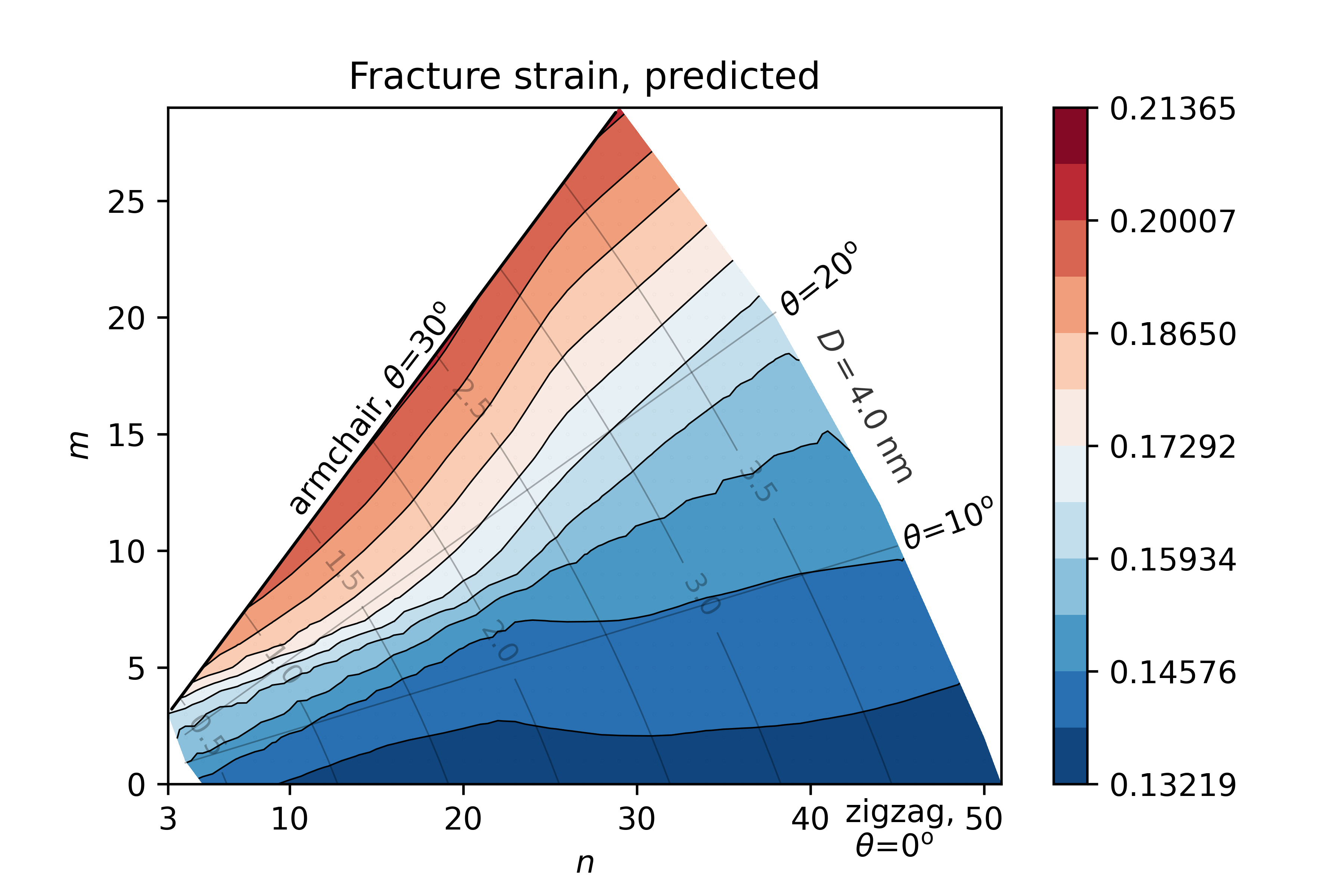}}
	\subcaptionbox {\label{fig:App_f25IF2_ML_FS}}  {\includegraphics[scale=0.43]{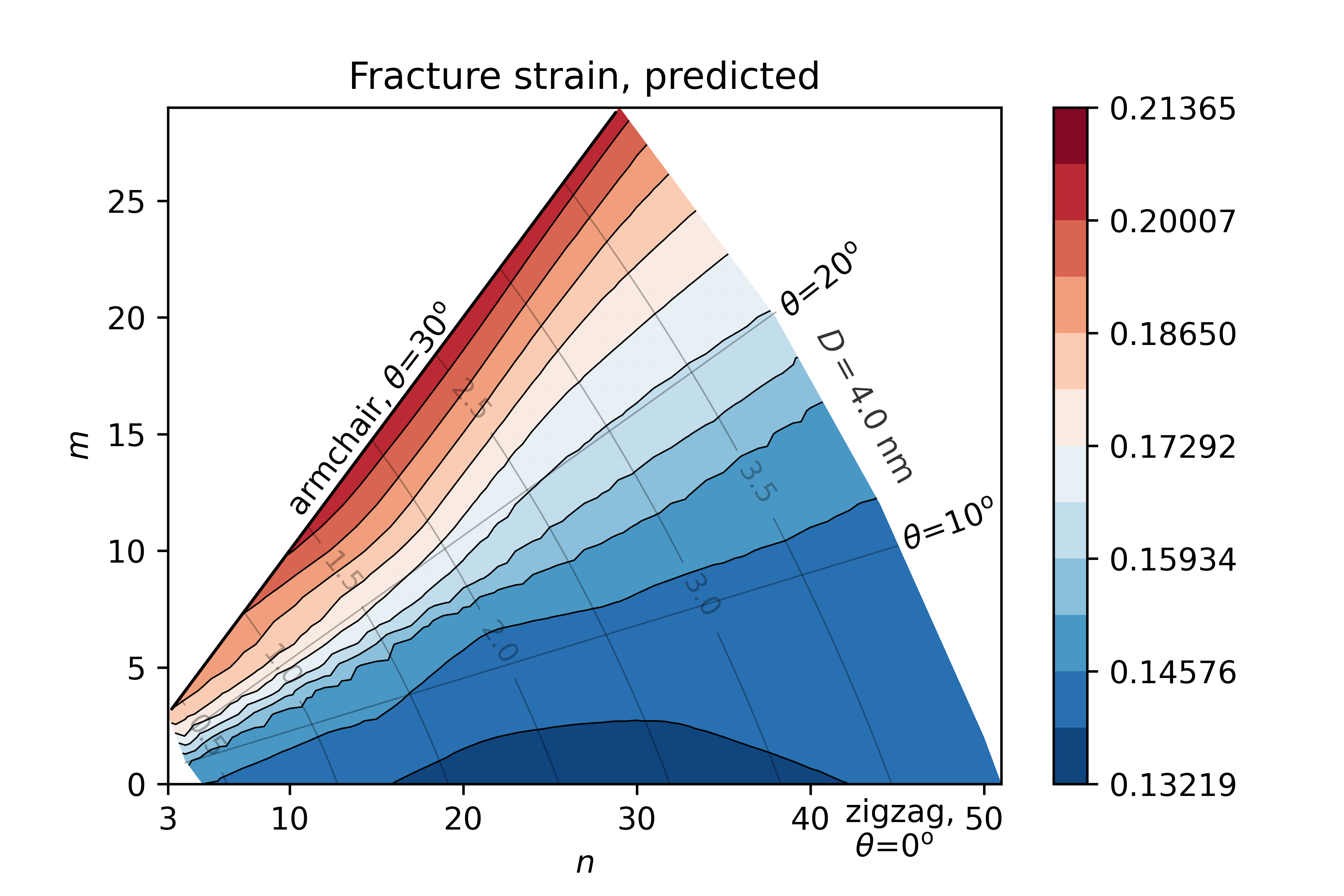}}
	\\	
	\subcaptionbox {\label{fig:App_f50IF2_ML_FS}}  {\includegraphics[scale=0.43]{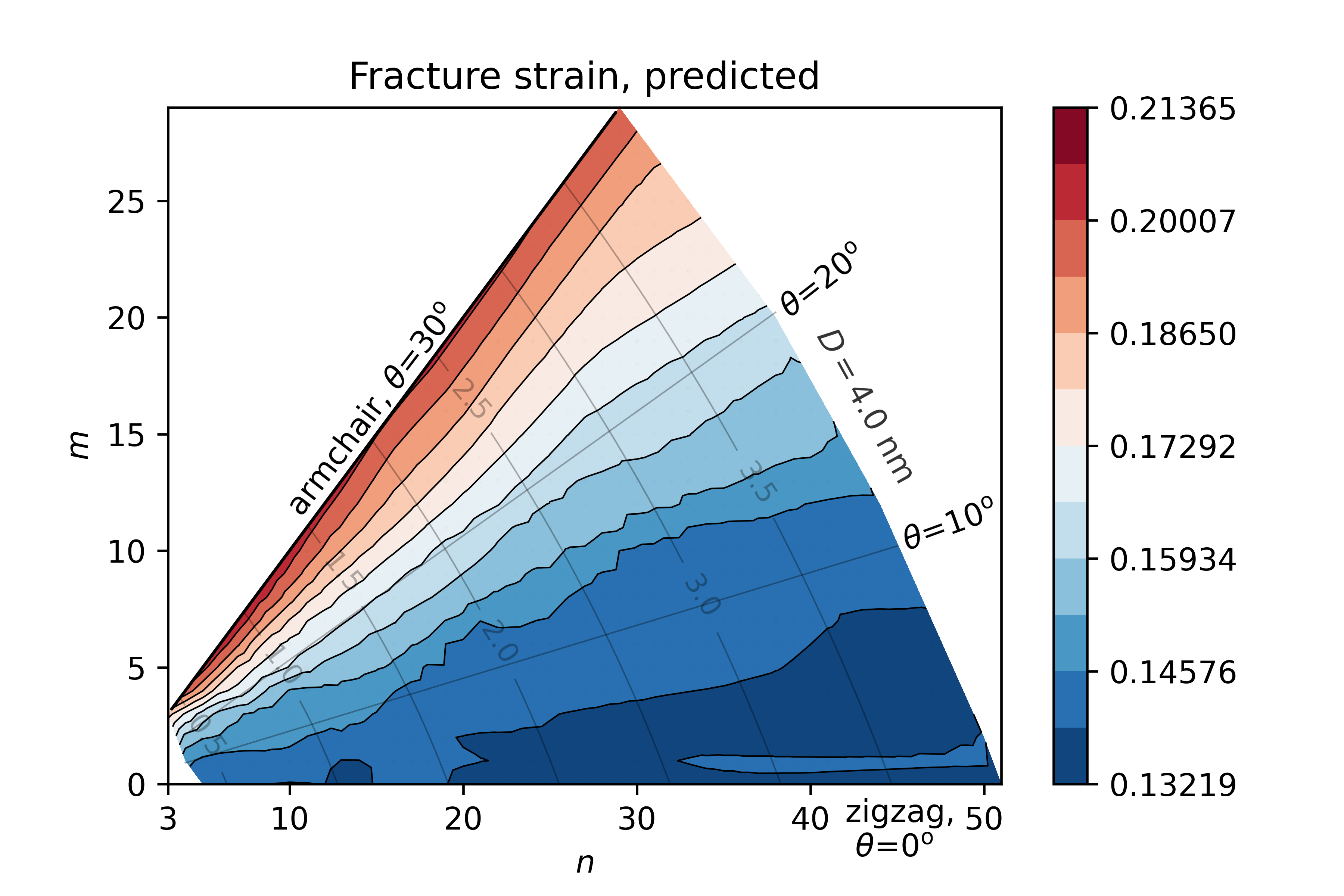}}
	\subcaptionbox {\label{fig:App_f75IF2_ML_FS}}  {\includegraphics[scale=0.43]{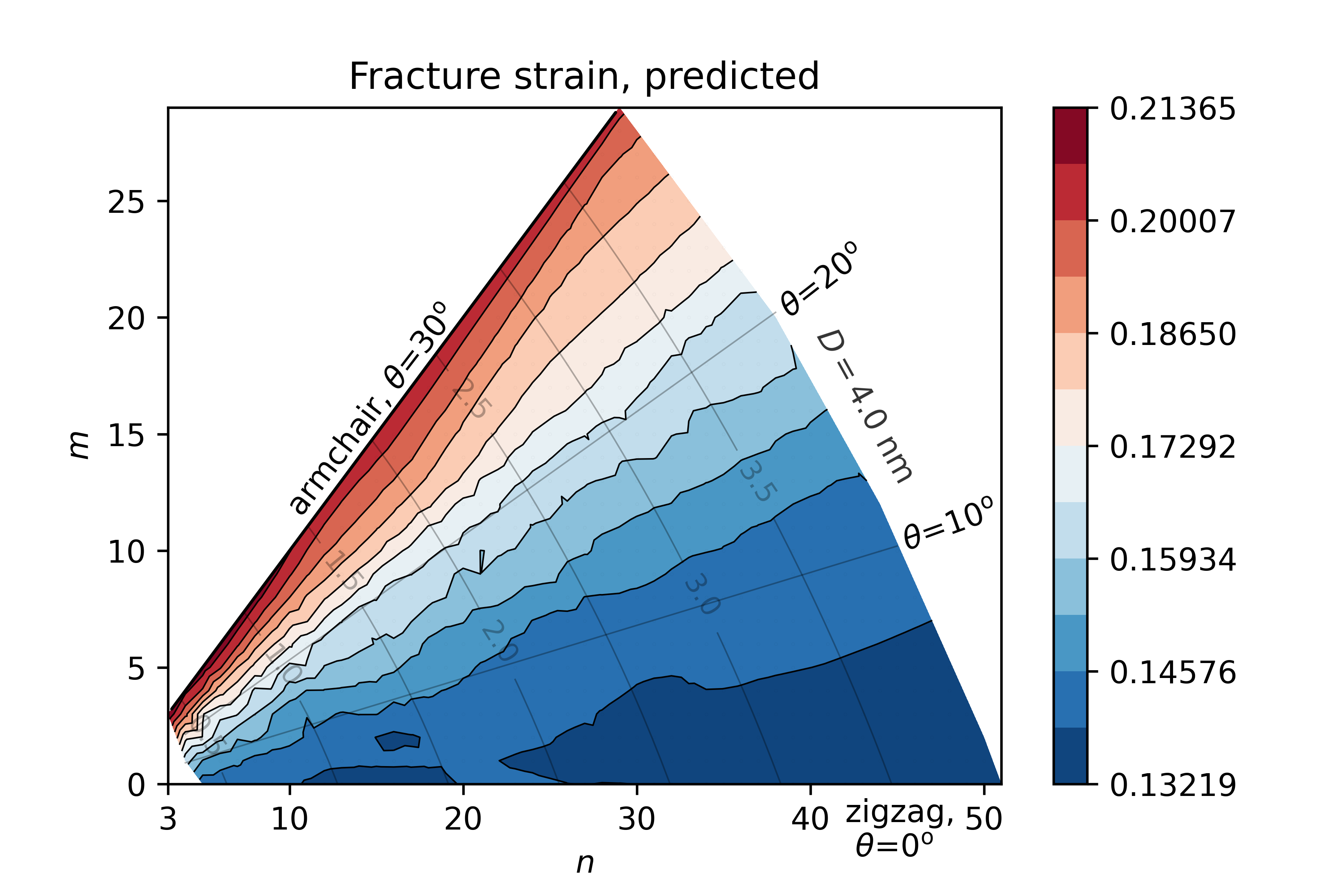}}
	\\
	\caption{Fracture strain of SWCNTs as predicted by DNN on a single set of results (Set 2) for different set sizes/fractions.  Contour limits are adjusted to MD averaged results, Fig.~\ref{fig:MD}. (a, b, c, d) Set fractions/$R^2$: 10\%/0.974, 25\%/0.985, 50\%/0.988, 75\%/0.977. }	
	\label{fig:AppfracML_FS}
\end{figure}

\begin{figure}[htp]
	\centering
	\subcaptionbox {\label{fig:App_f10IF2_ML_Eavg}}  {\includegraphics[scale=0.43]{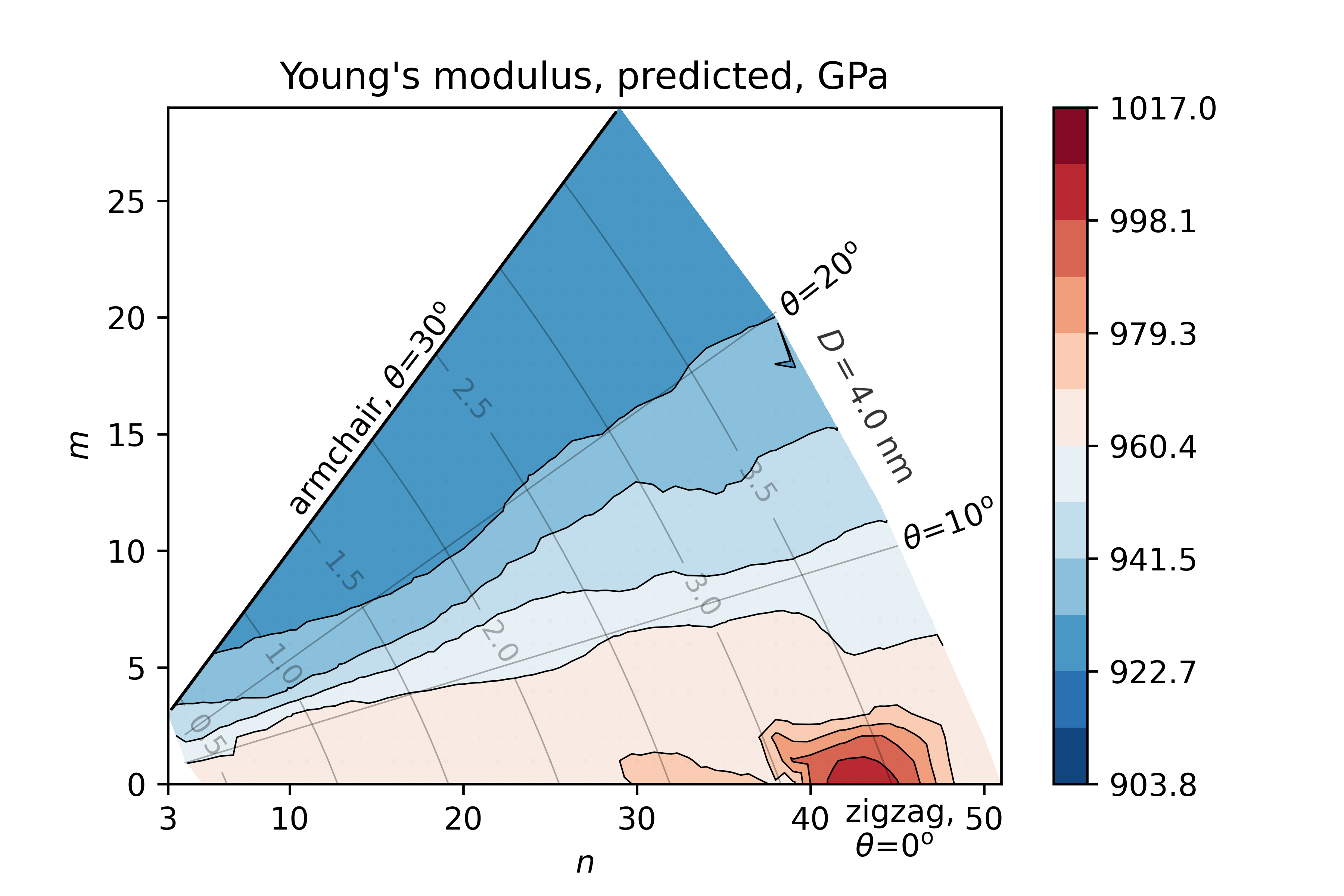}}	
	\subcaptionbox {\label{fig:App_f25IF2_ML_Eavg}}  {\includegraphics[scale=0.43]{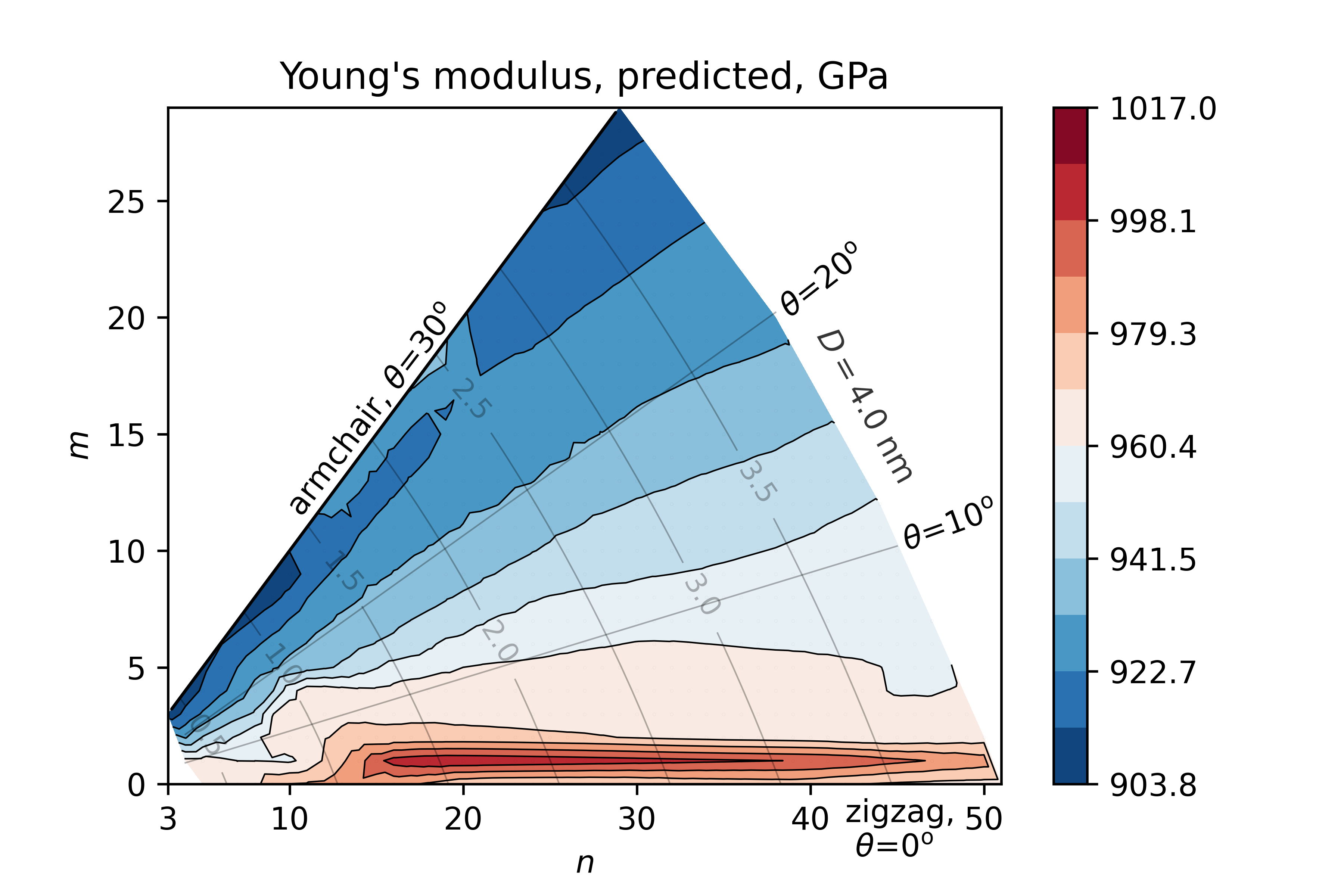}}
	\\
	\subcaptionbox {\label{fig:App_f50IF2_ML_Eavg}}  {\includegraphics[scale=0.43]{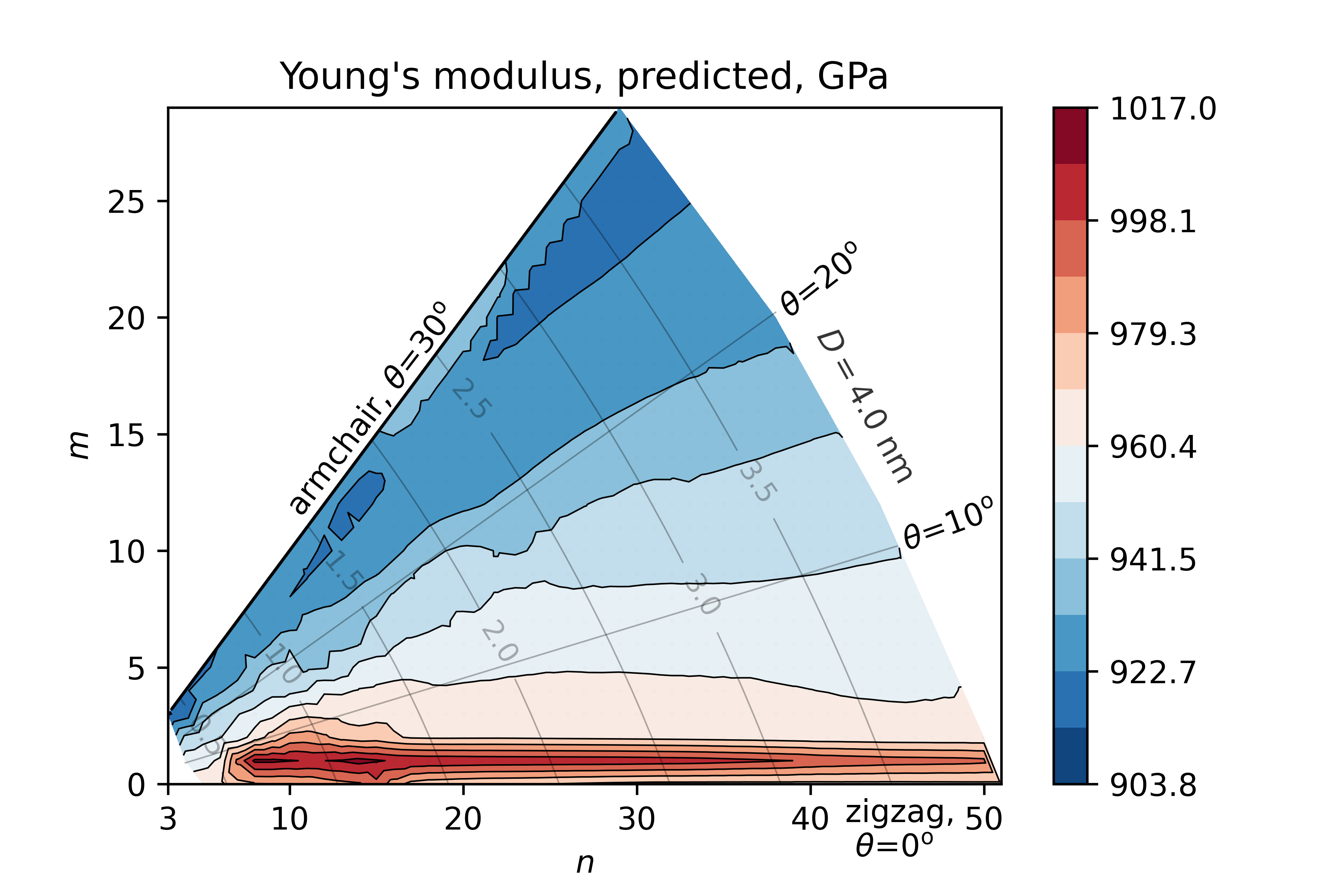}}
	\subcaptionbox {\label{fig:App_f75IF2_ML_Eavg}}  {\includegraphics[scale=0.43]{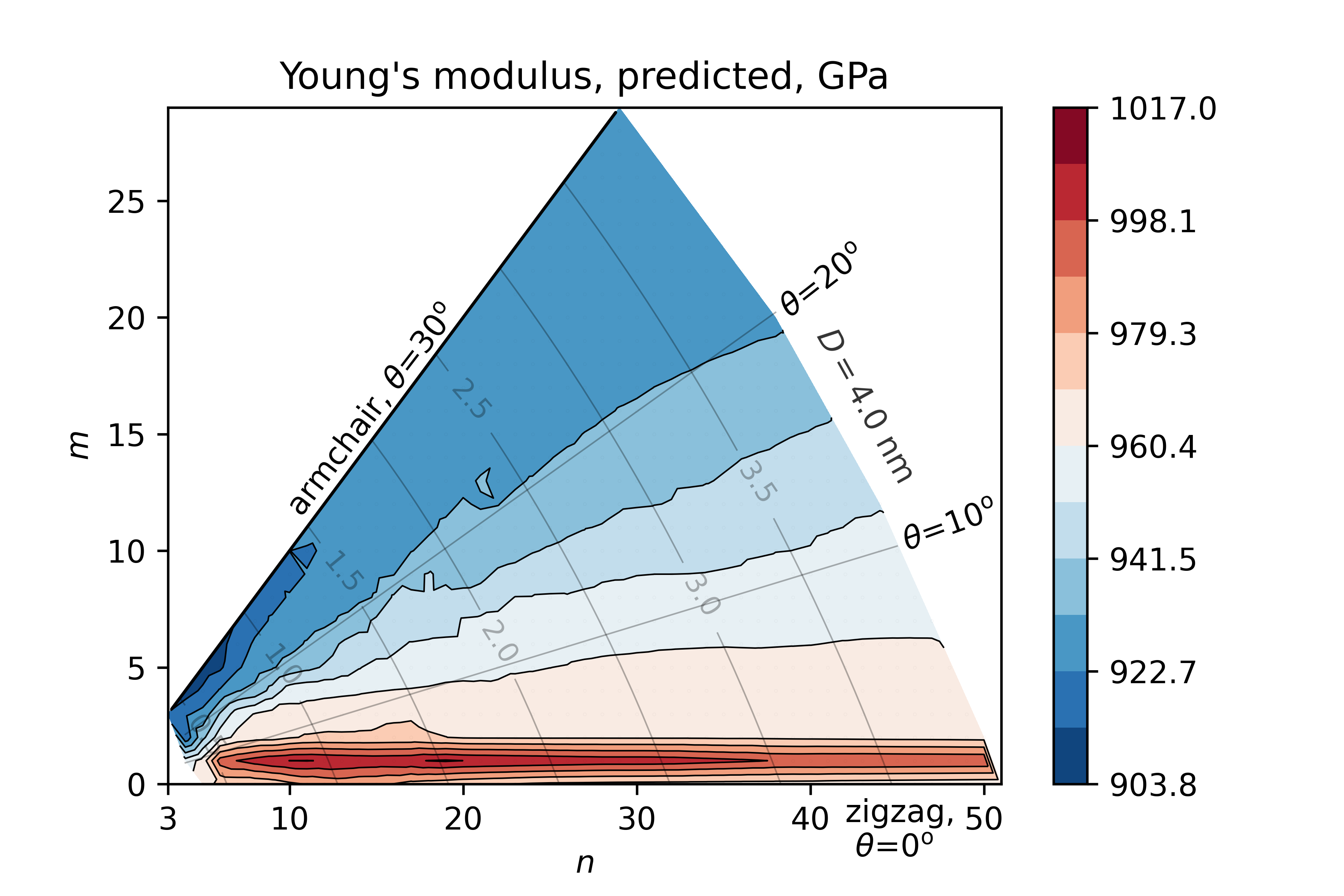}}
	\\
	\caption{Young's modulus of SWCNTs as predicted by DNN on the set with averaged results for different set sizes/fractions.  Contour limits are adjusted to MD averaged results, Fig.~\ref{fig:MD}. (a, b, c, d) Set fractions/$R^2$: 10\%/0.751, 25\%/0.899, 50\%/0.935, 75\%/0.948. }	
	\label{fig:AppfracML_YMavg}
\end{figure}

\begin{figure}[htp]
	\centering
	\subcaptionbox {\label{fig:App_f10IF2_ML_Pavg}}  {\includegraphics[scale=0.43]{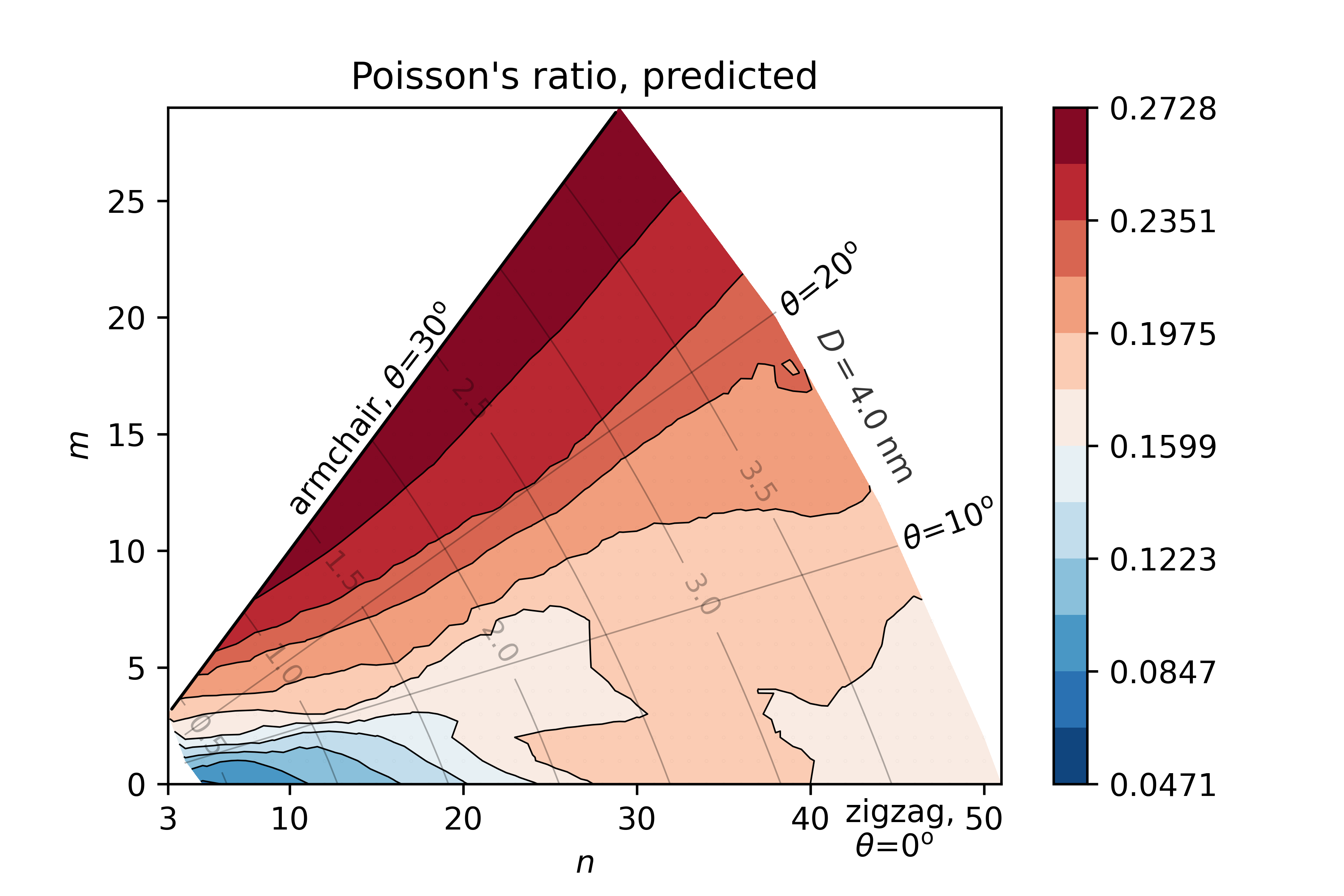}}
	\subcaptionbox {\label{fig:App_f25IF2_ML_Pavg}}  {\includegraphics[scale=0.43]{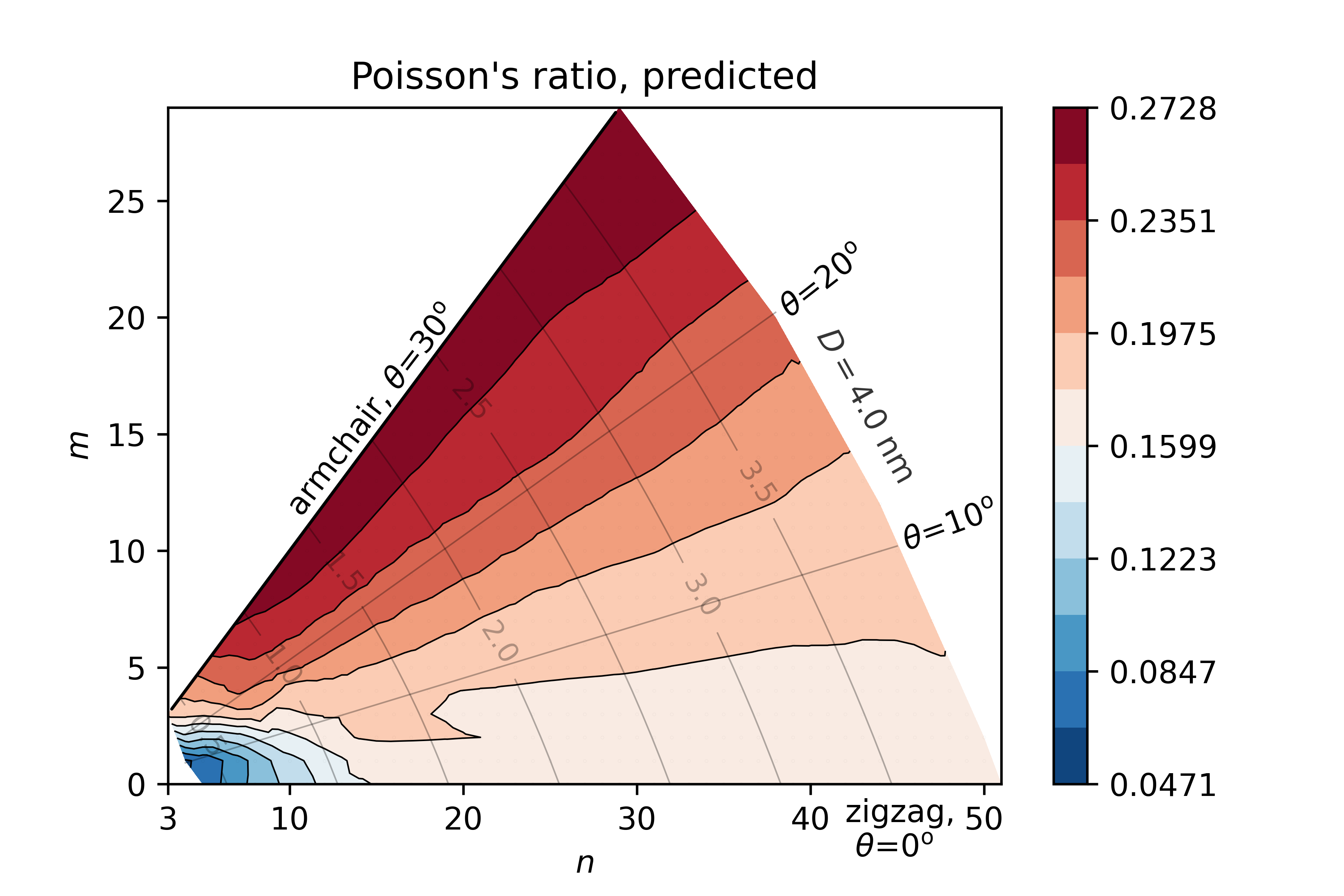}}
	\\	
	\subcaptionbox {\label{fig:App_f50IF2_ML_Pavg}}  {\includegraphics[scale=0.43]{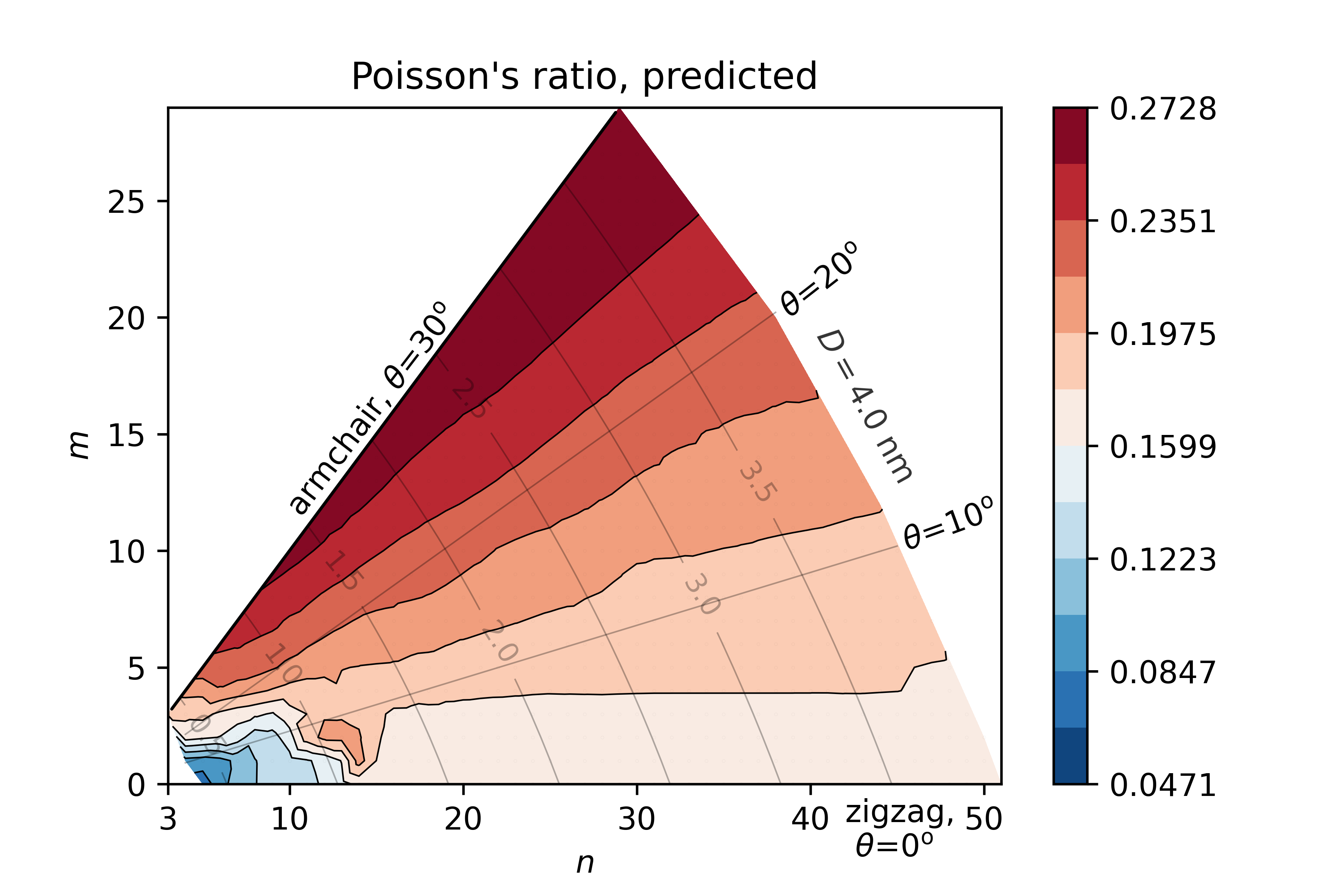}}
	\subcaptionbox {\label{fig:App_f75IF2_ML_Pavg}}  {\includegraphics[scale=0.43]{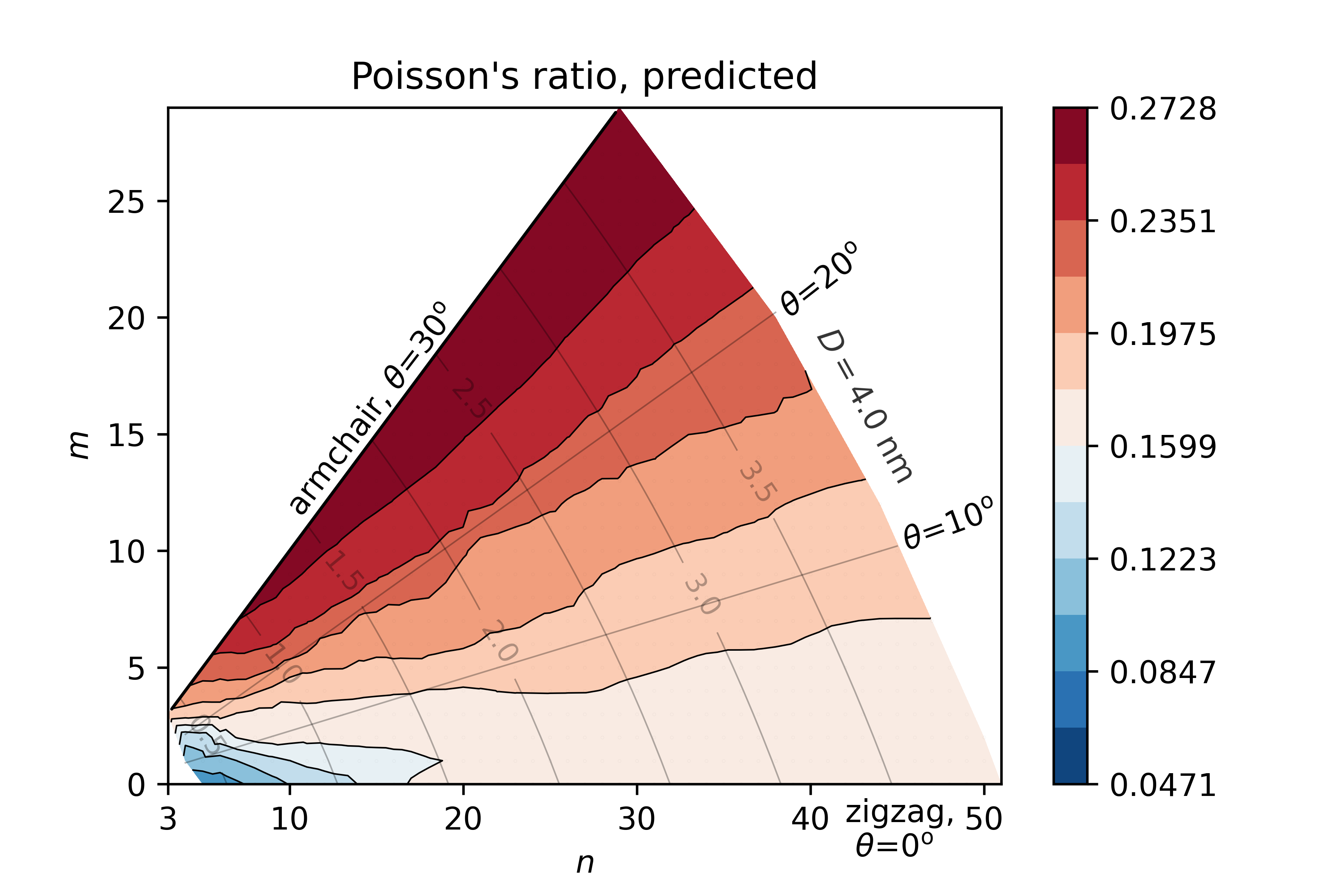}}
	\\
	\caption{Poisson's ratio of SWCNTs as predicted by DNN on the set with averaged results for different set sizes/fractions.  Contour limits are adjusted to MD averaged results, Fig.~\ref{fig:MD}. (a, b, c, d) Set fractions/$R^2$: 10\%/0.888, 25\%/0.951, 50\%/0.939, 75\%/0.949. }		
	\label{fig:AppfracML_Pavg}
\end{figure}

\begin{figure}[htp]
	\centering
	\subcaptionbox {\label{fig:App_f10IF2_MD_UTSavg}}  {\includegraphics[scale=0.43]{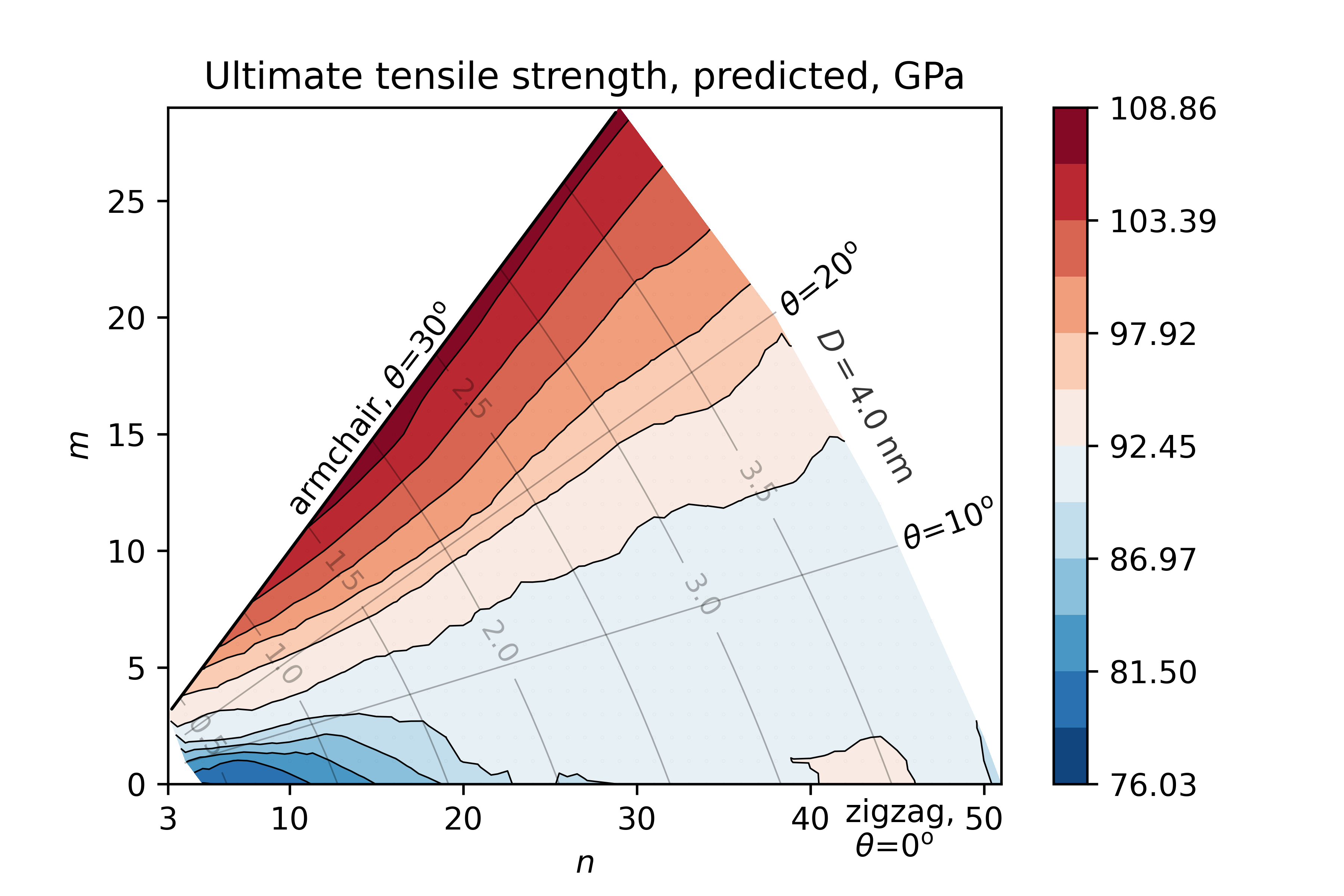}}
	\subcaptionbox {\label{fig:App_f25IF2_MD_UTSavg}}  {\includegraphics[scale=0.43]{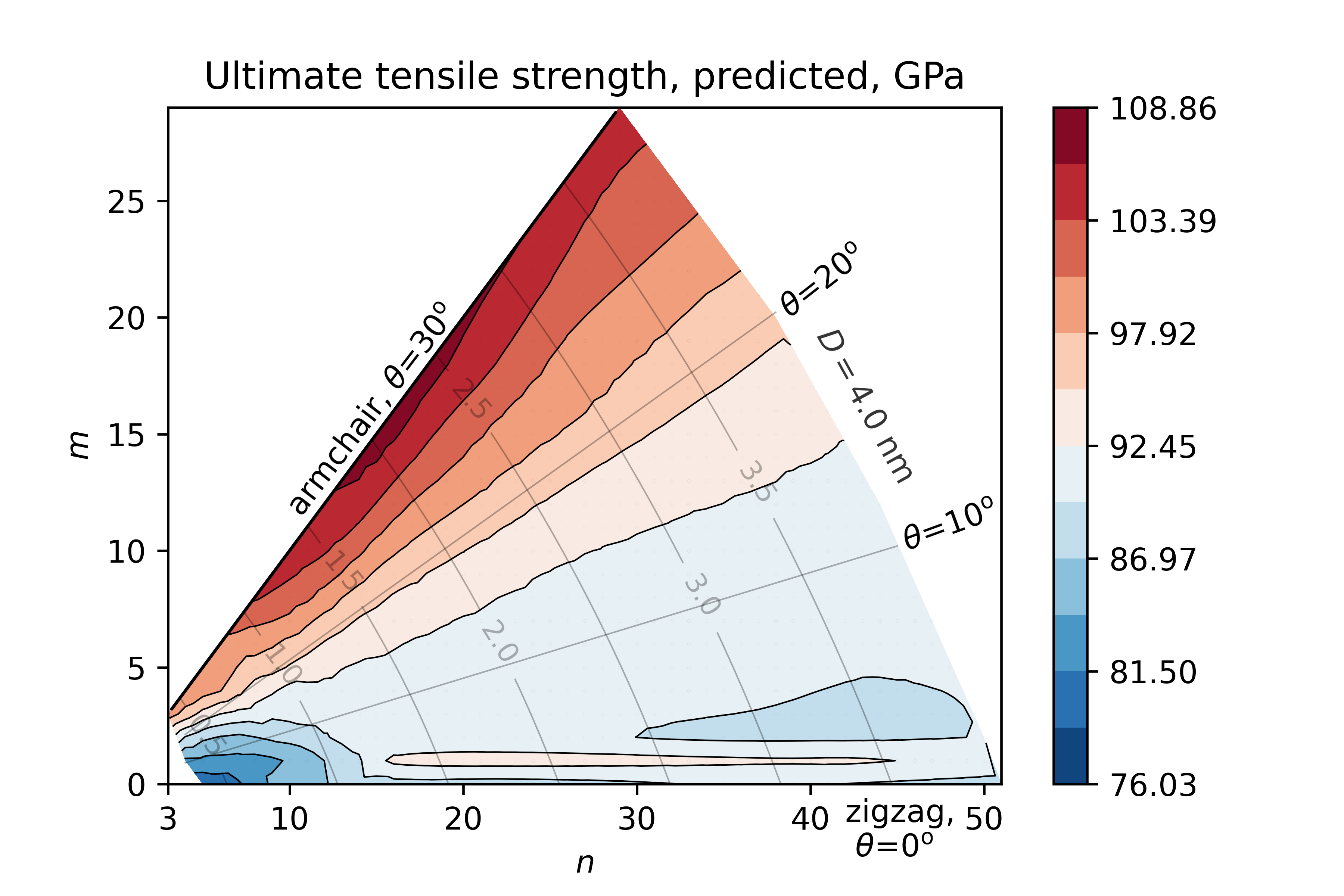}}
	\\	
	\subcaptionbox {\label{fig:App_f50IF2_ML_UTSavg}}  {\includegraphics[scale=0.43]{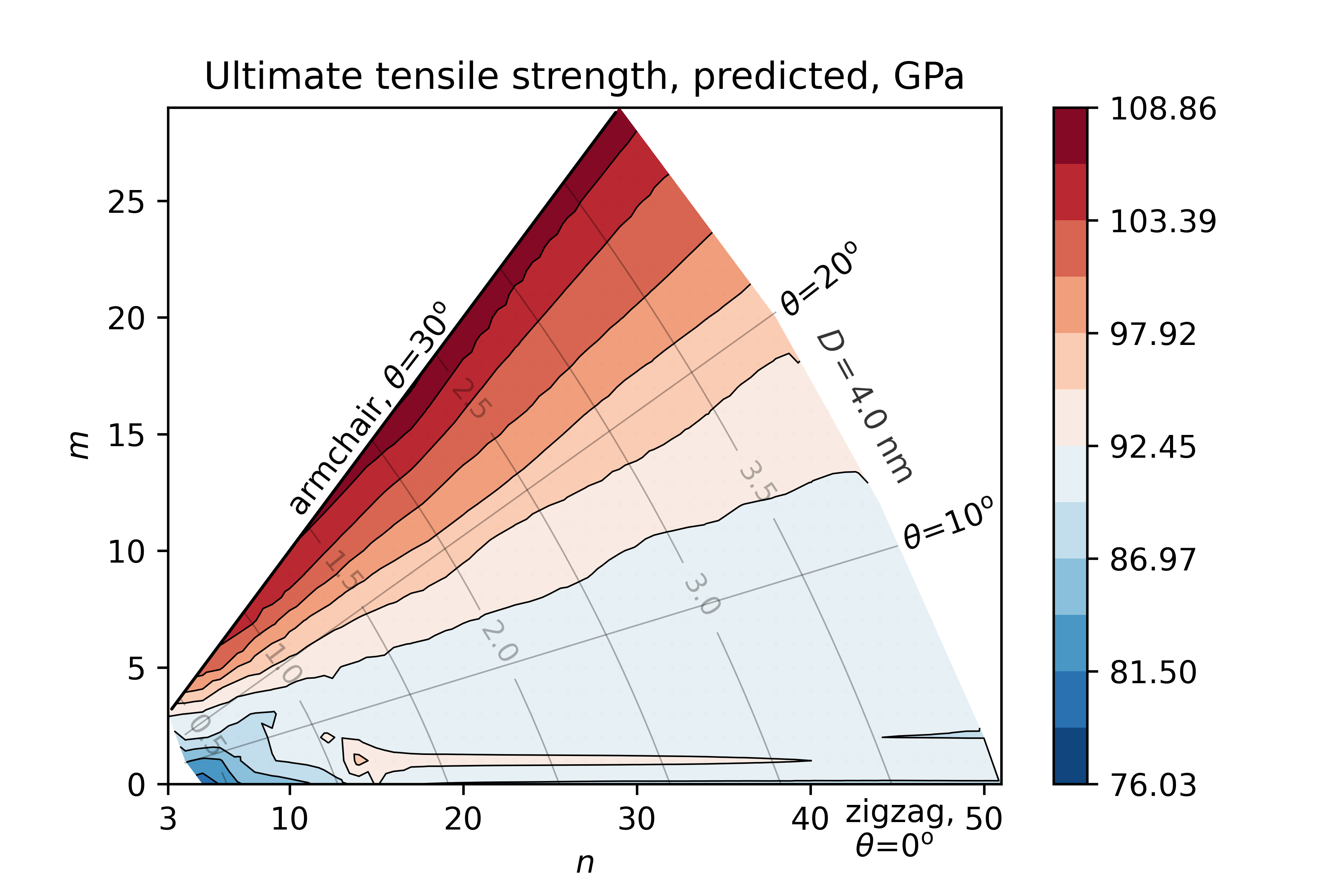}}
	\subcaptionbox {\label{fig:App_f75IF2_ML_UTSavg}}  {\includegraphics[scale=0.43]{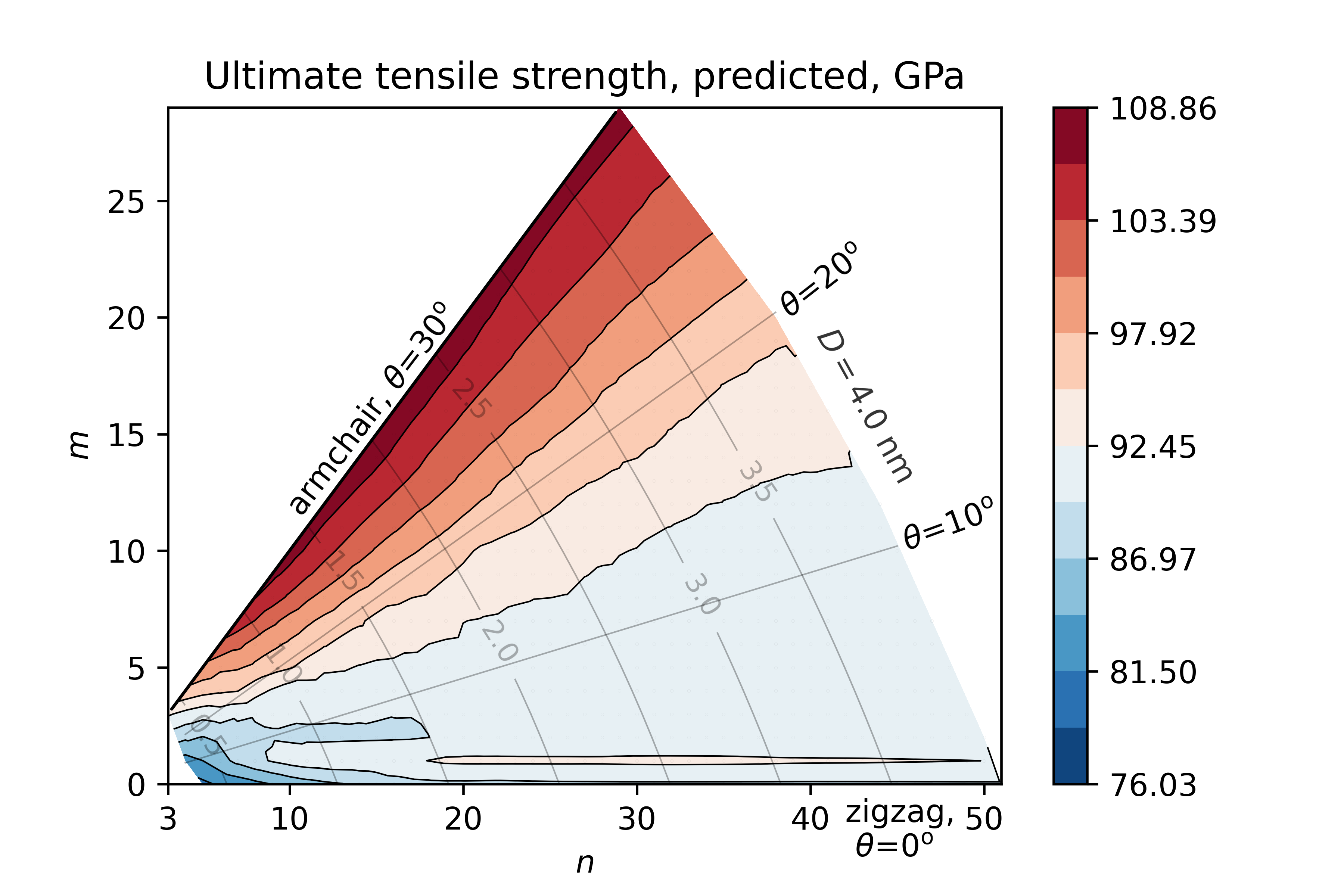}}
	\\
	\caption{Ultimate tensile strength of SWCNTs as predicted by DNN on the set with averaged results for different set sizes/fractions.  Contour limits are adjusted to MD averaged results, Fig.~\ref{fig:MD}. (a, b, c, d) Set fractions/$R^2$: 10\%/0.933, 25\%/0.965, 50\%/0.980, 75\%/0.987. }	
	\label{fig:AppfracML_UTSavg}
\end{figure}

\begin{figure}[htp]
	\centering
	\subcaptionbox {\label{fig:App_f10IF2_ML_FSavg}}  {\includegraphics[scale=0.43]{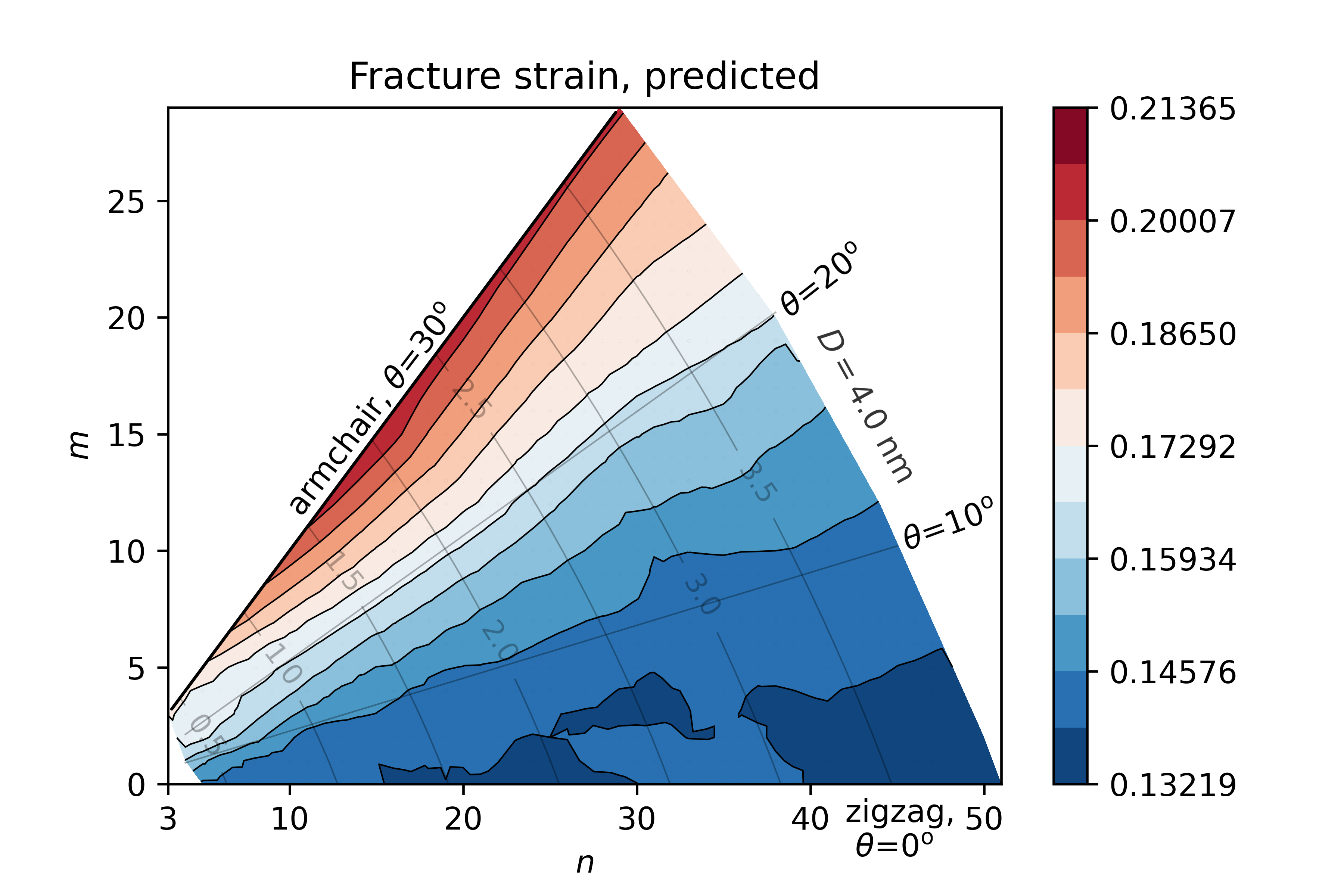}}
	\subcaptionbox {\label{fig:App_f25IF2_ML_FSavg}}  {\includegraphics[scale=0.43]{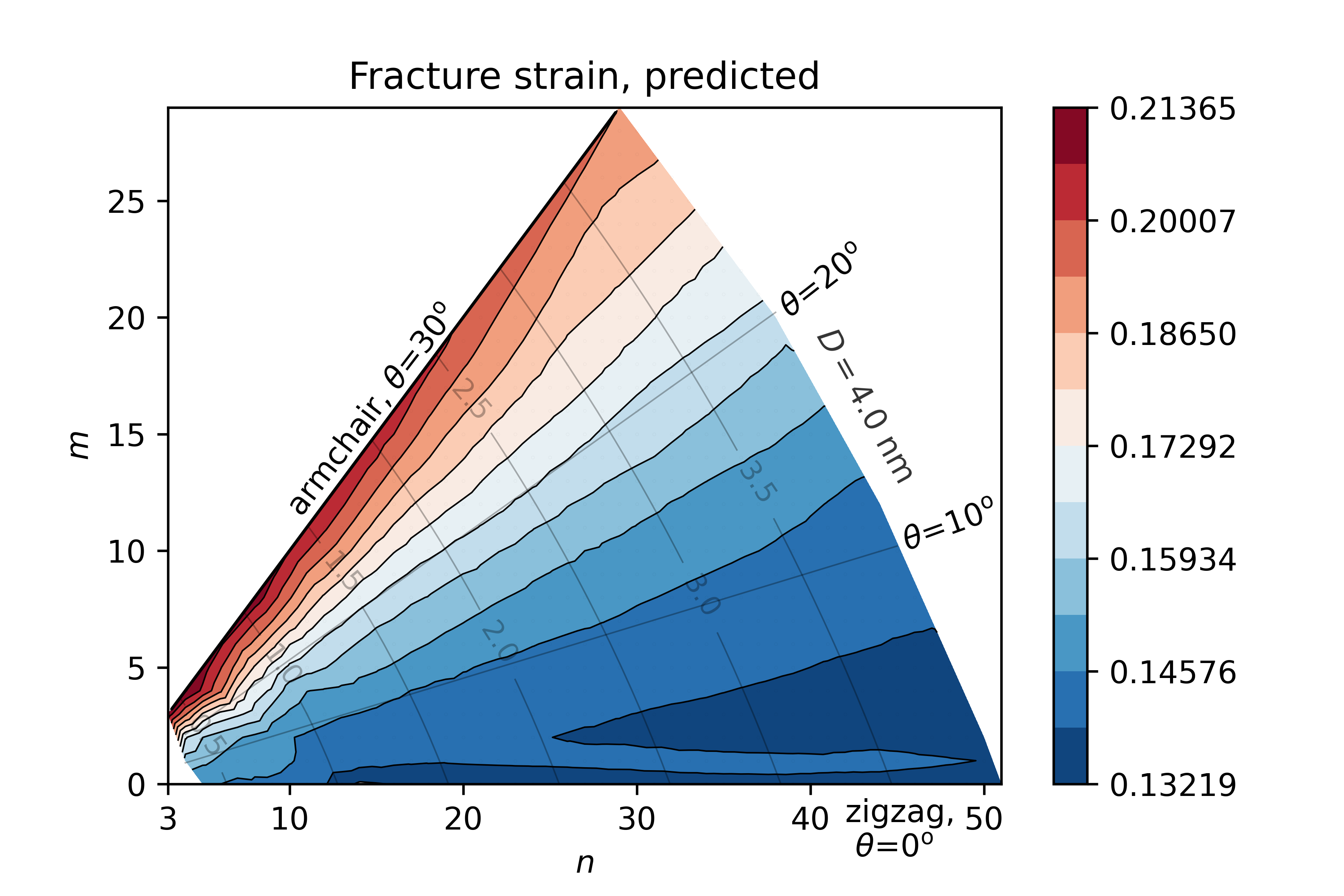}}
	\\	
	\subcaptionbox {\label{fig:App_f50IF2_ML_FSavg}}  {\includegraphics[scale=0.43]{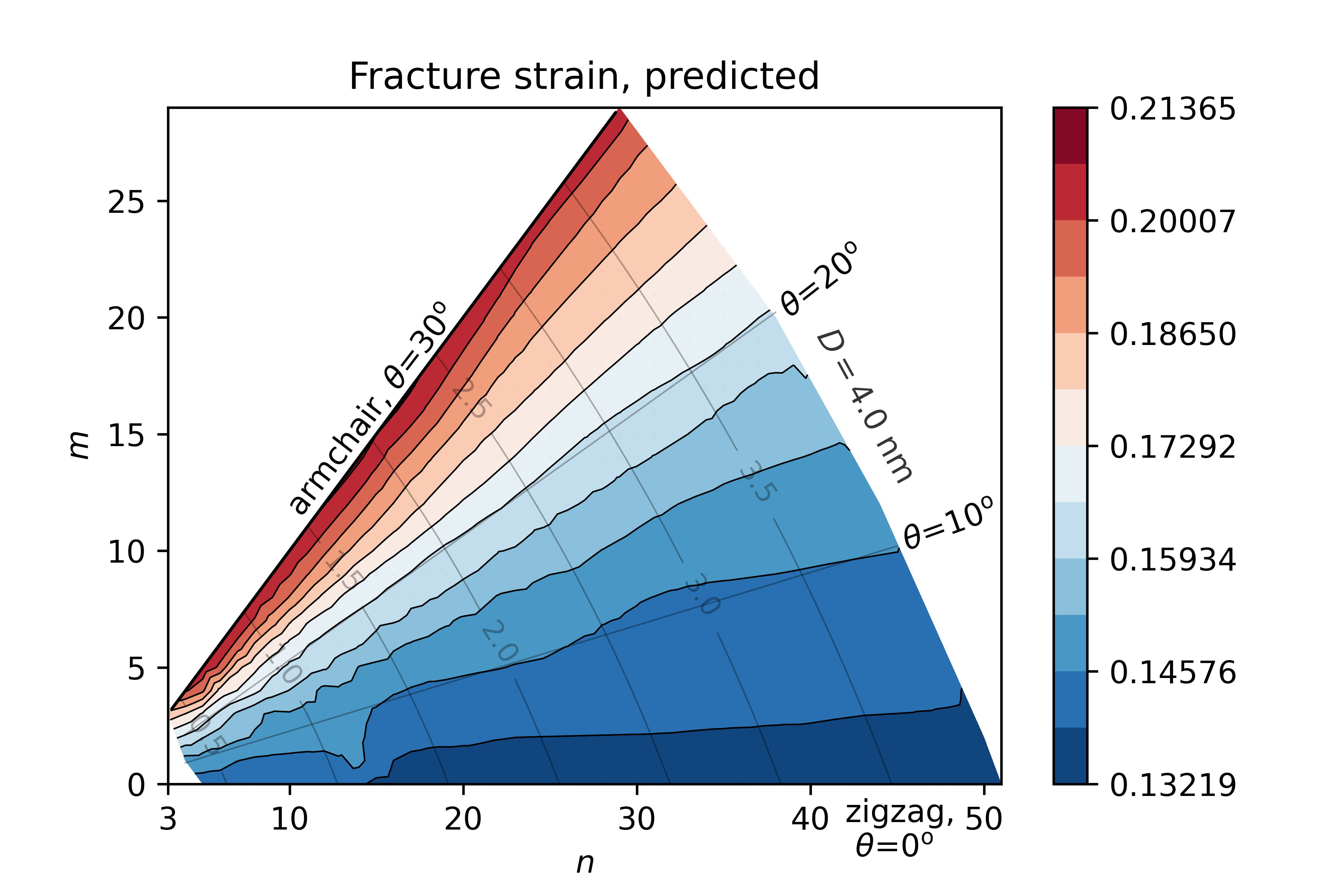}}
	\subcaptionbox {\label{fig:App_f75IF2_ML_FSavg}}  {\includegraphics[scale=0.43]{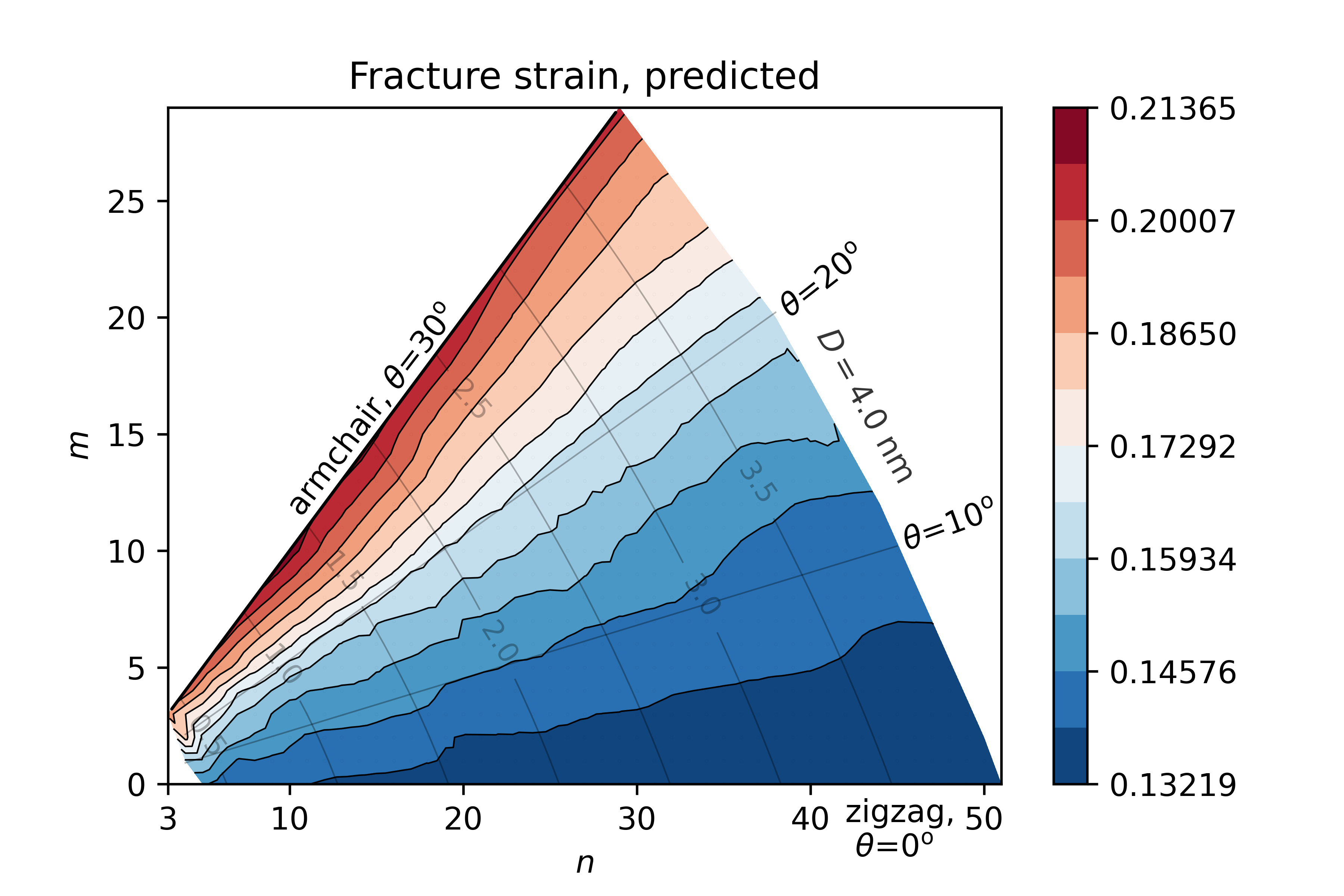}}
	\\
	\caption{Fracture strain of SWCNTs as predicted by DNN on the set with averaged results for different set sizes/fractions.  Contour limits are adjusted to MD averaged results, Fig.~\ref{fig:MD}. (a, b, c, d) Set fractions/$R^2$:10\%/0.963, 25\%/0.986, 50\%/0.984, 75\%/0.986.}
	\label{fig:AppfracML_FSavg}	
\end{figure}

\begin{figure}[htp]
	\centering
	\subcaptionbox {\label{fig:AppR2_E_comparison}}  {\includegraphics[scale=0.43]{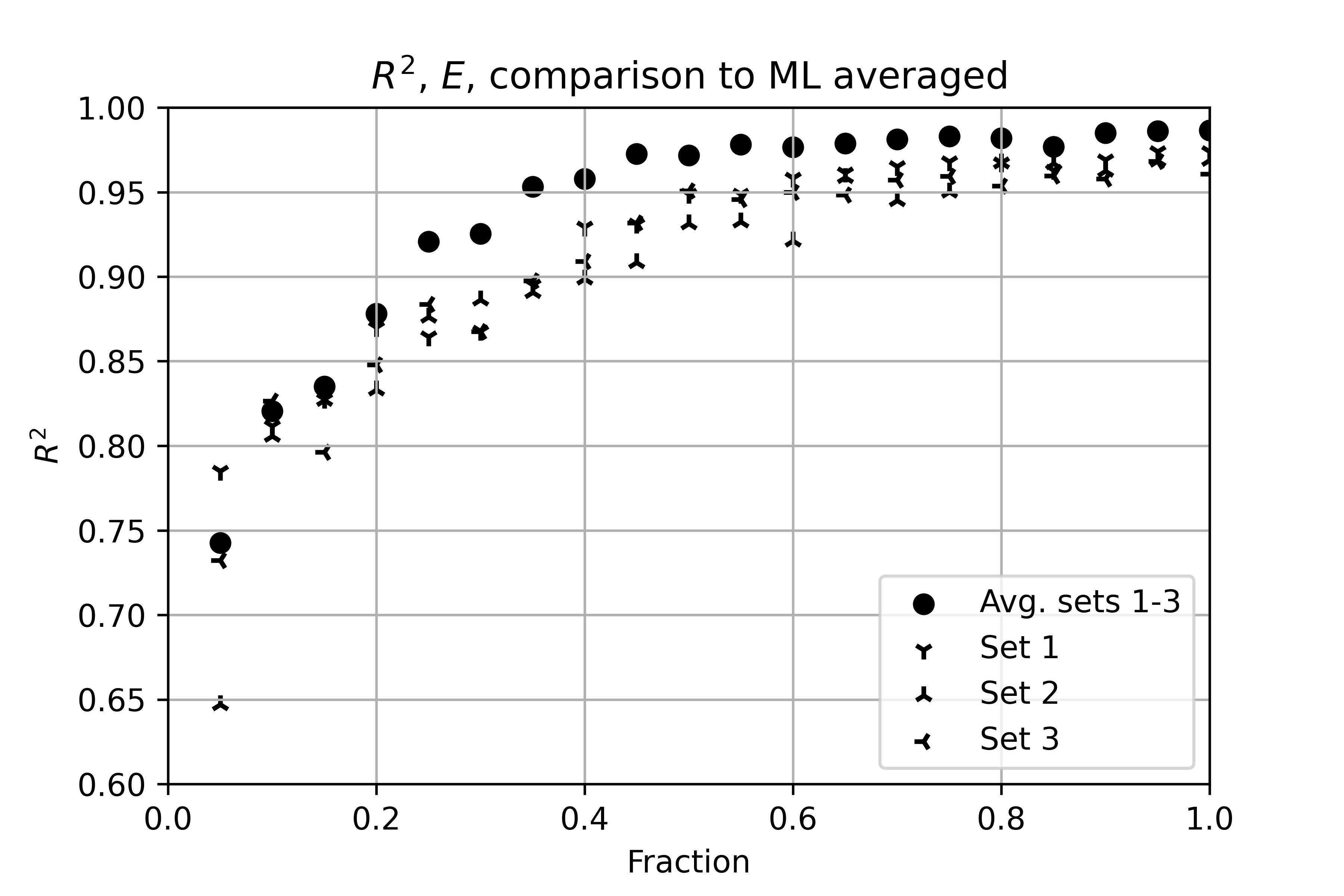}}
	\subcaptionbox {\label{fig:AppR2_P_comparison}}  {\includegraphics[scale=0.43]{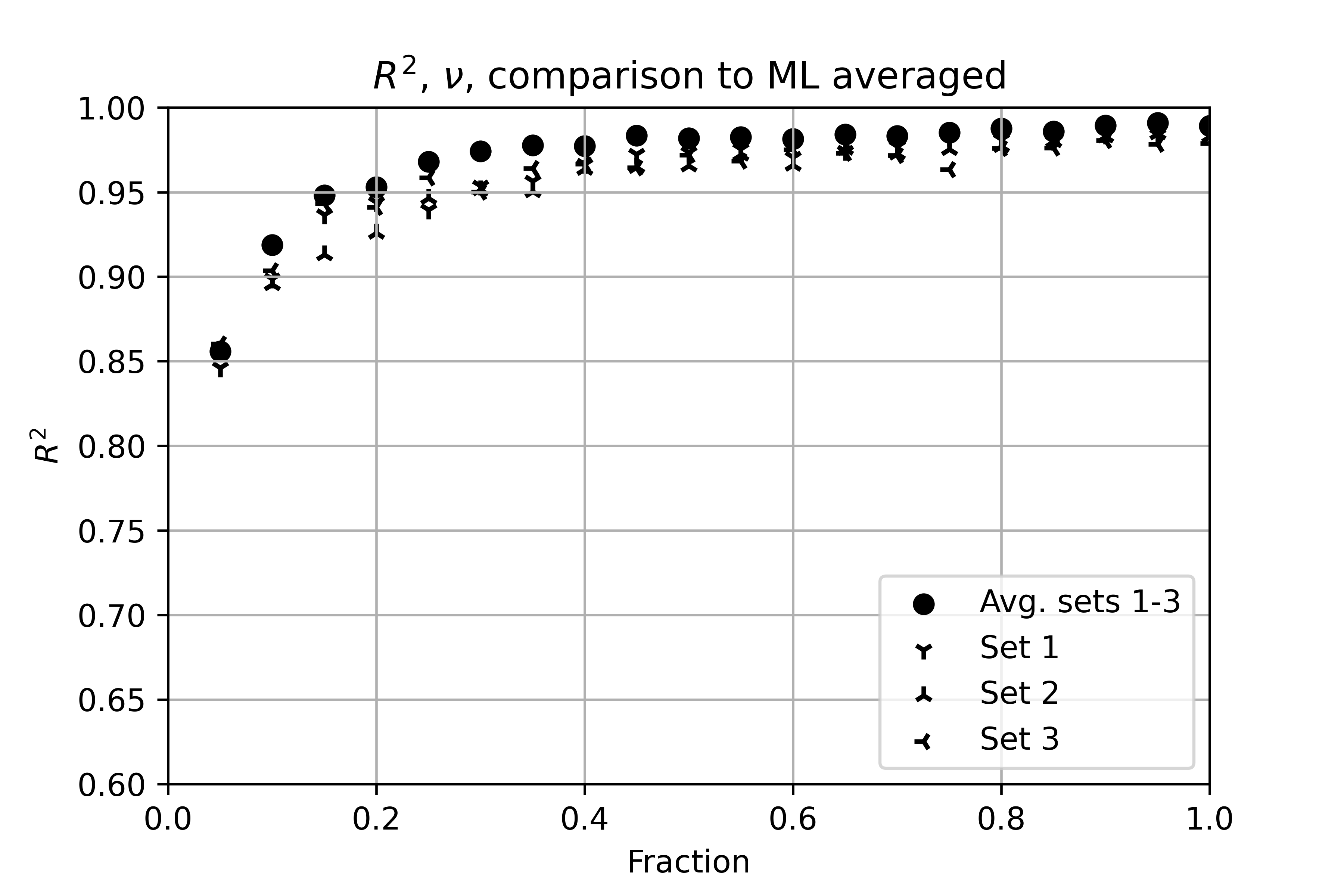}}
	\\
	\subcaptionbox {\label{fig:AppR2_UTS_comparison}}  {\includegraphics[scale=0.43]{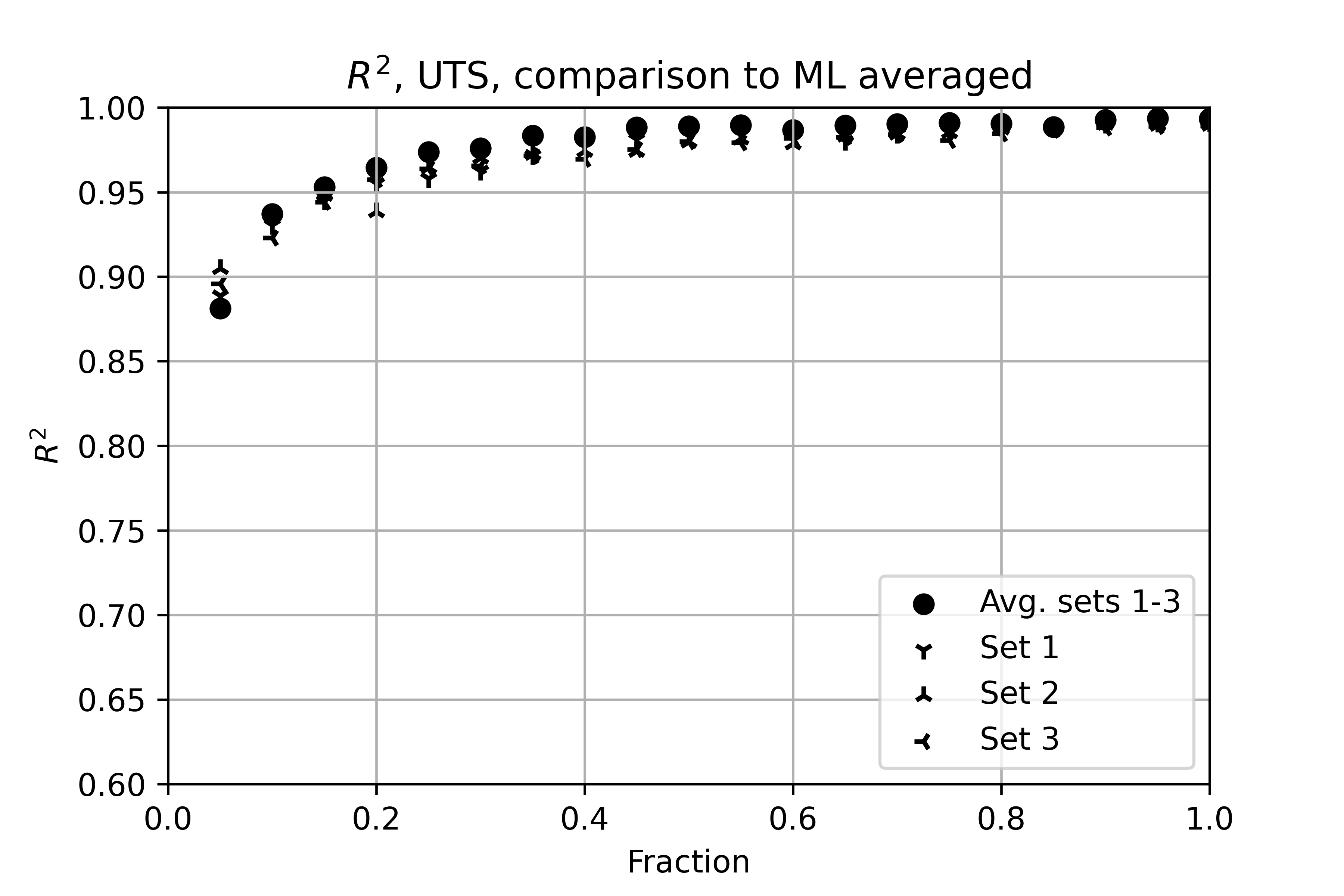}}
	\subcaptionbox {\label{fig:AppR2_FS_comparison}}  {\includegraphics[scale=0.43]{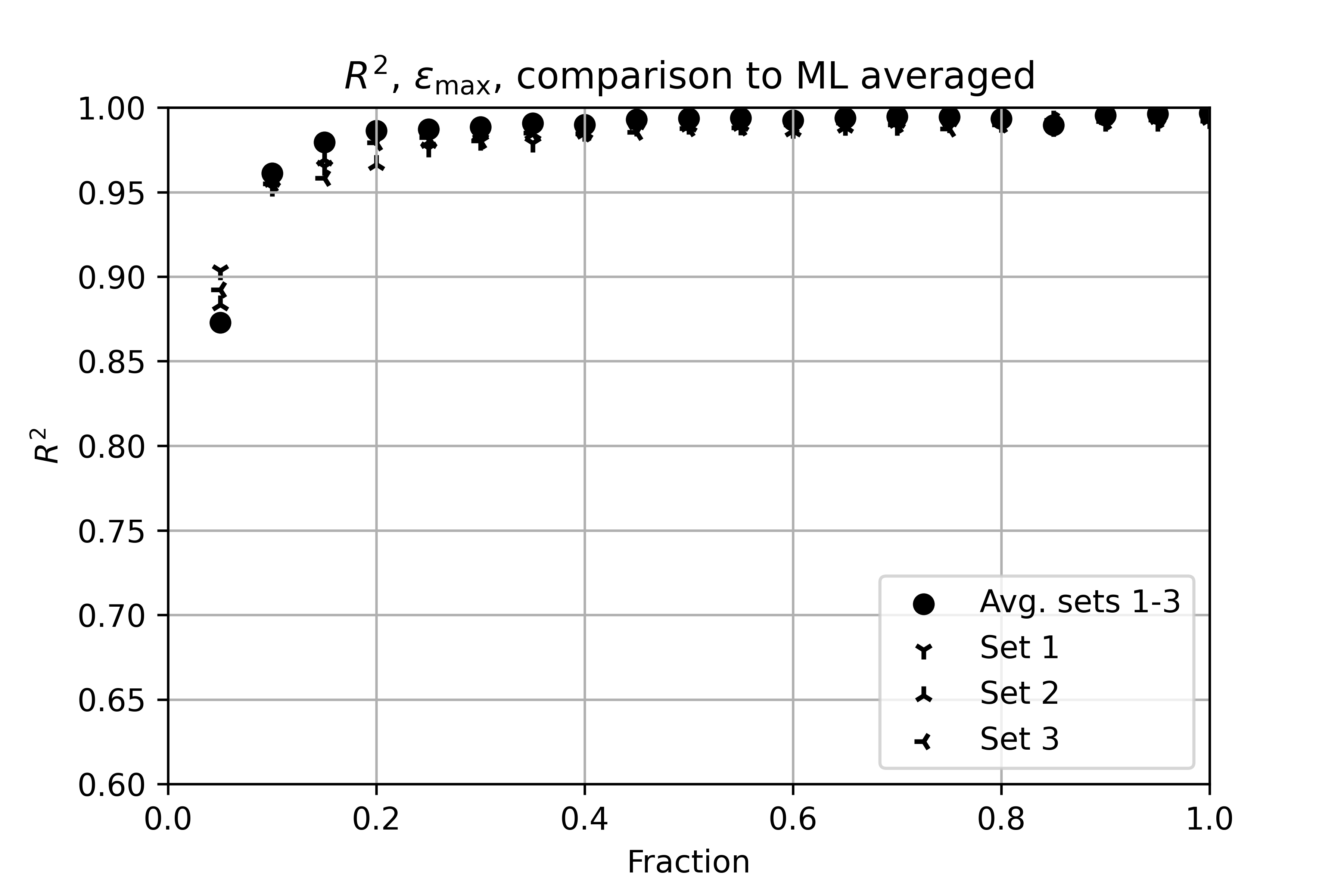}}
	\\
	\caption{Comparison of $R^2$ for different analysis set sizes to predicted DL results obtained with averaged set and individual sets 1-3. Each point is obtained as a mean value of 10 different fraction sets. (a) Young's modulus, (b) Poisson's ratio, (c) Ultimate tensile strength, (d) Fracture strain.}	
	\label{fig:AppR2_ML}
\end{figure}

\bigskip

\noindent

\bibliographystyle{elsarticle-num}

\bibliography{SmallSizeParameter}

\end{document}